\pdfoutput=1
\documentclass[a4paper,fleqn,usenatbib]{mnras}

\usepackage[T1]{fontenc}
\usepackage{ae,aecompl}

\usepackage{graphicx}	
\usepackage{amsmath}	
\usepackage{amssymb}	
\usepackage{hyperref}
\usepackage{caption}
\usepackage{multirow}
\usepackage{pdflscape}
\usepackage{makecell}

\title[(Un)-obscured star formation in cluster cores]{Quantifying the suppression of the (un)-obscured star formation in galaxy cluster cores at 0.2$\lesssim$$z$$\lesssim$0.9}

\author[L. Rodr\'{\i}guez-Mu\~noz et al.]{L. Rodr\'{\i}guez-Mu\~noz$^{1}$\thanks{E-mail: lucia.rodriguezmunoz@unipd.it},
G. Rodighiero$^{1}$,
C. Mancini$^{1}$,
P. G. P\'erez-Gonz\'alez$^{2,\,3}$,
\newauthor T. D. Rawle$^{4}$,
E. Egami$^{5}$,
A. Mercurio$^{6}$,
P. Rosati$^{7}$,
A. Puglisi$^{1,\,8}$,
\newauthor A. Franceschini$^{1}$,
I. Balestra$^{9}$,
I. Baronchelli$^{1,\,10}$,
A. Biviano$^{11}$,
H. Ebeling$^{12}$,
\newauthor A. C. Edge$^{13}$,
A. F. M. Enia$^{1}$,
C. Grillo$^{14,\,15}$,
C. P. Haines$^{16}$,
E. Iani$^{1}$,
\newauthor T. Jones$^{17,\,18}$,
M. Nonino$^{11}$,
I. Valtchanov$^{19}$,
B. Vulcani$^{1}$,
M. Zemcov$^{20}$\\
$^{1}$Dipartimento di Fisica e Astronomia ``G. Galilei'', Universit\`a degli Studi di Padova, Vicolo dell'Osservatorio 3, I-35122, Italy\\
$^{2}$Departamento de Astronom\'{\i}a y Astrof\'{\i}sica, Universidad Complutense de Madrid, Av. Complutense s/n, C.P. 28040, Madrid, Spain\\
$^{3}$Centro de Astrobiolog\'{\i}a, Instituto Nacional de T\'ecnica Aeroespacial, Carretera de Ajalvir km\,4, Torrej\'on de Ardoz, Madrid, E-28850, Spain\\
$^{4}$ESA/Space Telescope Science Institute (STScI), 3700 San Martin Drive, Baltimore, MD 21218, USA\\
$^{5}$Steward Observatory, University of Arizona, 933 N. Cherry Ave, Tucson, AZ 85721, USA\\
$^{6}$INAF-Osservatorio Astronomico di Capodimonte, via Moiariello 16, 80131 Napoli, Italy\\
$^{7}$Dipartimento de Fisica e Scienze della Terra, Universit\`a degli Studi di Ferrara, via Saragat 1, 44122 Ferrara, Italy\\
$^{8}$Laboratoire AIM-Paris-Saclay, CEA/DSM-CNRS-Universit\'e Paris Diderot, IRFU/Service d'Astrophysique, CEA Saclay,\\
Orme des Merisiers, F-91191 Gif-sur-Yvette, France\\
$^{9}$University Observatory Munich, Scheinerstrasse 1, D-81679 Munich, Germany\\
$^{10}$IPAC, Mail Code 314-6, Caltech, 1200 E. California Blvd., Pasadena, CA 91125, USA\\
$^{11}$INAF-Osservatorio Astronomico di Trieste, via G. B. Tiepolo 11, I-34131, Trieste, Italy\\
$^{12}$Institute for Astronomy, University of Hawaii, Honolulu, HI 96822, USA\\
$^{13}$Centre for Extragalactic Astronomy, Department of Physics, Durham University, South Road, Durham DH1 3LE, UK\\
$^{14}$Dark Cosmology Centre, Niels Bohr Institute, University of Copenhagen, Juliane Maries Vej 30, DK-2100 Copenhagen, Denmark\\
$^{15}$Dipartimento di Fisica, Universita degli Studi di Milano, via Celoria 16, I-20133 Milano, Italy\\
$^{16}$INAF - Osservatorio Astronomico di Brera, via Brera 28, I-20121 Milano, Italy\\
$^{17}$Department of Physics and Astronomy, PAB, 430 Portola Plaza, Box 951547, Los Angeles, CA 90095-1547, USA\\
$^{18}$Department of Physics, University of California Davis, 1 Shields Avenue, Davis, CA 95616, USA\\
$^{19}$Herschel Science Centre, European Space Astronomy Centre, ESA, E-28691 Villanueva de la Ca\~nada, Spain\\
$^{20}$Center for Detectors, School of Physics and Astronomy,
Rochester Institute of Technology, Rochester NY 14623, USA
}

\date{Accepted XXX. Received YYY; in original form ZZZ}

\pubyear{2018}

\begin{document}
\label{firstpage}
\pagerange{\pageref{firstpage}--\pageref{lastpage}}
\maketitle

\begin{abstract}

We quantify the star formation (SF) in the inner
cores ($\mathcal{R}$/$R_{200}$$\leq$0.3) of 24 massive galaxy
clusters at 0.2$\lesssim$$z$$\lesssim$0.9 observed by the {\it Herschel}
Lensing Survey and the Cluster Lensing and Supernova survey with {\it Hubble}.
These programmes, covering the rest-frame ultraviolet to far-infrared 
regimes, allow us to accurately characterize stellar mass-limited
($\mathcal{M}_{*}$$>$$10^{10}$\,$M_{\odot}$) samples of star-forming
cluster members (not)-detected in the mid- and/or far-infrared. We
release the catalogues with the photometry, photometric redshifts,
and physical properties of these samples. We also quantify the SF
displayed by comparable field samples from the Cosmic Assembly 
Near-infrared Deep Extragalactic Legacy Survey. 
We find that in intermediate-$z$ cluster cores, the SF activity is suppressed 
with respect the field in terms of both the fraction ($\mathcal{F}$) 
of star-forming galaxies (SFG)
and the rate at which they form stars ($\mathcal{SFR}$ and $s\mathcal{SFR} = \mathcal{SFR}/\mathcal{M}_{*}$). 
On average, the $\mathcal{F}$ of SFGs is a factor $\sim$$2$ smaller in cluster cores than in the field.
Furthermore, SFGs present average $\mathcal{SFR}$ and
$s\mathcal{SFR}$ typically $\sim$0.3\,dex smaller in the clusters than
in the field along the whole redshift range probed. Our results favour long
time-scale quenching physical processes as the
main driver of SF suppression in the inner cores of clusters since $z$$\sim$0.9, with shorter
time-scale processes being very likely
responsible for a fraction of the missing SFG population. 

\end{abstract}




\section{Introduction}

Galaxies appear to be distributed into two fairly distinct general groups
(e.g., \citealt{2003MNRAS.341...33K}, \citealt{2004ApJ...608..752B},
\citealt{2004ApJ...600..681B}, \citealt{2017A&A...605A...4H}):
a population of relatively red, quiescent galaxies (i.e., where
the star formation activity has already been quenched), which are
characterized by spheroid-dominated morphologies; and a population
of rather blue, star-forming galaxies (SFGs), with disk-dominated
morphologies. Understanding the nature of the processes that make a
galaxy a member of either category at any cosmological epoch is one of
the longest standing unsolved problems in astrophysics.

The fraction of
red/quiescent/early-type galaxies among the whole population scales with
the stellar mass ($\mathcal{M}_{*}$) of the galaxies up to 
$z$$\sim$4 (e.g., \citealt{2004ApJ...600..681B,2006MNRAS.373..469B}),
and with the density of the environments they inhabit at least
up to $z$$\sim$1 (e.g., \citealt{1980ApJ...236..351D},
\citealt{2002MNRAS.334..673L}).  Hence, different works have claimed
that this \textit{dichotomy} between (still)
star-forming and quenched galaxies, should be driven (independently;
\citealt{2010ApJ...721..193P}) by the impact on the evolution of galaxies
of two kind of processes: those somehow related to the stellar mass of
the galaxies they quench, and therefore, responsible for the so-called
\textit{mass quenching}; and those linked to physical processes taking
place in high density environments, responsible for the so-called 
\textit{environmental quenching}. The physical nature of these quenching 
processes and its evolution with redshift remains controversial.

A plethora of works have studied the star formation (SF) activity within
galaxy clusters at different redshifts as to quantify the environmental
influence on galaxy evolution (e.g., \citealt{1997ApJ...490..577D},
\citealt{1999ApJ...518..576P,2003Ap&SS.285..121P},
\citealt{2007MNRAS.374..809D}, \citealt{2008ApJ...685L.113S},
\citealt{2010ApJ...720...87F}, \citealt{2011MNRAS.412..246V}). This
large body of work gives evidence for a significant transformation of
galaxy populations in clusters since z$\sim$1. Already three decades ago,
\citet[see also \citealt{1978ApJ...219...18B}]{1984ApJ...285..426B} found
that the fraction of blue cluster members increases from zero in the local
universe to $\sim$20\% by $z$$\sim$0.4. This rapid evolution over the last
5 billion years can only be explained by the existence of a population of
field SFGs entering the cluster environment, which eventually is capable
of turning them into passively evolving systems. This scenario is also
favoured by the standard hierarchical cosmological model, which predicts
a peak in the rate of field galaxies entering the cluster environment
at $z$$\sim$0.4 \citep{1995MNRAS.274..153K}. 

In clusters, SFGs are not only less numerous than in the field, but they
seem to present also different properties with respect their isolated
counterparts. For instance, rich environments host a high fraction
of \textit{post-starburst} (PSB; e.g., \citealt{2009ApJ...693..112P},
\citealt{2014ApJ...796...65M}, \citealt{2017ApJ...838..148P}), and
\textit{jellyfish} galaxies (e.g., \citealt{2010MNRAS.408.1417S},
\citealt{2017ApJ...844...48P}). Also, first CO observations in
$z$$\sim$0.4-0.5 by \citet{2013A&A...557A.103J} show that cluster
members contain less molecular gas than field galaxies at the
same redshift. 

Works such as \citet{2009ApJ...705L..67P}, \citet{2010ApJ...710L...1V},
\citet{2013ApJ...775..126H}, or \citet{2016ApJ...816L..25P} 
find a different distribution of star formation rate (SFR), and
specific star formation rate ($s\mathcal{SFR}$; defined as the ratio
between the $\mathcal{SFR}$ and the $\mathcal{M}_{*}$ of a galaxy)
in the inner regions of clusters (i.e., within the virial radius,
$\mathcal{R}_{\mathrm{virial}}$) with respect to the field, with values
typically $\sim$0.2-0.3\,dex smaller for the former. 
This offset translates into a shift in the tight relation between
the $\mathcal{SFR}$ and $\mathcal{M}_{*}$ found for the star-forming
field galaxies up to $z$$\sim$4 (e.g, \citealt{2007ApJ...660L..43N},
\citealt{2011ApJ...739L..40R}, \citealt{2012ApJ...754L..29W},
\citealt{2017A&A...599A.134S}). Such a correlation is commonly known
as the \textit{main sequence} (MS) of SFGs. The existence of the MS
is interpreted as the proof for a typical mode in which the galaxies
form stars (e.g., \citealt{2015ApJ...801L..29R}). The tightness of the correlation
(0.3\,dex scatter; e.g., \citealt{2012ApJ...754L..29W}) is interpreted as a possible
consequence of the short time-scale of the dominant quenching process
(\citealt{2010ApJ...721..193P}) moving the field SFGs out of the MS. As
a consequence, the displacement of the cluster members MS towards
lower $\mathcal{SFR}$ values could imply that the dominant quenching
mechanisms in rich environments are different (e.g., slow quenching
mechanisms could populate the region below the MS with transition
galaxies on their way to be turned off; \citealt{2015ApJ...806..101H},
\citealt{2013ApJ...775..126H}, \citealt{2016ApJ...816L..25P}).
However, other
works such as \citet{2010ApJ...721..193P}, \citet{2010ApJ...720...87F},
\citet{2012MNRAS.423.3679W}, or \citet{2013ApJ...773...86T} find
the same $\mathcal{SFR}$ distribution in clusters as in the field
at intermediate redshifts. These discrepancies appear to be due to a
combination of different factors such as observational biases (e.g.,
$\mathcal{SFR}$ detection limit), different sample selection functions, 
and cluster-to-cluster differences (e.g., \citealt{2006ApJ...649..661G}, 
\citealt{2016ApJ...825...72A}).

A variety of mechanisms have been proposed as the responsible for
environmental quenching (see reviews by, e.g.,
\citealt{2006PASP..118..517B} and \citealt{2007MNRAS.381....7H}):
gravitational interactions with the potential well of nearby galaxies
or the cluster itself, also known as \textit{harassment}
(\citealt{1996Natur.379..613M}); removal and thermal heating of
the interstellar medium of the galaxies by the interaction with the
intra-cluster medium (ICM), the so-called \textit{ram-pressure stripping}
(RPS; \citealt{1972ApJ...176....1G}, \citealt{2017ApJ...844...48P});
the removal of the hot gas reservoirs of the halo of galaxies, or
\textit{strangulation}, and subsequent halt of the supply of material
needed to sustain the SF, leading up to the eventual \textit{starvation}
(\citealt{1980ApJ...237..692L}). These mechanisms shape the evolution
of galaxies in different time-scales, probably with different efficiency
depending on the properties of both galaxies and clusters, and the
particular circumstances under which the infall takes place (see,
e.g., \citealt{2006PASP..118..517B}, \citealt{2009ApJ...690.1292B}).
Furthermore, it has also been proposed that the environmental impact
on these SFGs starts in early stages of the infall if the accreted
galaxies are bound up in small groups (\textit{pre-processing}; e.g.,
\citealt{2015ApJ...806..101H}). Distinguishing among these mechanisms
remains challenging, and relies on the detailed study and accurate
quantification of the changes suffered by the SF processes
and structural properties of the galaxies in rich environments.

Recently, a number of state-of-the-art surveys have targeted massive
galaxy clusters at intermediate redshift with the main goal of exploring
low-luminosity galaxies at high redshift taking advantage of the
gravitational lensing phenomenon (e.g., \textit{Hubble} Frontier Fields,
\citealt{2017ApJ...837...97L}). In this work, we aim at shedding light
on the impact of environment on the star-forming activity in galaxies
populating clusters by using these surveys to study the cluster inhabitants
themselves.

We focus our analysis on 24 X-ray selected (i.e., with total
masses $\sim$5 to $\sim$30$\times$10$^{14}$$M_{\odot}$)
clusters targeted by the \textit{Herschel} Lensing Survey
(HLS; \citealt{2010A&A...518L..12E}), a far-infrared (FIR) and
sub-millimetre survey using the ESA \textit{Herschel} Space Observatory,
and the Cluster Lensing and Supernova survey with Hubble (CLASH;
\citealt{2012ApJS..199...25P}), a deep optical and near-infrared (NIR)
\textit{Hubble} Space Telescope program, as well as by other NIR and
mid-infrared (MIR) \textit{Spitzer} programs.  The sample extends
between 0.187$\leq$$z$$\leq$0.890, thus, covering a particularly
interesting cosmic epoch for the study of environmental quenching.

The wealth and quality of this optical-to-NIR photometric dataset 
allows us to identify cluster
galaxies applying a methodology based on photometric redshifts to
complement the spectroscopic membership assignment. Furthermore,
combining the whole multi-wavelength data we can accurately quantify the
average (un)-obscured SF hosted by $\mathcal{M}_{*}$-selected samples of
cluster SFGs. The use of \textit{Herschel} observations complementing
optical and NIR data guarantees a proper quantification of the SF
shrouded by dust. 

Indeed, SFGs detected in the MIR and/or FIR (M-FIR)
often have optical colours consistent with those of passively evolving
galaxies and therefore, they are easily missed by studies limited
to the optical or NIR regimes. Not quantifying the contribution of
these obscured processes can lead to an under estimation of the true
level of SF by a factor $\sim$10 \citep{2002A&A...382...60D}. This
can extremely affect high density environments studies where,
despite the overall reduced SF activity observed, a population
of dusty star-forming cluster galaxies has been detected at a
wide range of redshifts (e.g., \citealt{2002A&A...382...60D},
\citealt{2000A&A...361..827F}, \citealt{2006ApJ...649..661G},
\citealt{2007ApJ...654..825M}, \citealt{2008ApJ...685L.113S},
\citealt{2009ApJ...693.1840B}, \citealt{2009ApJ...693..140D},
\citealt{2009ApJ...704..126H}, \citealt{2010A&A...518L..14R},
\citealt{2011A&A...532A..77B}, \citealt{2011A&A...532A.145P},
\citealt{2011ApJ...736...38K}, \citealt{2011MNRAS.416..680C},
\citealt{2012ApJ...756..106R}, \citealt{2014MNRAS.437..437A},
\citealt{2016ApJ...825...72A}). 

Ultimately, we systematically quantify
the suppression of the formation activity in galaxy cluster cores
with respect the field. For this end, we consistently build reference field
samples across the same redshift range by applying \textit{the same}
analysis to the optical-to-FIR publicly available photometry on three
of the fields targeted by the Cosmic Assembly Near-infrared Deep
Extragalactic Legacy Survey (CANDELS; \citealt{2011ApJS..197...35G},
\citealt{2011ApJS..197...36K}).

This article is organized as follows: Section~\ref{clusters_sample}
describes the cluster sample and corresponding
data. Section~\ref{multi-wavelength-photo} describes our approach to
combining the different photometric data and building the multi-wavelength
catalogue we use to derive photometric redshifts (Section~\ref{zphot})
and physical properties of galaxies through a SED-fitting approach
(Section~\ref{rainbow}). In Section~\ref{members_selection}, we
detail our procedure to select cluster members using spectroscopic and
photometric redshifts estimations. The final cluster members samples of
SFGs are presented in Section~\ref{clusters_members_samples} and further characterized in Section~\ref{SMF}. The quantification of the SF activity in the core of 
these clusters
is discussed in Section~\ref{sf_process}. Finally, an interpretation of our results is 
given in Section~\ref{discussion}, and a summary and the main
conclusions of this work are given in Section~\ref{summary}.

Throughout this work we assume a flat $\Lambda$CDM cosmology with
$H_0$$=$$70$\,kms$^{-1}$Mpc$^{-1}$, $\Omega_{m}$$=$0.3, and 
$\Omega_{\Lambda}$$=$0.7. Star-formation rates and stellar masses 
are based on a \citet{1955ApJ...121..161S} initial mass function (IMF).

The catalogues of star-forming cluster members associated to
this paper, including multi-wavelength photometry, photometric
redshifts, and physical properties, can be downloaded
from the public flavour of the \textsc{Rainbow} Cosmological
Database\footnote{\href{http://rainbowx.fis.ucm.es}{http://rainbowx.fis.ucm.es}}
(\citealt{2008ApJ...675..234P},
\citealt{2011ApJS..193...13B,2011ApJS..193...30B}).

\section{Galaxy Clusters Sample \& Data}\label{clusters_sample}
\begin{table*}
	\centering
	\caption{Description of the galaxy cluster sample. We display the following information: (1) Cluster ID; (2-3) Coordinates of the cluster centre as in \protect\citet{2012ApJS..199...25P}; (4) redshift \protect\citealt{2012ApJS..199...25P}; (5) Velocity dispersion (we use the value $\sigma_{\mathrm{cl}}$$=$1600\,km\,s$^{-1}$ when no observational estimation was found in the literature); (6) Radius within which the mean density is 200 times the critical density at the redshift where the cluster is located ($\sim$$R_{\mathrm{virial}}$ according to the simulations of \protect\citealt{1996ApJ...469..494E}; we use $R_{200}$$=$2000~kpc, see for instance \protect\citet{2014ApJ...795..163U}, for those cases for which no precise value was found in the literature); (7) The SF activity of the BCG as quantified through the emission of the UV, corrected for extinction ($\mathcal{SFR}_{\mathrm{BCG, UV, corr.}}$; \protect\citealt{2015ApJ...805..177D}), and the emission in the FIR ($\mathcal{SFR}_{\mathrm{BCG, TIR}}$; \protect\citealt{2012ApJ...747...29R}); (8) cool-core tracer $\mathcal{C}$ parameter as published by \citet{2016ApJ...819...36D}; (9) number of spectroscopic redshifts within the area covered by the CLASH catalogue ($\sim$0.0015~deg$^2$). Note: $^{+}$ 0.209 according to \protect\citet{2003A&A...397..431M}; 
$^{a}$ \protect\citet{2014ApJ...783...52G}; 
$^{b}$ \protect\citet{2003A&A...397..431M}; 
$^{c}$ \protect\citet{2012AJ....144...79G}; 
$^{d}$ \protect\citet{2016ApJS..224...33B}; 
$^{e}$ \protect\citet{2013A&A...558A...1B}; 
$^{f}$ \protect\citet{2007ApJ...661L..33E}; 
$^{g}$ \protect\citet{2016A&A...585A.160A}; 
$^{h}$ \protect\citet{2013ApJ...765...24N}; 
$^{i}$ \protect\citet{2014Msngr.158...48R}; 
$^{j}$ \protect\citet{2012ApJ...757...22C}; 
$^{k}$ \protect\citet{2015A&A...574A..11K}; 
$^{l}$ \protect\citet{2012ApJS..199...26H}; 
$^{m}$ \protect\citet{2014ApJS..211...21E}; 
$^{n}$ \protect\citet{2015ApJ...812..114T} and \protect\citet{2014ApJ...782L..36S}; 
$^{o}$ \protect\citet{2002ApJ...577..133R}; 
$^{p}$ \protect\citet{2002ApJ...573..524C}; 
$^{q}$ \protect\citet{1996ApJ...470..172S}; 
$^{r}$ \protect\citet{2009ApJS..182..543A}; 
$^{s}$ $\sigma_{\mathrm{cl}}$ and $\mathcal{R}_{200}$ derived using the 
value of the mass within $\mathcal{R}_{200}$ ($\mathcal{M}_{200}$)
from \protect\citet{2014ApJ...795..163U}.}
	\label{tab:ancillary}
	\begin{tabular}{lccccccccc} 
		\hline
		\rotatebox{0}{ID} & RA & Dec & $z$ & $\sigma_{\mathrm{cl}}$ & $R_{200}$ & $\mathcal{SFR}_{\mathrm{BCG,\,UV corr.}}$/$_{\mathrm{TIR}}$ & $\mathcal{C}$ & \#$z_{\mathrm{spec}}$\\
		 & [J2000] & [J2000] &   & [km\,s$^{-1}$] & [kpc] & [$M_{\odot}$yr$^{-1}$] &  &  \\
		(1) & (2) & (3) & (4) & (5) & (6) & (7) & (8) & (9) \\
		\hline
		A0383 & 02:48:03.40 & -03:31:44.9 & 0.187\,\,\, & 931$^{+59}_{−59}$$^{a}$ &  1220$^{+10}_{-10}$$^{a}$ & 3.3$\pm$0.4 / 4.0$\pm$0.2 & 0.525 & 37$^{a,h}$ \\
		A0209 & 01:31:52.54 & -13:36:40.4 & 0.206$^{+}$ & 1394$^{+88}_{-99}$$^{b}$ & 2130$^{+50}_{-50}$$^{g}$ & 1.2$\pm$1.1 / \,\,\,\,\,\,\,\,\,--\,\,\,\,\,\,\,\,\, & 0.167 & 73$^{b,i,g}$  \\
		A2261 & 17:22:27.18 & \,\,32:07:57.3 & 0.224\,\,\, & 1524$^{s}$ & 1942$^{s}$ & 3.3$\pm$2.8 / \,\,\,\,\,\,\,\,\,--\,\,\,\,\,\,\,\,\, & 0.331 & 5$^{j}$ \\
		RBS1748 & 21:29:39.94 & \,\,00:05:18.8 & 0.234\,\,\, & 1600 & 2000 & 2.9$\pm$0.4 / \,\,\,\,\,\,\,\,\,--\,\,\,\,\,\,\,\,\, & 0.426 & --  \\
		A0611 & 08:00:56.82 & \,\,36:03:23.6 & 0.288\,\,\, & 1316$^{s}$ & 1760$^{+97}_{-89}$$^{h}$ & 0.9$\pm$1.7 / \,\,\,\,\,\,\,\,\,--\,\,\,\,\,\,\,\,\, & 0.335 & 23$^{h}$  \\
		MS2137 & 21:40:15.18 & -23:39:40.7 & 0.313\,\,\, & 1257$^{s}$ & 1318$^{+140}_{-107}$$^{h}$ & 5.6$\pm$0.7 / \,\,\,\,\,\,\,\,\,--\,\,\,\,\,\,\,\,\, & 0.589 & --  \\
		AS1063 & 22:48:43.96 & -44:31:51.3 & 0.348\,\,\, & 1660$^{+230}_{-150}$$^{c}$ & 2376$^{s}$ & 2.3$\pm$0.5 / \,\,\,\,\,\,\,\,\,--\,\,\,\,\,\,\,\,\, & 0.194 & 136$^{i,k}$  \\
		MACS1931 & 19:31:49.66 & -26:34:34.0 & 0.352\,\,\, & 1339$^{s}$ & 1641$^{s}$ & 83.1$\pm$2.3 / \,\,\,\,\,\,\,\,\,\,--\,\,\,\,\,\,\,\,\,\,\,\, & 0.545 & --  \\
		MACS1115 & 11:15:51.90 & \,\,01:29:55.1 & 0.355\,\,\, & 1364$^{s}$ & 1668$^{s}$ & 6.4$\pm$0.5 / \,\,\,\,\,\,\,\,\,--\,\,\,\,\,\,\,\,\, & 0.430 & --  \\
		RXJ1532 & 15:32:53.78 & \,\,30:20:58.7 & 0.363\,\,\,  & 1031$^{s}$ & 1278$^{s}$ & 48.6$\pm$2.6 / \,\,\,\,\,\,\,\,\,\,--\,\,\,\,\,\,\,\,\,\,\,\, & 0.571 & 1$^{l}$ \\
		MACS1720 & 17:20:16.95 & \,\,35:36:23.6 & 0.387\,\,\, & 1296$^{s}$ & 1569$^{s}$ & 6.1$\pm$0.7 / \,\,\,\,\,\,\,\,\,--\,\,\,\,\,\,\,\,\, & 0.417 & --  \\
		MACS0416 & 04:16:09.39 & -24:04:03.9 & 0.397\,\,\, & 996$^{+12}_{-36}$$^{d}$ & 1820$^{+110}_{-110}$$^{d}$ & 3.5$\pm$0.8 / \,\,\,\,\,\,\,\,\,--\,\,\,\,\,\,\,\,\, & 0.091 & 219$^{d,n,m}$ \\
		MACS0429 & 04:29:36.05 & -02:53:06.1 & 0.399\,\,\, & 1140$^{s}$ & 1385$^{s}$ & 20.1$\pm$2.1 / \,\,\,\,\,\,\,\,\,\,--\,\,\,\,\,\,\,\,\,\,\,\, & 0.531 & --  \\
		MACS1206 & 12:06:12.15 & -08:48:03.4 & 0.440\,\,\, & 1087$^{+53}_{-55}$$^{e}$ & 1980$^{+100}_{-100}$$^{e}$ & 6.8$\pm$3.0 / \,\,\,\,\,\,\,\,\,--\,\,\,\,\,\,\,\,\, & 0.223 & 81$^{e}$  \\
		MACS0329 & 03:29:41.56 & -02:11:46.1 & 0.450\,\,\, & 1165$^{s}$ & 1386$^{s}$ & 31.0$\pm$2.4 / \,\,\,\,\,\,\,\,\,\,--\,\,\,\,\,\,\,\,\,\,\,\, & 0.488 & --  \\
		RXJ1347 & 13:47:30.59 & -11:45:10.1 & 0.451\,\,\, & 1710$^{s}$ & 1987$^{s}$ & 16.5$\pm$1.8 / \,\,\,\,\,\,\,\,\,\,--\,\,\,\,\,\,\,\,\,\,\,\, & 0.506 & 42$^{o,p,q}$  \\
		MACS1311 & 13:11:01.67 & -03:10:39.5 & 0.494\,\,\, & 1600 & 2000 & 5.8$\pm$1.9 / \,\,\,\,\,\,\,\,\,--\,\,\,\,\,\,\,\,\, & 0.488 & --  \\
		MACS1149 & 11:49:35.69 & \,\,22:23:54.6 & 0.544\,\,\, & 1840$^{+120}_{-170}$$^{f}$ & 2352$^{s}$ & 2.1$\pm$0.7 / \,\,\,\,\,\,\,\,\,--\,\,\,\,\,\,\,\,\, & 0.111 & 378$^{m}$  \\
		MACS0717 & 07:17:32.63 & \,\,37:44:59.7 & 0.545\,\,\, & 1660$^{+120}_{-130}$$^{f}$ & 2358$^{s}$ & 5.4$\pm$1.4 / \,\,\,\,\,\,\,\,\,--\,\,\,\,\,\,\,\,\, & 0.055 & 143$^{l,m}$  \\
		MACS1423 & 14:23:47.76 & \,\,24:04:40.5 & 0.545\,\,\, & 1300$^{+120}_{-170}$$^{f}$ & 2000 & 16.7$\pm$1.2 / 46.5$\pm$0.8 & 0.555 & 96$^{m}$  \\
		MACS2129 & 21:29:26.06 & -07:41:28.8 & 0.570\,\,\, & 1400$^{+120}_{-180}$$^{f}$ & 2000 & 1.6$\pm$0.1 / \,\,\,\,\,\,\,\,\,--\,\,\,\,\,\,\,\,\, & 0.211 & 85$^{m}$  \\
		MACS0647 & 06:47:50.27 & \,\,70:14:55.0 & 0.584\,\,\, & 900$^{+170}_{-200}$$^{f}$ & 1442$^{s}$ & 2.1$\pm$0.3 / \,\,\,\,\,\,\,\,\,--\,\,\,\,\,\,\,\,\, & 0.242 & --  \\
		MACS0744 & 07:44:52.82 & \,\,39:27:26.9 & 0.686\,\,\, & 1101$^{+130}_{-150}$$^{f}$ & 1521$^{s}$ & 8.5$\pm$3.1 / \,\,\,\,\,\,\,\,\,--\,\,\,\,\,\,\,\,\, & 0.365 & --  \\
		CLJ1226 & 12:26:58.37 & \,\,33:32:47.4 & 0.890\,\,\, & 1600 & 2000 & 2.7$\pm$1.5 / \,\,\,\,\,\,\,\,\,--\,\,\,\,\,\,\,\,\, & 0.245 & 9$^{l,r}$  \\
		\hline
	\end{tabular}
\end{table*}
\begin{table}
	\centering
	\caption{In this table we show an overview of the photometric bands used in this work: (1) name of the observing band and instrument; (2) effective wavelength of the filter; (3) median FWHM of the PSF in arcseconds; (4) name of the project to which the data belongs. ($^{*}$) 
Spitzer Programs \#17 (PI: Fazio), \#83 (PI: Rieke), \#545 (PI: Egami), \#40652 (PI: Kocevski), \#50393 (PI: Kocevski), \#60034 (PI: Egami), \#80168 (PI: Bouwens). ($^{+}$) Spitzer Programs \#83 (PI: Rieke), \#40652 (PI: Kocevski), \#40872 (PI: Smith), \#50393 (PI: Kocevski).} 
	\label{tab:photometry}
	\begin{tabular}{lrccc} 
		\hline
		Band & $\lambda_{\mathrm{eff}}$ & FWHM & Project \\
		(1) & (2) & (3) & (4) \\
		\hline
		WFC3-F225W & 237.84\,nm & 0$\arcsec$.08 & CLASH \\
		WFC3-F275W & 271.47\,nm & 0$\arcsec$.08 & CLASH \\
		WFC3-F336W & 335.86\,nm & 0$\arcsec$.07 & CLASH \\
		WFC3-F390W & 393.22\,nm & 0$\arcsec$.07 & CLASH \\
		ACS-F435W & 436.33\,nm & 0$\arcsec$.08 & CLASH \\
		ACS-F475W & 475.05\,nm & 0$\arcsec$.08 & CLASH \\
		ACS-F606W & 596.11\,nm & 0$\arcsec$.08 & CLASH \\
		ACS-F625W & 630.97\,nm & 0$\arcsec$.08 & CLASH \\
		ACS-F775W & 770.59\,nm & 0$\arcsec$.08 & CLASH \\
		ACS-F814W & 807.31\,nm & 0$\arcsec$.09 & CLASH \\
		ACS-F850LP & 905.26\,nm & 0$\arcsec$.09 & CLASH \\
		WFC3-F105W & 1.06\,$\mu$m & 0$\arcsec$.13 & CLASH \\
		WFC3-F110W & 1.15\,$\mu$m & 0$\arcsec$.13 & CLASH \\
		WFC3-F125W & 1.25\,$\mu$m & 0$\arcsec$.14 & CLASH \\
		WFC3-F140W & 1.40\,$\mu$m & 0$\arcsec$.14 & CLASH \\
		WFC3-F160W & 1.54\,$\mu$m & 0$\arcsec$.15 & CLASH \\
		IRAC-3.6\,$\mu$m & 3.56\,$\mu$m & 2$\arcsec$.1\,\,\, & $^{*}$ \\
		IRAC-4.5\,$\mu$m & 4.50\,$\mu$m & 2$\arcsec$.1\,\,\, & $^{*}$ \\
		IRAC-5.8\,$\mu$m & 5.74\,$\mu$m & 2$\arcsec$.2\,\,\, & $^{*}$ \\
		IRAC-8.0\,$\mu$m & 7.93\,$\mu$m & 2$\arcsec$.2\,\,\, & $^{*}$ \\
		MIPS-24\,$\mu$m & 23.84\,$\mu$m & 5$\arcsec$\,\,\,\,\,\,\, & $^{+}$ \\
		PACS-100\,$\mu$m & 102.25\,$\mu$m & 8$\arcsec$\,\,\,\,\,\,\, & HLS \\
		PACS-160\,$\mu$m & 165.59\,$\mu$m & 12$\arcsec$\,\,\,\,\,\,\,\,\, & HLS \\
		SPIRE-250\,$\mu$m & 253.13\,$\mu$m & 18$\arcsec$\,\,\,\,\,\,\,\,\, & HLS \\
		SPIRE-350\,$\mu$m & 355.87\,$\mu$m & 25$\arcsec$\,\,\,\,\,\,\,\,\, & HLS \\
		SPIRE-500\,$\mu$m & 511.19\,$\mu$m & 36$\arcsec$\,\,\,\,\,\,\,\,\, & HLS \\
		\hline
	\end{tabular}
\end{table}

The \textit{Herschel Lensing Survey} \citep[HLS;][]{2010A&A...518L..12E}
is a large imaging survey of galaxy clusters in the far-infrared
(FIR) and sub-millimetre using the ESA \textit{Herschel Space
Observatory} \citep{2010A&A...518L...1P}.  HLS provides deep PACS
\citep{2010A&A...518L...2P} and SPIRE \citep{2010A&A...518L...3G}
imaging (see Section~\ref{herschel}) for a sample of 65 X-ray-luminous
(i.e., massive) clusters of galaxies in the redshift range between
0.2$\lesssim$$z$$\lesssim$0.9. The primary aim of HLS is to observe
the most effective gravitational lenses available, probing beyond the
confusion limit of the Herschel instruments to observe intrinsically
faint, high-redshift sources \citep[e.g.,][]{2010A&A...518L..13R,
2010A&A...518L..14R}.  However, the HLS is also a remarkable survey for
the study of SF processes taking place within high density environments
\citep[e.g.,][]{2016MNRAS.459.1626R, 2014MNRAS.442..196R}.  On the
one hand, it targets a significant number of clusters, which avoids
deriving misleading results due to cluster-to-cluster variations
(e.g., \citealt{2016ApJ...825...72A}).  On the other hand, the clusters
targeted by the HLS span over a redshift range in which these systems
are thought to undergo a major evolution due to the transformation
of infalling star-forming field galaxies into passive objects (e.g.,
\citealt{1995MNRAS.274..153K}, \citealt{2015ApJ...806..101H}).

Among the fields targeted by the HLS, we focus our work on a subsample
of 24 clusters (see Table~\ref{clusters_sample}) also observed
by the \textit{Cluster Lensing and Supernova survey with Hubble}
\citep[CLASH; ][]{2012ApJS..199...25P}. CLASH is a Multi-Cycle Treasury
Program with the aim of providing ultra-deep photometry of 25 X-ray
selected, massive ($\sim$5 to $\sim$$30\times10^{14}\,M_{\odot}$)
galaxy clusters in a total of 16 passbands using HST ACS/WFC,
WFC3/UVIS, and WFC3/IR (see Section~\ref{hst} for details). CLASH
clusters are drawn heavily from the Abell and MACS cluster catalogues
(\citealt{1958ApJS....3..211A}, \citealt{1989ApJS...70....1A},
\citealt{2001ApJ...553..668E}, \citealt{2007ApJ...661L..33E}, \citealt{2010MNRAS.407...83E}, \citealt{2012MNRAS.420.2120M}).

The wealth of photometric and spectroscopic data available for
this galaxy clusters sample, that we call CLASH+HLS, enables the
accurate identification and characterization of their galaxy
population (e.g., \citealt{2016A&A...585A.160A},
\citealt{2016A&A...590A.108M}, \citealt{2016ApJS..224...33B}). Indeed,
CLASH+HLS clusters have been extensively studied in previous
works. CLASH photometry together with spectroscopy from different
surveys (see Section~\ref{spec}) have provided strong constraints
on the cluster inner mass distributions and profiles (e.g.,
\citealt{2015ApJ...801...44Z}, \citealt{2013A&A...558A...1B},
\citealt{2014A&A...571A..80A}). Also, their dynamical state and
substructures have been analyzed through different techniques, such
as the Sunyaev-Zel'dovich effect (SZ; \citealt{1972CoASP...4..173S},
\citealt{2016MNRAS.460..569R}) and X-ray surface brightness analysis (see
\citealt{2016MNRAS.460..569R} and references therein), as well as lensing
(e.g., \citealt{2013ApJ...762L..30Z}, \citealt{2015ApJ...800...38G}) and
kinematics of galaxy populations (e.g., \citealt{2015A&A...579A...4G}).
Despite the X-ray selection, that generally favours highly relaxed
clusters, the sample is found to be not homogeneously dynamically relaxed
(\citealt{2012ApJS..199...25P}, \citealt{2016MNRAS.460..569R}).  Finally,
a number of works have studied in detail the brightest cluster galaxies
(BCG) of the CLASH+HLS systems. For instance, \citet{2015ApJ...805..177D}
and \citet{2016ApJ...819...36D} carried out a study on the morphology
and SF activity of these peculiar galaxies, using the rest-frame
UV imaging provided by CLASH. Furthermore, they also characterized
the intra cluster gas in the vicinity of the BCGs and beyond, by
analysing the X-ray emission of the inner cluster cores. Complementary,
\citet{2012ApJ...747...29R} studied the obscured SF activity undergone by
the BCGs of the massive clusters observed by HLS, and its dependence with
the X-ray gas cooling times for cool-core (CC) clusters\footnote{Cool-core
clusters are defined as those systems with X-ray cooling times $<$1\,Gyr
\citep{1994ARA&A..32..277F}.}.

In the following subsections, we describe the photometric
and spectroscopic datasets available on the cluster fields (see
Table~\ref{tab:photometry} \& \ref{tab:photometry_ir} for a summary of
their main characteristics), as well as other ancillary data found in
the literature.

\subsection{\textit{Hubble} optical and near-infrared photometry}\label{hst}

In this work, we use the 
CLASH\footnote{\href{https://archive.stsci.edu/prepds/clash/}{https://archive.stsci.edu/prepds/clash/}} 
photometric dataset published by
\citet{2012ApJS..199...25P}. This data release contains the photometry
performed on the \textit{HST} ACS/WFC (F435W, F475W, F606W, F625W,
F775W, F814W, and F850LP), WFC3/UVIS (F225W, F275W, F336W, and F390W),
and WFC3/IR (F105W, F110W, F125W, F140W, and F160W) deep imaging
of 25 massive intermediate redshift clusters.  Object detection and
photometry is accomplished using SExtractor \citep{1996A&AS..117..393B}
in dual image mode using a weighted sum of the ACS/WFC and WFC3/IR images
(see \citealt{2012ApJS..199...25P} for details on the \textit{HST} data
reduction, catalogue build up, and main characteristics). These catalogues
cover an area of $\sim$5\,arcmin$^2$, limited by the WFC3/IR images
($\sim$2.0$\times$2.3\,arcmin$^2$), and therefore, they mainly sample the
very inner cluster cores. An angular distance of 2.0\,arcmin corresponds
to 375\,kpc and 932\,kpc for the lowest and largest redshifts
in the sample, respectively. The total area covered, including the 24
clusters, is $\sim$135\,arcmin$^{2}$. The exposure times of the frames
vary between 2000 and 5000\,s, reaching average (5$\sigma$)
limiting AB magnitudes of $\sim$26. A summary of the properties of the
dataset is shown in Table~\ref{tab:photometry}.

\subsection{\textit{Spitzer} near and mid-infrared photometry}\label{spitzer}

A series of programs with \textit{Spitzer} have covered all
CLASH clusters with IRAC 3.6 and 4.5$\mu$m bands. Furthermore,
40\% of them have also been observed with IRAC 5.8 and
8.0$\mu$m channels, and 50\% has been covered by MIPS 24$\mu$m
band. These data were extracted from the \textit{Spitzer} Heritage archive
\footnote{\href{http://irsa.ipac.caltech.edu/applications/Spitzer/SHA}{http://irsa.ipac.caltech.edu/applications/Spitzer/SHA}}.
\textit{Spitzer} images reduction, source detection, and photometry
were carried out as described in \citet{2005ApJ...630...82P}
and \citet{2008ApJ...675..234P}, for MIPS and IRAC, respectively.
Briefly, the data reduction was carried out with MOPEX (Mosaicking and
Point-source Extraction), the package provided by the \textit{Spitzer}
Science Center for reducing and analysing imaging data.  In the case
of IRAC, the source detection and photometry were carried out with
SExtractor \citep{1996A&AS..117..393B}, using the same procedure as
\citet{2004ApJS..154...44H}.  Photometry was performed using a small
circular aperture, and an aperture correction was applied to get the total
flux. IRAC beam sizes are 2.1, 2.1, 2.2, and 2.2$\arcsec$ respectively
for increasing wavelengths. The average sensitivities reached at
5$\sigma$ are 1.4, 1.5, 4.5, 4.2\,$\mu$Jy.
In the case of MIPS images, characterized by a larger point-spread
function, the photometry was extracted by PSF fitting. Several detection
passes are used in order to make catalogues as complete as possible,
in spite of the significant source confusion. The MIPS 24$\mu$m
beam size is 5$\arcsec$.  The average MIPS 24$\mu$m limiting flux
at 5$\sigma$ is 234\,$\mu$Jy. In Table~\ref{tab:photometry} and
\ref{tab:photometry_ir} we summarize the properties of these photometric
catalogues. We report the heterogeneous sensitivities reached by IRAC
and MIPS imaging on the different CLASH clusters.  In particular, MIPS
24$\mu$m limiting fluxes vary between 77 and  852\,$\mu$Jy.

\subsection{\textit{Herschel} far-infrared photometry} \label{herschel}

This study employs the PACS 100, 160$\mu$m, and SPIRE
250, 350, 500$\mu$m imaging provided by HLS for all the
clusters. We use the catalogues created by the HLS team following
the methodology presented by \citet{2010A&A...518L..15P} and
\citet{2010A&A...518L..14R,2016MNRAS.459.1626R}.  Source catalogues
and photometry in all bands were obtained with standard PSF fitting
methodology, relying on a set of fixed IRAC and MIPS prior position
catalogues. PACS imaging at 100 and 160$\mu$m has mean 5$\sigma$ flux limits
of 4.7 and 8.7\,mJy, while in the three SPIRE bands, the
typical 5$\sigma$ limits are 19.4, 15.3, and 13.7\,mJy,
respectively for the 250, 350, and 500$\mu$m bands. The beam sizes for the
five \textit{Herschel} bands (sorted by increasing effective wavelength)
are 8, 12, 18, 25, and 36$\arcsec$, respectively.

\begin{table*}
	\centering
	\caption{Limiting fluxes (5$\sigma$) of the 
\textit{Spitzer} and \textit{Herschel} photometric catalogues used in this work.}
	\label{tab:photometry_ir}
	\begin{tabular}{lccccccccccc} 
		\hline
		 & \multicolumn{5}{c}{$\mathcal{F}_{\mathrm{lim}}$ [$\mu$Jy]} & & \multicolumn{5}{c}{$\mathcal{F}_{\mathrm{lim}}$ [mJy]} \\
		\cline{2-6} \cline{8-12}
		 & \multicolumn{4}{c}{$Spitzer$/IRAC} & \multicolumn{1}{c}{$Spitzer$/MIPS} & & \multicolumn{2}{c}{$Herschel$/PACS} & \multicolumn{3}{c}{$Herschel$/SPIRE}\\
\rotatebox{0}{Cluster} & \rotatebox{0}{3.6$\mu$m} & \rotatebox{0}{4.5$\mu$m} & \rotatebox{0}{5.8$\mu$m} & \rotatebox{0}{8.0$\mu$m} & \rotatebox{0}{24$\mu$m} & & \rotatebox{0}{100$\mu$m} & \rotatebox{0}{160$\mu$m} & \rotatebox{0}{250$\mu$m} & \rotatebox{0}{350$\mu$m} & \rotatebox{0}{500$\mu$m}\\
(1) & (2) & (3) & (4) & (5) & (6) & & (7) & (8) & (9) & (10) & (11) \\
		\hline
		A0209 & 2.0 & 1.7 & 5.2 & 5.4 & 268.7 & & 4.6 & \,\,\,9.1 & 14.6 & 14.0 & 10.7\\
		A0383 & 2.7 & 2.2 & 6.7 & 6.3 & 317.6 & & 4.8 & \,\,\,9.4 & 14.8 & 13.7 & 10.8\\
		MACS0329 & 1.3 & 1.3 & -- & -- & -- & & 4.5 & \,\,\,8.5 & 19.9 & 15.8 & 14.7\\
		MACS0416 & 1.2 & 1.2 & -- & -- & -- & & 4.7 & \,\,\,8.5 & 19.2 & 14.9 & 13.9\\
		MACS0429 & 1.3 & 1.3 & -- & -- & -- & & 4.5 & \,\,\,8.3 & 21.0 & 16.7 & 14.4\\
		MACS0647 & 1.1 & 1.3 & -- & -- & -- & & 4.8 & 10.8 & 23.1 & 20.3 & 14.7\\
		MACS0717 & 1.7 & 1.9 & -- & -- & 133.3 & & 4.7 & \,\,\,9.2 & 17.8 & 15.9 & 12.0\\
		MACS0744 & 2.2 & 3.0 & 1.2 & 1.8 & -- & & 4.4 & \,\,\,8.4 & 14.5 & 14.1 & 11.3\\
		A0611 & 1.0 & 1.0 & -- & -- & 380.6 & & 4.8 & \,\,\,8.4 & 15.0 & 13.9 & 11.1\\
		MACS1115 & 1.3 & 1.4 & -- & -- & -- & & 4.7 & \,\,\,8.7 & 20.4 & 16.1 & 14.5\\
		MACS1149 & 0.9 & 0.9 & -- & -- & -- & & 4.7 & \,\,\,8.6 & 15.1 & 15.3 & 14.9\\
		MACS1206 & 1.1 & 1.1 & -- & -- & 305.7 & & 4.5 & 10.3 & 25.8 & 21.9 & 18.3\\
		CLJ1226 & 3.6 & 3.6 & 2.0 & 1.8 & 131.7 & & 6.5 & 11.3 & 22.2 & 18.3 & 18.6\\
		MACS1311 & 1.2 & 1.4 & -- & -- & -- & & 4.7 & \,\,\,8.4 & 20.1 & 15.6 & 14.2\\
		RXJ1347 & 1.7 & 1.5 & 4.5 & 2.7 & 143.7 & & 4.3 & \,\,\,7.8 & 21.1 & 18.6 & 18.5\\
		MACS1423 & 1.4 & 1.8 & -- & -- & \,\,\,95.5 & & 5.2 & \,\,\,9.5 & 14.2 & 12.6 & 10.3\\
		RXJ1532 & 1.2 & 1.2 & -- & -- & 180.3 & & 4.8 & \,\,\,8.4 & 18.3 & 14.5 & 13.5\\
		MACS1720 & 0.9 & 0.8 & -- & -- & -- & & 4.7 & \,\,\,8.7 & 19.6 & 14.9 & 13.0\\
		A2261 & 1.9 & 1.9 & 5.8 & 4.6 & 108.5 & & 4.6 & \,\,\,8.9 & 20.0 & 16.1 & 14.0\\
		MACS1931 & 3.6 & 2.7 & -- & -- & 851.9 & & 4.5 & \,\,\,8.7 & 19.5 & 15.2 & 13.4\\
		MACS2129 & 1.9 & 1.7 & 1.1 & 1.6 & 112.6 & & 5.2 & 13.7 & 33.5 & 28.3 & 29.2\\
		RBS1748 & 1.2 & 1.2 & -- & -- & 311.8 & & 4.5 & \,\,\,9.4 & 15.3 & 14.5 & 11.4\\
		MS2137 & 1.8 & 1.6 & 7.1 & 7.7 & \,\,\,97.7 & & 5.1 & \,\,\,9.4 & 14.5 & 13.3 & 11.1\\
		AS1063 & 2.2 & 1.7 & 6.6 & 6.0 & \,\,\,76.9 & & 4.8 & \,\,\,7.7 & 14.7 & 14.6 & 10.9\\
		\hline
	\end{tabular}
\end{table*}

\subsection{Spectroscopic Data}\label{spec}

One of the programs with a greater contribution to our spectroscopic
redshift sample is the spectroscopic survey carried out on the 13
southern CLASH clusters with the Visible Multi-Object Spectrograph
\citep[VIMOS;][]{2003SPIE.4841.1670L} mounted on the Very Large Telescope
(VLT), the so-called CLASH-VLT survey (CLASH-VLT Large Programme
186.A0.798; P.I.: P. Rosati; \citealt{2014Msngr.158...48R}).  We refer
the reader to \citet{2013A&A...558A...1B} and \citet{2016ApJS..224...33B}
for details on spectroscopic data, target selection, and performance
statistics of the mentioned project. We also make use of spectroscopic
redshift measurements from the Grism Lens Amplified Survey from
Space \citep[GLASS;][]{2014ApJ...782L..36S,2015ApJ...812..114T},
a large \textit{Hubble Space Telescope} program aimed at
obtaining grism spectroscopy of the HFF. Besides these, 
we also gather spectroscopic redshifts from other surveys 
(see Table~\ref{tab:ancillary} for a complete list of the works included).
Finally, we also retrieve redshifts through NASA/IPAD Extragalactic 
Database (NED), mainly from the 2MASS Redshift Survey 
\citep{2012ApJS..199...26H}, and the Seventh Data Release of the 
Sloan Digital Sky Survey \citep{2009ApJS..182..543A}. 
In Section~\ref{zphot} we describe the properties of the final 
spectroscopic sample.

\section{Multi-wavelength photometry}\label{multi-wavelength-photo}

We merge the photometric datasets described in the previous section
to obtain UV-to-FIR SEDs for all the sources in the catalogues released by
CLASH.
To this end, we use the \textsc{Rainbow}
Cosmological Database (\citealt{2008ApJ...675..234P},
\citealt{2011ApJS..193...13B,2011ApJS..193...30B}) and associated
software package. We use CLASH catalogues as parent catalogues to take
advantage of the high resolution of HST imaging. However, this requires
taking special care of the inevitable blending of sources in bands with
poorer resolution, as well as possible counterpart misidentification.

In the following subsections, we describe the strategy that we
use for the build-up of our multi-wavelength photometric catalogue.

\subsection{Cross-matching catalogues}

Initially, \textsc{Rainbow} searches for counterparts of our parent
catalogue in the rest of the bands. In practice, each catalogue is
cross-matched to the CLASH positions. \textsc{Rainbow} takes into
account possible astrometry offsets between the bands by re-aligning
each pair of them using the positions of several sources in small
1$\arcmin$$\times$1$\arcmin$ boxes around a given source.  The search
radii we use to find counterparts candidates are 1$\arcsec$.5,
2$\arcsec$.5, 2$\arcsec$.5, 4$\arcsec$.0, 9$\arcsec$.0, 9$\arcsec$.0,
and 12$\arcsec$.0 for IRAC, MIPS 24$\mu$m, PACS 100 and 160$\mu$m,
and SPIRE 250, 350, and 500$\mu$m catalogues. These values are chosen
in order to cope with the typical WCS offsets between different
images, as well as uncertainties in the determination of the center
for faint MIPS and \textit{Herschel} sources. We note, however, that a
comparison of the CLASH vs MIPS/\textit{Herschel} coordinates for secure
(i.e., bright) mid- and far-IR sources points out that the typical WCS
uncertainty is $\sim$0$\arcsec$.2 for IRAC, $\sim$0$\arcsec$.4 for
MIPS, $\sim$0$\arcsec$.4 for PACS, and $\sim$1$\arcsec$.3 for SPIRE. In
Section~\ref{counterpart} we take into account both the search radius
and the WCS accuracy measurements to discuss how many \textit{HST}
counterparts we find for each M- and FIR source, and how we select the
most likely among the former.

\subsection{IRAC fluxes deblending}

The IRAC photometry is recomputed on CLASH positions following a
deconvolution method detailed in \citet{2011ApJS..193...13B}.  The
procedure is similar to that used in, e.g., \citet{2006A&A...449..951G},
\citet{2008ApJ...682..985W}, \citet{2009ApJ...691.1879W}, or
\citet{2010ApJS..187..251W}, and briefly consists on the convolution of
the PSF of the higher resolution image to the IRAC PSF and a subsequent
scaling of the flux of each source in a way that the total flux equals
the emission of the blended source in the lower resolution image.

\subsection{M- and FIR counterpart assignment}\label{counterpart}

Given the larger beam sizes of the M/FIR bands, a simple cross-correlation
of the optical/NIR and M/FIR catalogues frequently assigns the same M/FIR
source to different optical/NIR counterparts (especially when using
\textit{HST} images). On average, the relaxed search
radii we use to cross-match catalogues lead to the assignation of each MIPS
24$\mu$m, PACS, and SPIRE source to 2, 5, and 32 optical/NIR sources,
respectively. However, within the WCS accuracy measurements there are,
on average, 1 optical/NIR source for each detection in MIPS~24$\mu$m,
PACS, and SPIRE~250$\mu$m and 250$\mu$m, and 2 optical/NIR sources for
each SPIRE~500$\mu$m source. These latter values are more informative of
the level of uncertainty in our cross-matching procedure and reliability
of the counterparts identification, as well as possible blending affecting
the low resolution bands.

Due to the large difference between the resolution of CLASH and M/FIR
bands, it is not advisable to apply a deblending procedure such
as it was done on IRAC photometry. Instead, we limit our approach
to the identification of the \textit{most likely} counterpart, or
dominant contributor to the M/FIR fluxes, among the multiple short
wavelength counterparts assigned to the same M/FIR sources. The fact
that the FIR catalogues are built using IRAC and MIPS 24$\mu$m priors
guarantees a consistent framework to link the photometry across the
whole wavelength range. Different studies have addressed the task
of identifying counterparts of FIR/Sub-millimetre galaxies in shorter
wavelengths \citep[e.g., ][]{2013MNRAS.431..194A}, avoiding using simply
the shortest distance match with the aim of achieving a more physically
driven identification. Our approach \textit{steps
through} the N-to-FIR wavelength range and evaluates which of the IR
SEDs of the multiple candidates is most likely to be associated with
the M/FIR detection.

We first set local and average SNR limits in the FIR bands. These
limits are 2$\sigma$ and 3$\sigma$ for MIPS and \textit{Herschel}
bands (see Table~\ref{tab:photometry_ir}, where we show the flux values
corresponding to the 5$\sigma$ detection in each band and cluster). \textcolor{black}{ {The
2$\sigma$ is used to maximize the information available to identify
the FIR counterparts, however, we clarify that we do not consider MIPS
24$\mu$m fluxes below 3$\sigma$ detections in the rest of the work.}}
Then, we select as the optical/NIR counterpart of each MIPS 24$\mu$m
source the brightest candidate in the reddest IRAC band available. Then,
we shift this methodology to larger wavelength bands. We select as the
optical/NIR counterpart of each PACS source the brightest candidate in
MIPS 24$\mu$m. When MIPS is not available, we use the reddest IRAC band
in which the source is detected. Finally, we select as the optical/NIR
counterpart of each SPIRE source, the brightest candidate in the reddest
PACS band available, if any. Otherwise, MIPS 24$\mu$m and IRAC bands
are used. If different optical/NIR candidates present very similar
fluxes (within 1$\sigma$) in the band that is used to identify the
counterpart, we impose a criterion of minimum distance, and therefore,
we select as the optical/NIR counterpart the galaxy with the closest
position to the M/FIR source. In all cases described, the MIPS, PACS,
and SPIRE fluxes of the CLASH sources that are not identified as real
counterparts are flagged and they are not used subsequently. Therefore,
each M/FIR source is assigned to a single optical/NIR source.  We note
that using IRAC as a tracer of PACS or SPIRE emitters can lead to spurious
associations. This is because NIR and FIR trace different components and
processes in the galaxies. In the clusters with MIPS coverage, the average
fraction of \textit{Herschel} sources' optical counterparts identified
by their IRAC fluxes is 20\% and 32\% for PACS and SPIRE,
respectively. These values increase, however, in those fields without
MIPS photometry, reaching 91\% and 49\%, respectively. These
cases are flagged for further check. After a thorough visual inspection
of the output of our procedure, we detect only obvious mismatch cases
in galaxies located in the border of the \textit{HST}/WFC3 images. We
have identified a number of galaxies suffering from over-deblending in
the CLASH catalogues, which means that the photometry of these galaxies
are divided into different sources. In these cases, the flux of the MIR
and FIR catalogues are generally assigned to source corresponding to
the central region of the galaxy.

\section{Photometric redshifts}\label{zphot}

\textcolor{black}{ {Photometric redshifts ($z_{\mathrm{phot}}$) are computed using the
\textsc{EAZY} code \citep{2008ApJ...686.1503B}, specifically conceived 
for this task. \textsc{EAZY} is a template-fitting code
based on $\chi^2$ minimization between observed photometry and a set of 
6 SED templates. Among them, 5 templates are generated following the 
\citet{2007AJ....133..734B} \textit{non-negative matrix factorization} algorithm 
with PEGASE stellar population synthesis models 
(\citealt{1997A&A...326..950F}) and a calibration set of synthetic photometry 
derived from semi-analytic models. The last one is a dusty 
starburst model, and it is added to the set in order to compensate for
the lack of dusty galaxies in the calibration photometric sample.}}

The achievable quality of photometric redshifts depends strongly on the
quality of the photometric dataset itself, and the wavelength domain
it covers \citep[e.g., ][]{2012MNRAS.421.2002P}.  In particular,
it benefits from high-quality photometry sampling strong continuum
features (e.g., Lyman or Balmer breaks). In this sense, the 16
CLASH broadband photometric points enable high levels of accuracy in
the photometric redshift estimation (\citealt{2014A&A...562A..86J},
\citealt{2017MNRAS.470...95M}, \citealt{2017ApJ...848...37C}).  In order
to make use of the whole potential of our dataset, we fit not only
the whole wavelength range covered by CLASH, but also the IRAC
photometric points.  Furthermore, for those clusters with available
spectroscopic samples we perform a zero-point fine-tuning (following
the methodology by \citealt{2011ApJS..193...13B,2011ApJS..193...30B})
to account for mismatches between the CLASH colours and the SED-fitting
template library colours, or other hypothetical systematic problems. The
median absolute zero-points used are 3\% and 5\% for CLASH and IRAC bands, 
respectively.

\begin{figure}
        \centering
        \includegraphics[width=1\linewidth]{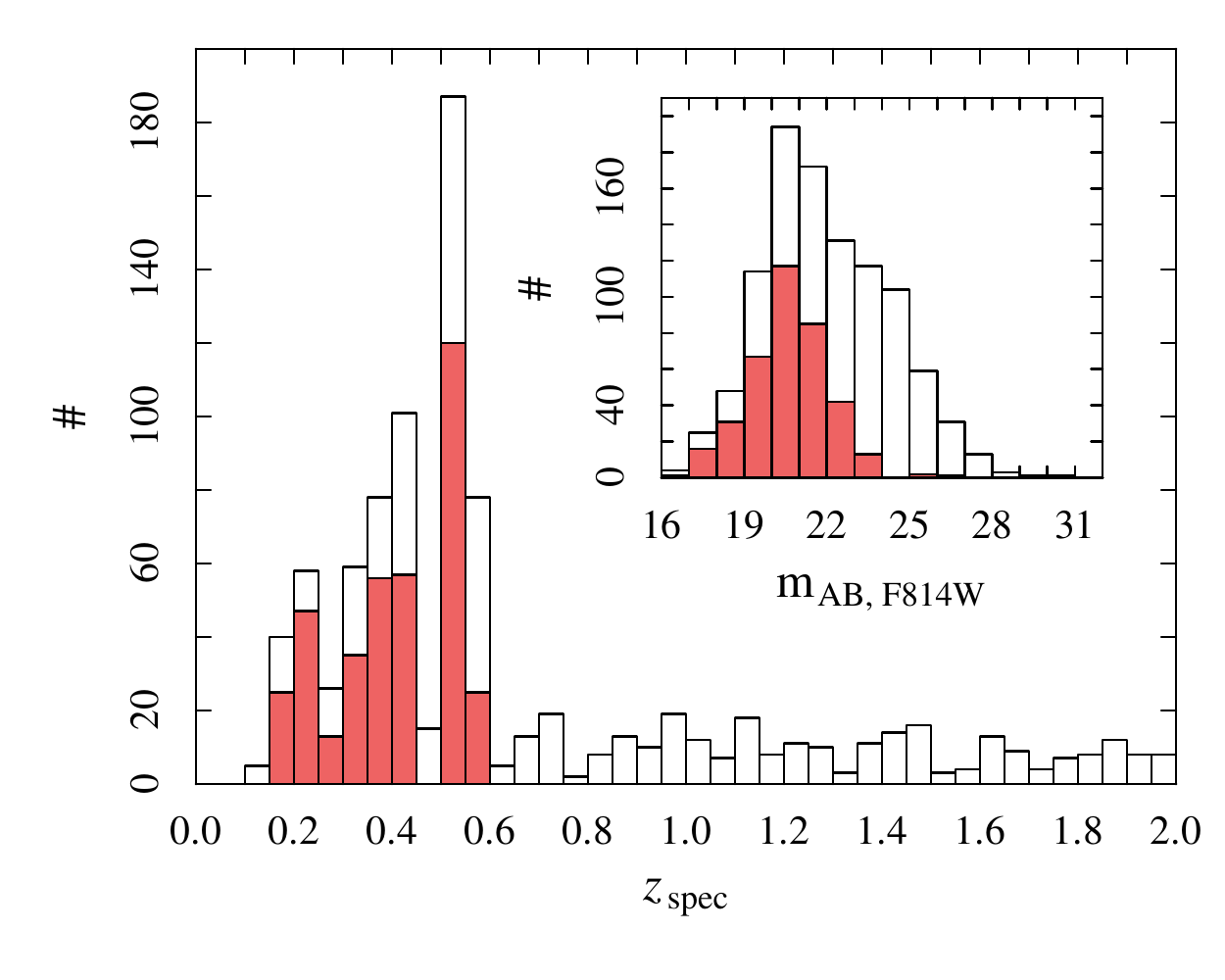}
        \caption{Distribution of $z_{\mathrm{spec}}$ for our spectroscopic 
sample (1034 galaxies; empty histogram). The distribution of the redshifts
of the 378 spectroscopically confirmed cluster members is given in red. In
this figure, we show the distribution up to $z$$=$2, which contains
90\% of the sample. The nested panel shows the corresponding distribution
of magnitudes in the ACS/F814W band.}
        \label{fig:charact_zspec}
\end{figure}

\begin{figure}
        \centering
        \includegraphics[width=1\linewidth]{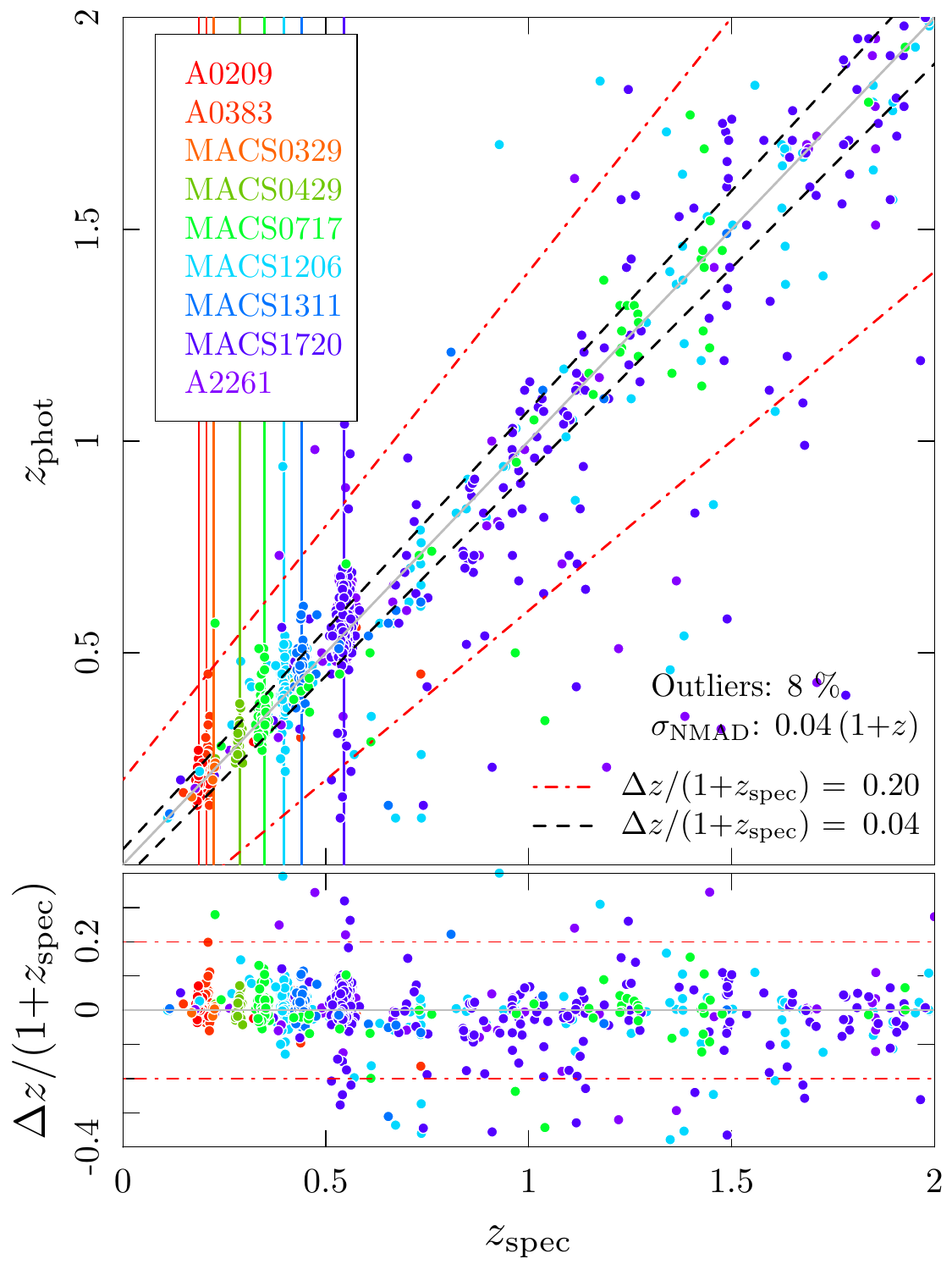}
        \caption{Evaluation of the $z_{\mathrm{phot}}$ quality. 
The black and red dashed lines show, respectively, the accuracy reached by
our results considering the whole spectroscopic sample and the definition
of outlier. The vertical lines mark the redshift of each cluster (Table~\ref{clusters_sample}).}
        \label{fig:zphot_zspec}
\end{figure}

\subsection{Photometric redshifts quality}\label{zphot_q}

We assess the quality of the $z_{\mathrm{phot}}$ obtained for each cluster by
comparing them against the available and reliable\footnote{The \textit{
reliability} of the $z_{\mathrm{spec}}$ is given by the spectroscopic
surveys in the form of a quality flag normally linked to the number
and SNR of the spectral features identified on the spectrum, that are
used to calculate the redshift.} $z_{\mathrm{spec}}$. We cross-correlate
CLASH dataset with the spectroscopic catalogues using a radius of
0$\arcsec$.5. The total reference spectroscopic sample is composed
of 1034 spectroscopically confirmed galaxies within the area of the
WFC3 imaging (i.e. the area covered by the photometric catalogues) over
the 24 CLASH+HLS clusters we analyse. This sample is by definition
inhomogeneous, as can be expected of the combination of studies designed
with different scientific objectives and selection criteria. It extends
between 0.1$<$$z$$\lesssim$9, with the 90\% of the galaxies at $z$$<$2. 
Figure~\ref{fig:charact_zspec} displays the distribution of
$z_{\mathrm{spec}}$ (empty histogram), and the distribution of magnitudes
in the ACS/F814W band (empty histogram; nested panel).

A number of quantities have
been used in the literature to quantify the behaviour of the data points in
this diagram (see, e.g., \citealt{2009A&A...508.1173P}), either in terms
of scatter, as well as the presence of outliers and systematic offsets.
In the last decade, the normalized median absolute deviation
($\sigma_{\mathrm{NMAD}}$; \citealt{1983ured.book.....H}) of the
difference between the $z_{\mathrm{phot}}$ and the $z_{\mathrm{spec}}$
($\Delta z = z_{\mathrm{phot}}-z_{\mathrm{spec}}$) has been frequently used to
characterize the scatter of the distribution of $z_{\mathrm{phot}}$
(e.g., \citealt{2009ApJ...690.1236I}). A typical photometric redshift
error distribution has tails that clearly depart from a pure Gaussian
distribution, in addition to a relatively large fraction of outliers. The
$\sigma_{\mathrm{NMAD}}$ estimator manages to achieve a stable estimate
of the spread of the core of the $z_{\mathrm{phot}}$ distribution without being
affected by the mentioned tails. It is defined as
\begin{equation}
\centering
\sigma_{\mathrm{NMAD}} = 1.48 \times \mathrm{median} \left( \frac{\left| \Delta z - \mathrm{median}\left(\Delta z\right) \right|}{1+z_{\mathrm{spec}}} \right). 
\centering
\end{equation}
Following the notation by 
\citet{2011ApJS..193...30B}, we consider the fraction of catastrophic 
outliers, $\eta$, defined as those cases for which 
\begin{equation}
\centering
|\Delta z| / (1+z_{\mathrm{spec}}) > 0.2. 
\centering
\end{equation}
Finally, in order to characterize the systematic offsets of the 
photometric redshifts obtained, $\delta$, we use the expression
\begin{equation}
\centering
\delta = \Delta z / (1+z_{\mathrm{spec}}). 
\centering
\end{equation}
\noindent When compared with the spectroscopic sample, our photometric
redshift estimations present $\sigma_{\mathrm{NMAD}}$$=$0.04,
and 8\% of catastrophic outliers 
(see Figure~\ref{fig:zphot_zspec}). The outliers
are typically either faint sources with noisy photometry in \textit{HST}
and/or IRAC bands (e.g., high redshift galaxies, objects located in the
border of the CLASH catalogues) or galaxies for which the IRAC photometry
seems to be contaminated by bright nearby objects.  We do not identify
systematic effects, with an average $\delta$$=$$-0.01$.  These values
are comparable with those published by \citet{2014A&A...562A..86J}
for CLASH clusters.

As we are using the $z_{\mathrm{phot}}$ to select cluster members,
we also assess their quality using only a subsample of
spectroscopic members. We follow the selection criteria used by
\citet[][see Section~4.2]{2017MNRAS.470...95M} in order to be able
to compare our results with theirs. The cluster members reference
spectroscopic sample is formed by galaxies for which the difference
between its $z_{\mathrm{spec}}$ and the cluster redshift ($\Delta
z_{\mathrm{cl}}$) fulfills $\left|\Delta z_{\mathrm{cl}}\right| \leq
0.01$. Also, in order to guarantee an optimal sampling of the optical
and NIR SED, only galaxies detected at least on 14 CLASH bands are
considered. Using these criteria we select 378 galaxies (see red histogram
in Figure~\ref{fig:charact_zspec}). In this case, our photometric redshift
estimations present $\sigma_{\mathrm{NMAD}}$$=$0.03, and 2\% of 
catastrophic outliers. These values are comparable with
to those obtained by \citet{2017MNRAS.470...95M}: 
$\sigma_{\mathrm{NMAD}}$$=$0.02, and $\eta$$<$3\%. 
Neither in this case we identify systematic
effects, with an average deviation $\delta$$=$0.01.

\section{Spectral energy distribution fitting with \textsc{Rainbow}}\label{rainbow}

In order to derive the physical properties of the galaxies found
on CLASH+HLS fields, we apply a SED-fitting analysis to the
entire dataset gathered and described in the previous sections. We
use the \textsc{Rainbow} Cosmological Database software package
\citep{2008ApJ...675..234P,2011ApJS..193...13B,2011ApJS..193...30B}
to fit, on the one hand, the optical/NIR photometry (CLASH \& IRAC),
and on the other hand, the M/FIR photometry (MIPS \& \textit{Herschel}).
In both cases, we fix the redshifts derived with \textsc{EAZY} or, 
when available, the $z_{\mathrm{spec}}$.

In particular, the optical/NIR fitting code performs a $\chi^2$
minimization between the observed data and a set of semi-empirical
template SEDs computed from spectroscopically confirmed
galaxies modeled with PEGASE stellar population synthesis models
(\citealt{1997A&A...326..950F}). In particular, we use the templates
generated by \citealt{2008ApJ...675..234P} (see their Appendix B)
assuming a single stellar population with a exponentially declining
star formation history (SFH; $\mathcal{SFR}(t) \propto e^{-t/\tau}$)
with a time-scale ($\tau$) varying between 1\,Myr (instantaneous burst)
and 100\,Gyr (constant SFH) and an age that can take values between 1\,Myr
and 13.5\,Gyr. We also assume a \citet{1955ApJ...121..161S} IMF spanning
stellar masses from 0.1 to 100 $M_{\odot}$, metallicity ($\mathcal{Z}$)
values 0.005, 0.0.02, 0.2, 0.4, 1.0, 2.5, and 5.0 $\mathcal{Z}_{\odot}$,
extinction between 0 and 5~mag, and a \citet{2000ApJ...533..682C}
attenuation law. We complement the set of templates with QSO and 
AGN empirical templates drawn from \citet{2007ApJ...663...81P} 
that account for the galaxies whose UV-to-NIR emission is domitated by an
AGN. In the case of the M/FIR SED-fitting, the $\chi^2$ minimization is
performed between the observed photometry and the typical dust emission
models by \citet{2001ApJ...556..562C}, \citet{2002ApJ...576..159D},
\citet{2009ApJ...692..556R}, and \citet{2007ApJ...657..810D}. 

\subsection{Stellar masses}

The $\mathcal{M}_{*}$ of each galaxy 
is estimated by \textsc{Rainbow} from the average scale factor required 
to match the template monochromatic luminosities to the observed fluxes, 
weighted with the photometric errors. \textcolor{black}{ {The random uncertainty
of the $\mathcal{M}_{*}$ is derived from the dispersion in the 
mass-luminosity rations in the different bands.}} The average expected 
uncertainty in the estimations of $\mathcal{M}_{*}$ taking into account 
variations in $\mathcal{Z}$, SFH, or IMF are within 0.3~dex 
(\citealt{2008ApJ...675..234P}). 

\subsection{Star formation rates}

We take advantage of our rich dataset to analyse the SF activity
undergone by the galaxies in these fields in terms of total
$\mathcal{SFR}$ ($\mathcal{SFR}_{\mathrm{TOT}}$). Similarly to previous
works (see \citealt{2012ARA&A..50..531K} and references therein), we
consider that the total SF activity of a galaxy can be derived from the
combination of (1) the UV luminosity emitted by young stars that is
able to escape from the inter-stellar medium (ISM), and (2) 
the UV luminosity that is absorbed by the ISM and re-emitted in the M/FIR regime. 
We use the recipe of \citet{2005ApJ...625...23B},
which is based on the calibration of \citet{1998ARA&A..36..189K}:
\begin{align}
\mathcal{SFR}_{\mathrm{TOT}} =& \,\mathcal{SFR}_{\mathrm{TIR}} + \mathcal{SFR}_{\mathrm{UV}} \label{eq:SFR_tot}\\
\mathcal{SFR}_{\mathrm{TIR}}/M_{\odot}yr^{-1} =& \,1.8 \times 10^{-10} \mathcal{L}_\mathrm{TIR}/L_{\odot} \label{eq:SFR_TIR}\\
\mathcal{SFR}_{\mathrm{UV}}/M_{\odot}yr^{-1} =& \,5.9 \times 10^{-10} \mathcal{L}_\mathrm{2800\AA}/L_{\odot} \label{eq:SFR_UV}
\end{align}
\noindent where $\mathcal{L}_{\mathrm{TIR}}$ is the integrated total IR luminosity and $\mathcal{L}_{2800}$
is the rest-frame monochromatic luminosity at 2800\,\AA\, (uncorrected
for extinction). 

\textcolor{black}{ {We compute $\mathcal{L}_{\mathrm{TIR}}$ by integrating
the best-fit \citet{2007ApJ...657..810D} dust emission templates
between 8 to 1000\,$\mu$m. As we mentioned previously, we use four
different libraries of dust emission models in our analysis. The main
differences between these models are the prominence of the PAHs
and their dependence with the total IR luminosity, as well as the ratio
between the mass of hot and cold dust. A discussion on these properties
is beyond the scope of this paper, nevertheless, we use all
these template sets to include the differences between the assumptions
made by them in the uncertainty of the total IR luminosity. Therefore,
the $\mathcal{L}_{\mathrm{TIR}}$ values given in this work are derived
from the \citet{2007ApJ...657..810D} libraries, whereas the uncertainties
are the RMS of the $\mathcal{L}_{\mathrm{TIR}}$ estimations using the 4
template libraries. We have checked that the differences between the
luminosities given by the best fitting templates of each library are
of the order of $\lesssim$20\%.}}

We calculate $\mathcal{L}_{2800}$ 
interpolating the best fitted optical/NIR empirical template at
2800\,\AA (rest-frame). This wavelength is covered by observational
data over the whole redshift range of interest.

Obviously, this formalism can only be used in the case of galaxies
detected in the M/FIR. For those galaxies not detected by MIPS
or \textit{Herschel}, we compute $\mathcal{SFR}_{\mathrm{TOT}}$ by
correcting the UV luminosities (i.e., $\mathcal{SFR}_{\mathrm{UV}}$)
for dust attenuation ($\mathcal{A}_{\mathrm{UV}}$) following
the expression 
\begin{equation} 
\mathcal{SFR}_{\mathrm{TOT}} = \,\mathcal{SFR}_{\mathrm{UV, corr.}} = \,\mathcal{SFR}_{\mathrm{UV}} \times 10^{0.4\,\mathcal{A}_{\mathrm{UV}}} \label{eq:SFR_uvcorr}
\end{equation} 
\noindent where the $\mathcal{SFR}_{\mathrm{UV}}$ is obtained using Equation~\ref{eq:SFR_UV}.

\citet{1999ApJ...521...64M} demonstrate that local
starburst galaxies exhibit a relatively tight, monotonic relation
between the ratio between the UV and the
TIR luminosity ($\mathcal{IRX}$) and the UV slope 
($\beta$\footnote{The UV continuum slope is defined
by assuming that the UV regime of the SED of a galaxy can be described by
a power law ($\propto \lambda^{\beta}$; \citealt{1994ApJ...429..582C},
\citealt{1999ApJ...521...64M}).}). Through this relationship, they
derive a relation between the extinction of the UV (in particular,
the attenuation at 1600\AA) and the $\beta$ itself, providing a simple
relation that can be applied to correct UV luminosities. However,
this and other typical attenuation recipes based on the UV slope
(e.g., \citealt{1994ApJ...429..582C}) are derived for extreme
starburst galaxies, while the sources for which we need the
correction (i.e. those not-detected in the M/FIR) are less extreme
SFGs. Thus, using those expressions can lead to an overestimation
of the extinction and an overcorrection of the UV luminosity. 
Therefore, we derive an extinction correction optimized for our work 
(see Appendix~\ref{uv_cor}).

In what follows, the values of the $\mathcal{SFR}_{\mathrm{TOT}}$ refer to
the $\mathcal{SFR}_{\mathrm{UV, corr.}}$ (Equation~\ref{eq:SFR_uvcorr},
in which we use our own $\mathcal{A}_{\mathrm{UV}}$), except in those
cases when the M/FIR is available, where we consider the addition of
the $\mathcal{SFR}_{\mathrm{TIR}}$ and the $\mathcal{SFR}_{\mathrm{UV}}$
(Equation~\ref{eq:SFR_tot}).

\section{Cluster Members Selection}\label{members_selection}

\begin{figure}
\includegraphics[width=1\linewidth]{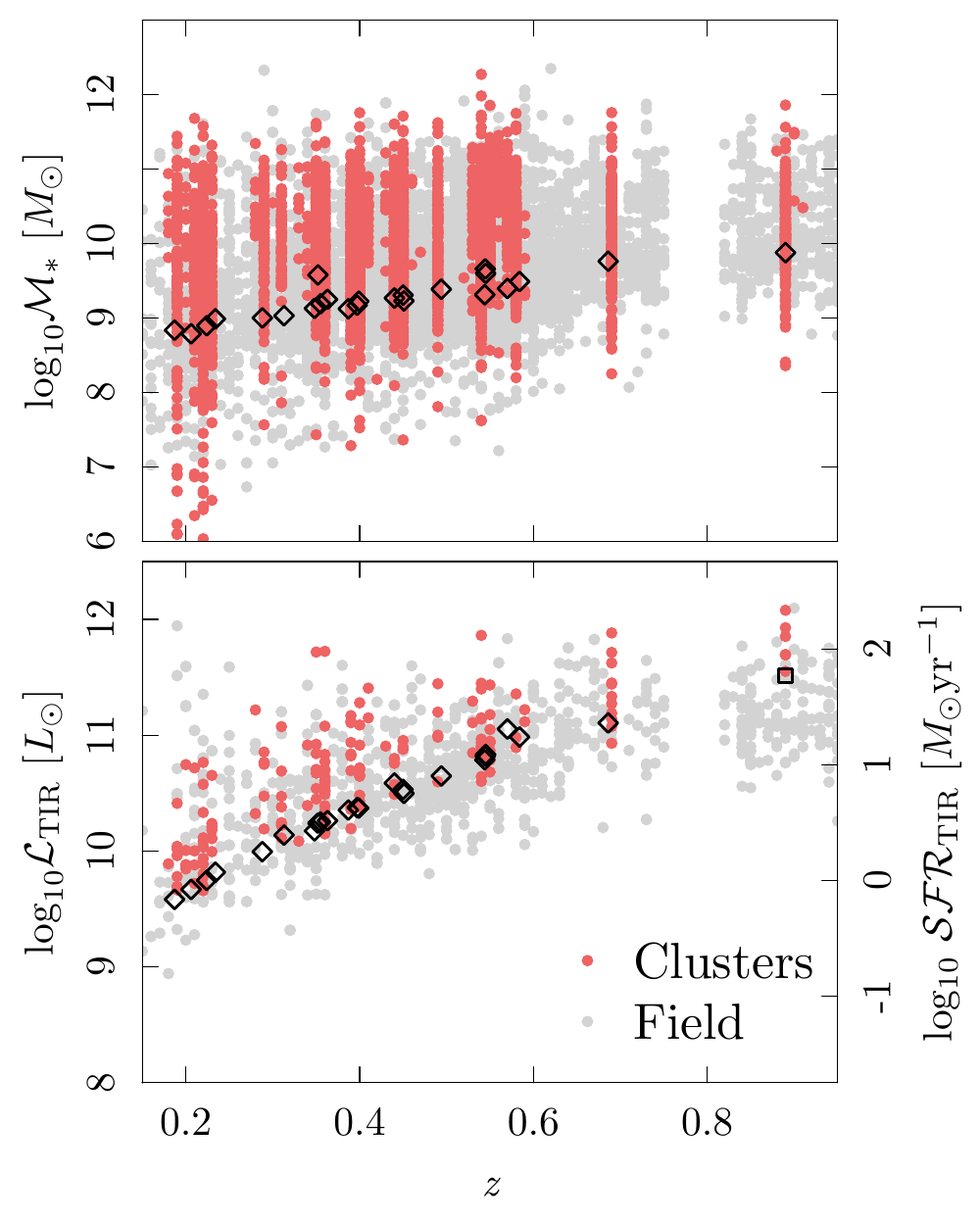}
\caption{\textit{Top panel}: Distribution of 
$\mathcal{M}_{*}$ of the cluster members and field galaxies samples 
(IRAC 4.5$\mu$m 3$\sigma$ detection) with redshift. The mass
limits obtained for each cluster as it is described in the text
are marked with black diamonds. \textcolor{black}{ {\textit{Bottom panel}: 
Variation with redshift of the total infrared luminosities (left axis) and
$\mathcal{SFR}_{\mathrm{TIR}}$ (right axis) of the parent samples of 
M-FIR detected sources. In the case of the field, we show the sample prior to the 
$\mathcal{SFR}_{\mathrm{TIR}}$ cuts described in the text. 
Black symbols indicate the $\mathcal{L}_{\mathrm{TIR}}$ 
and $\mathcal{SFR}_{\mathrm{TIR}}$ corresponding
to the limiting flux of MIPS 24$\mu$m (squares) or PACS 100$\mu$m
(diamonds). For each cluster only the deepest limit is represented.
In both panels, red symbols indicate cluster galaxies, grey symbols
indicate field galaxies. The number of field galaxies in this plot has 
been down-sampled to 30\% of the original sample size, for visualization 
purposes.}}} \label{fig:limits}
\end{figure}

\begin{table}
	\centering
	\caption{Summary of some of the quantities 
used for the identification of cluster members and an evaluation of the
technique: (1) Cluster ID; (2) number of spectroscopic members as defined
by Equation~\ref{eq:zspec}; (3) $\sigma_{\mathrm{NMAD}}$ derived for the
individual clusters; (4) number of $\sigma_{\mathrm{NMAD}}$ to be used
in the integration of the $\mathcal{P}(z)$; (5) membership probability
threshold; (6) completeness level (\%); (7) fraction of interlopers (\%).}
	\label{tab:probability}
	\begin{tabular}{lcccccc} 
		\hline
		\rotatebox{0}{ID} & \#$z$ & $\sigma_{\mathrm{NMAD}}$ & $n$ & $\mathcal{P}_{\mathrm{thr}}$ & $\mathcal{K}$ & $\mathcal{I}$ \\
		(1) & (2) & (3) & (4) & (5) & (6) & (7) \\
		\hline
		A0383 & 33 & 0.02 & 3 & 0.30 & 91 & 8 \\
		A0209 & 50 & 0.04 & 3 & 0.75 & 92 & 7 \\
		A0611 & 21 & 0.03 & 3 & 0.55 & 95 & 5 \\
		AS1063 & 71 & 0.06 & 1 & 0.15 & 87 & 10 \\
		MACS0416 & 84 & 0.09 & 2 & 0.75 & 86 & 13 \\
		MACS1206 & 51 & 0.06 & 3 & 0.85 & 88 & 11 \\
		RXJ1347 & 13 & 0.07 & 1 & 0.25 & 85 & 13 \\
		MACS1149 & 160 & 0.12 & 2 & 0.85 & 91 & 9 \\
		MACS0717 & 83 & 0.05 & 3 & 0.75 & 89 & 10 \\
		MACS2129 & 11 & 0.09 & 1 & 0.70 & 64 & 27 \\
		\hline
	\end{tabular}
\end{table}

The most unambiguous way to identify cluster members relies on
accurate spectroscopic redshifts.  However, the acquisition of complete
$z_{\mathrm{spec}}$ samples remains infeasible except for a relatively
small and bright fraction of the galaxy population. Indeed, using
photometric redshifts to estimate the distances to galaxies has become
a fundamental aim of galaxy surveys conducted during recent years (e.g.,
\citealt{2009ApJ...690.1236I}, \citealt{2011ApJS..193...30B}). Although
less accurate than spectroscopic ones, photometric redshifts provide a way
to estimate distances for galaxies too faint for spectroscopy or samples
too large to be practical for complete spectroscopic coverage. Given
the incomplete and inhomogeneous spectroscopic coverage of our sample of
clusters we are forced to use criteria to select cluster members based
either on $z_{\mathrm{spec}}$ or $z_{\mathrm{phot}}$.

The spectroscopic cluster members are identified as those galaxies
with $z_{\mathrm{spec}}$ within the redshift range defined by the
redshift of the cluster, $z_{\mathrm{cl}}$, and its velocity dispersion,
$\sigma_{cl}$. In Table~\ref{tab:ancillary} we show the values we use and
the corresponding references. In practice, we use the following criteria
(see \citealt{2009A&A...495..707C}):
\begin{equation}\label{eq:zspec}
\left|z_{\mathrm{cl}}-z_{\mathrm{spec}}\right| < 3\times\sigma_{\mathrm{cl}}\times(1+z_{\mathrm{cl}})
\end{equation}

For those cases in which a $z_{\mathrm{spec}}$ is not available, our
member selection relies on the redshift probability distribution, $\mathcal{P}(z)$,
given by \textsc{EAZY} instead on the
individual $z_{\mathrm{phot}}$ associated to each galaxy. This approach captures
all the photometric redshift information, which can significantly reduce
the impact of the catastrophic errors in the $z_{\mathrm{phot}}$-$z_{\mathrm{spec}}$ plane
\citep[e.g.,][]{2002MNRAS.330..889F}. This is of key importance to our
work, as it translates into a smaller contamination with foreground and
background sources in our cluster members selection.  In particular, we
use the method developed by \citet{2009A&A...508.1173P} based exclusively
on photometric redshift estimates. This approach modifies the technique
presented by \citet{2000AJ....120.2851B} in order to take advantage of
the $\mathcal{P}(z)$. It calculates a probability
of being a cluster member ($\mathcal{P}_{\mathrm{member}}$) integrating
$\mathcal{P}(z)$ within a redshift range centred in the redshift of the
cluster $z_{cl}$ and with a width ($\Delta z$) related to the accuracy
of the photometric redshifts (see Section~\ref{zphot_q}).
\begin{equation}
\mathcal{P}_{\mathrm{member}} = \int^{z_{\mathrm{cl}}+\Delta z}_{z_{\mathrm{cl}}-\Delta z}{\mathcal{P}(z) dz}
\end{equation}

In our case, we use $\Delta z =
n\times\sigma_{\mathrm{NMAD}}\times(1+z_{\mathrm{cl}})$.  Applying this
technique to those galaxies for which we have a reliable spectroscopic
redshift we can calibrate the cluster member selection, which
means to find a probability threshold ($\mathcal{P}_{\mathrm{thr}}$)
over which a galaxy is considered to be a cluster member, given a
certain $n$.  Table~\ref{tab:probability} shows the values of $n$ and
$\mathcal{P}_{\mathrm{thr}}$ we find to maximize the completeness level
($\mathcal{K}$) and minimize the percentage of interlopers ($\mathcal{I}$)
for those clusters with spectroscopic members. Table~\ref{tab:probability}
also gives the values of $\mathcal{K}$ and $\mathcal{I}$ for each case. We
reach $\mathcal{K}$$>$80\% and $\mathcal{I}$$<$20\% (limiting values used
also by \citealt{2009A&A...508.1173P}) for 9 out of the 10 clusters with
more than 10 spectroscopic cluster members available. In the case of
MACS2129, the cluster with fewer spectroscopic members available (11),
we retrieve $\mathcal{K}$$=$64\% and $\mathcal{I}$$=$27\%. Still,
the members sample we derive for it includes 73\% of correct cluster
members. For those clusters for which less than 10 spectroscopic
redshifts were available, we use the average value of $n$, and the
probability threshold derived for the other individual clusters:
$n$$=$2, $\mathcal{P}_{\mathrm{thr}}$$=$0.5. The reader can find examples
of the application of a similar selection procedure in the works by
(e.g.) \citet{2008ApJ...684..905E}, \citet{2011MNRAS.412..246V}, and
\citet{2013ApJ...779..138B}.

Thorough studies of SED-fitting code performance have identified and
quantified their tendency to derive overconfident $\mathcal{P}(z)$. This
means that the confidence intervals derived for the $z_{\mathrm{phot}}$
are too narrow. Given that we base our photometric cluster members
identification on the $\mathcal{P}(z)$ provided by \textsc{EAZY}, we
perform a simple check to evaluate the impact of this effect on our
work. In practice, we check that the distribution of spectroscopic
redshifts in the cluster is comparable with the distribution
obtained combining the photometric redshifts $\mathcal{P}(z)$
\citep{2010MNRAS.403.2137S}. Additionally, we perform the check
described by \citet{2016MNRAS.457.4005W} through which we find that the
overconfidence of the $\mathcal{P}(z)$ we use can be corrected broadening
it by applying a convolution with a $\sigma$$=$$0.2$ gaussian. We
have checked that the impact of this effect on our work is negligible
in the final selection of cluster members, given that broadening the
$\mathcal{P}(z)$ leads to a different calibration of the membership
determination method with smaller $P_{\mathrm{thr}}$.

\section{Cluster members \& field reference samples}\label{clusters_members_samples}

The main objective of our study is to compare the SF activity that
takes place in the inner region of intermediate redshift clusters with
the typical observed in lower density environments (i.e., field). In this section,
we describe the different galaxy samples from which we derive the
results of this work. \textcolor{black}{ {In the rest of the article the samples are
frequently subdivided in three increasing redshift bins (0.2$<$$z$$<$0.4,
0.4$<$$z$$<$0.6, 0.6$<$$z$$<$0.9). The two first bins are chosen to have
equal number of clusters (11), while the last one includes only the
two highest redshift ones. Furthermore, the samples are divided into three
cluster-centric distance ($\mathcal{R}$) bins. 
The first bin ($\mathcal{R}/R_{200}$$<$0.1)
is the only one available across the whole redshift range. The
second one (0.1$<$$\mathcal{R}/R_{200}$$<$0.2) is
visible in the two highest redshift bins. Finally, the third one
(0.2$<$$\mathcal{R}/R_{200}$$<$0.3) is covered only
in the highest redshift clusters. Table~\ref{tab:samples_c:R1},
\ref{tab:samples_c:R2}, and~\ref{tab:samples_c:R3} show the
number counts and average properties of the various galaxy clusters
subsamples. Table~\ref{tab:samples_f} displays the number counts and
average properties of field galaxy samples.}}

\subsection{Samples of cluster members}\label{clusters_members}

For each CLASH+HLS field, we build a general \textit{cluster members
sample} out of the previously described CLASH parent catalogues. We
consider only sources with a $>$3$\sigma$ detection in IRAC\,4.5$\mu$m
band to avoid spurious and extremely faint systems, and 
fluxes larger than the average limiting fluxes at 3$\sigma$ level (see
Table~\ref{tab:photometry_ir} for the limiting fluxes at 5$\sigma$
detection level). Using the methodology
described in Section~\ref{members_selection}, we select a total of 3121
cluster members distributed into the 24 clusters analysed. This number
does not include the 259 galaxies for which the SED-fitting is not able
to derive an accurate value of mass: those sources fitted with a
template of an active galaxy and sources with fewer than 4 photometric
data points.

Figure~\ref{fig:limits} represents the distribution with redshift
of the $\mathcal{M}_{*}$ estimations derived through the SED-fitting
(Section~\ref{rainbow}) for the cluster members parent sample. We also
represent the $\mathcal{M}_{*}$ limits given the 3$\sigma$ IRAC\,4.5$\mu$m
limit fluxes for each cluster (see Table~\ref{tab:photometry_ir}).
\textcolor{black}{ {This conservative estimations are performed using the
same set of templates described in Section~\ref{rainbow} with solar
metallicity, $\tau$\,$=$\,1\,Myr, and an age that corresponds to the
age of the Universe at each redshift.}}

To create comparable galaxy samples at different
redshifts, we focus our analysis on cluster members with
log$_{10}\mathcal{M}_{*}/M_{\odot}$$>$10. Our final cluster members
sample contain 1518 galaxies.

We have performed a comparison between the cluster members we
select using our approach and the members catalogues published
by \citet{2017ApJ...848...37C} for all CLASH clusters.  On
average, 90$^{+3}_{-7}$\% of the galaxies with 
log$_{10}\mathcal{M}_{*}/M_{\odot}$$>$10 in each of our 
samples have a counterpart in their
general catalogues. Among them, 87$^{+9}_{-8}$\% are also
considered cluster members by \citet{2017ApJ...848...37C}. Finally,
only a 6$^{+14}_{-4}$\% of galaxies included in the cluster members
catalogues of their publication are not included in our cluster members
samples. Therefore, in this range of stellar masses the differences are 
within our estimated levels of completeness and contamination. 

\begin{figure}
\includegraphics[width=1\linewidth]{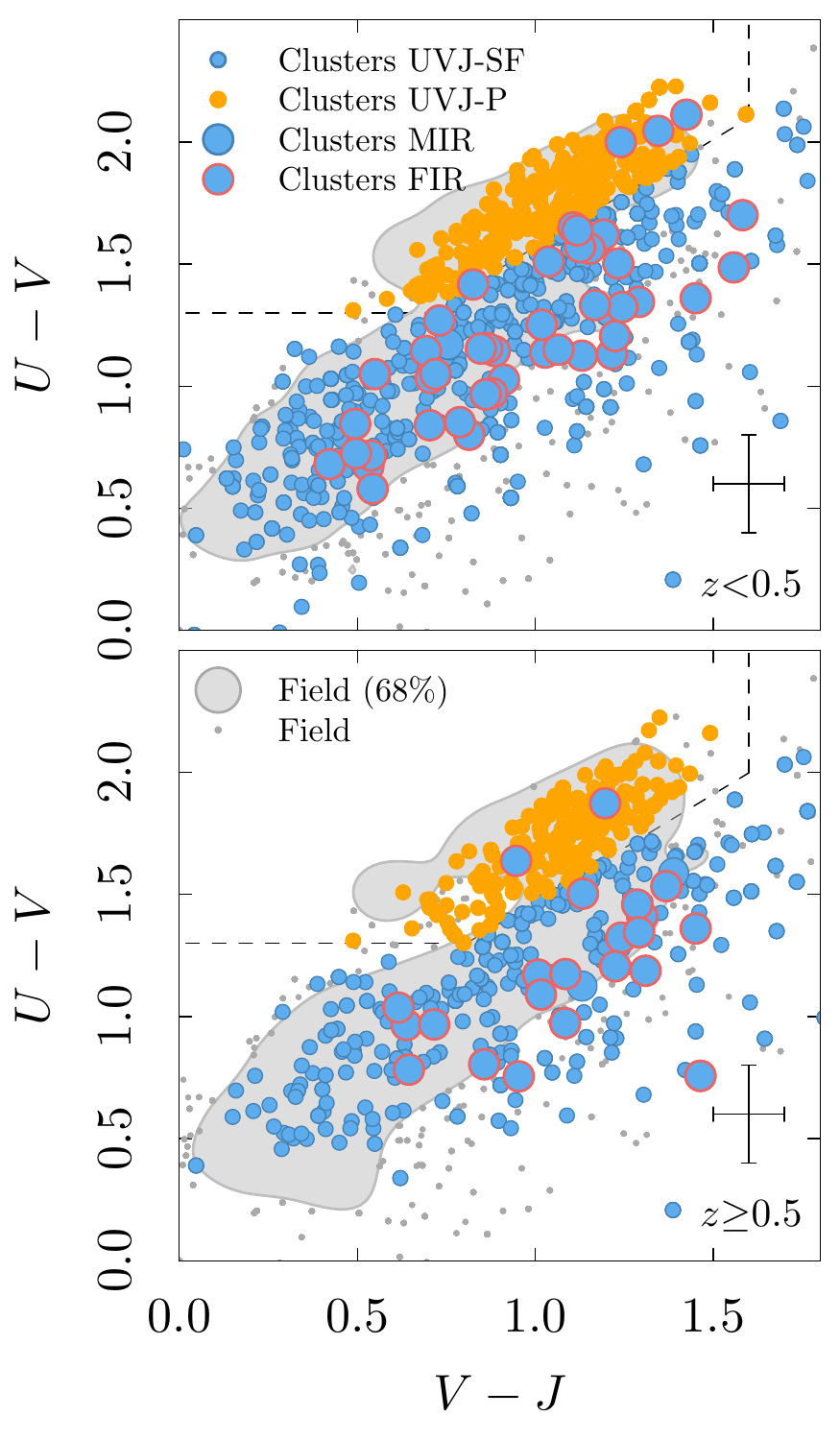}
\caption{$UVJ$-diagram for the cluster members (circles) and 
field galaxies (grey contours and points) in two redshift bins (top panel
0<$z$<0.5; bottom panel, 0.5<$z$<1.0). Dashed lines mark the corresponding
boundaries defined by \protect\citet{2009ApJ...691.1879W} to distinguish
between quiescent and SFGs. The circles that represent
those cluster members within the locus of the passive (star-forming)
galaxies are coloured in orange (blue). The cluster members detected in
the FIR are highlighted with larger blue
circles and a red border.} \label{fig:uvj}
\end{figure}

\begin{figure}
\includegraphics[width=1\linewidth]{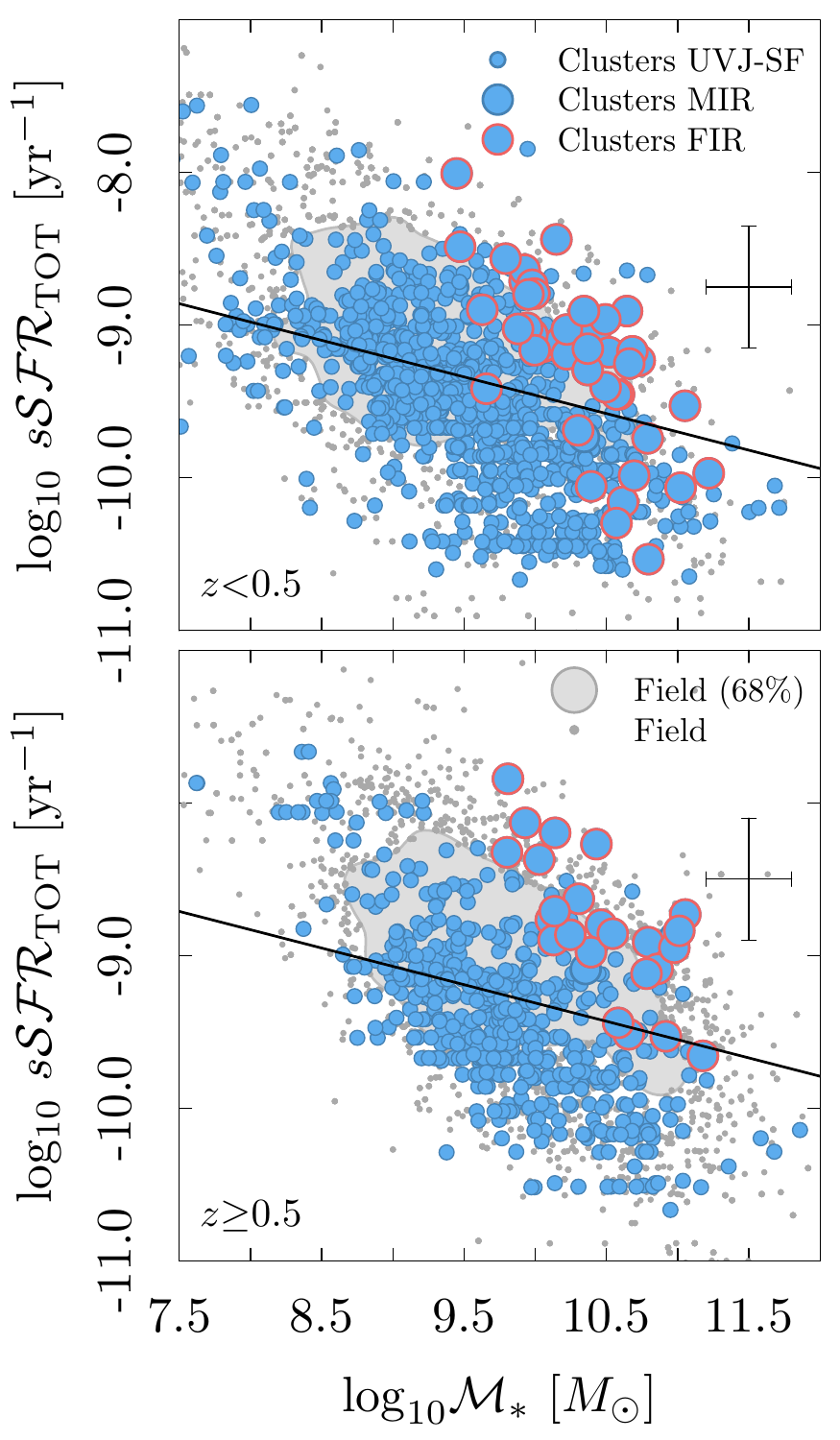}
\caption{$s\mathcal{SFR}_{\mathrm{TOT}}$ vs. $\mathcal{M}_{*}$ for 
the star-forming cluster members (blue points) and field galaxies (grey
contours and points) in the two redshift bins in Figure~\protect\ref{fig:uvj}.  The
cluster members detected in the FIR are highlighted with larger
blue circles and a red border. The black lines represent the MS by
\protect\citet{2015ApJ...801L..29R} scaled to the median redshift of
the corresponding bin considering an evolution with redshift of the
$s\mathcal{SFR}$ of the shape $(1+z)^{2.8\pm0.1}$
\protect\citep{2012ApJ...747L..31S}.} \label{fig:ms_s}
\end{figure}

\subsection{Samples of field galaxies}

In order to build a reference sample to which compare the properties of
the cluster members, we make use of the outstanding datasets available
on three of the CANDELS fields (\citealt{2011ApJS..197...35G},
\citealt{2011ApJS..197...36K}). In particular, we focus on both the
GOODS fields (\citealt{2004ApJ...600L..93G}; see Sections~\ref{goodss},
\ref{goodsn}) and COSMOS (\citealt{2007ApJS..172....1S}; see
Section~\ref{cosmos}). 

Using an analogous approach to that described in
Sections~\ref{multi-wavelength-photo},~\ref{zphot}, and~\ref{rainbow},
we create multi-wavelength catalogues and derive the photometric redshifts
and physical properties (e.g., $\mathcal{M}_{*}$, $\mathcal{SFR}$) 
of the galaxies in CANDELS catalogues. Then,
we apply the same spectroscopic and photometric redshift criteria to 
select a \textit{field sample} corresponding
to each cluster members sample in terms of redshift range. Then, for each 
field sample, we select only the galaxies with a $>$3$\sigma$ detection 
in IRAC\,4.5$\mu$m band and a IRAC\,4.5$\mu$m flux larger than the 
3$\sigma$ detection limit of each corresponding cluster sample. 
Figure~\ref{fig:limits} represents the distribution of the field samples in the $\mathcal{M}_{*}$-$z$ plane. 

The final field parent sample contains 7466 systems with
log$_{10}\mathcal{M}_{*}/M_{\odot}$$>$10. We exclude the 360 galaxies
without a robust mass estimation (see previous section).

\subsection{Samples of star-forming and passive galaxies}
\begin{figure*}
\includegraphics[width=0.9\linewidth]{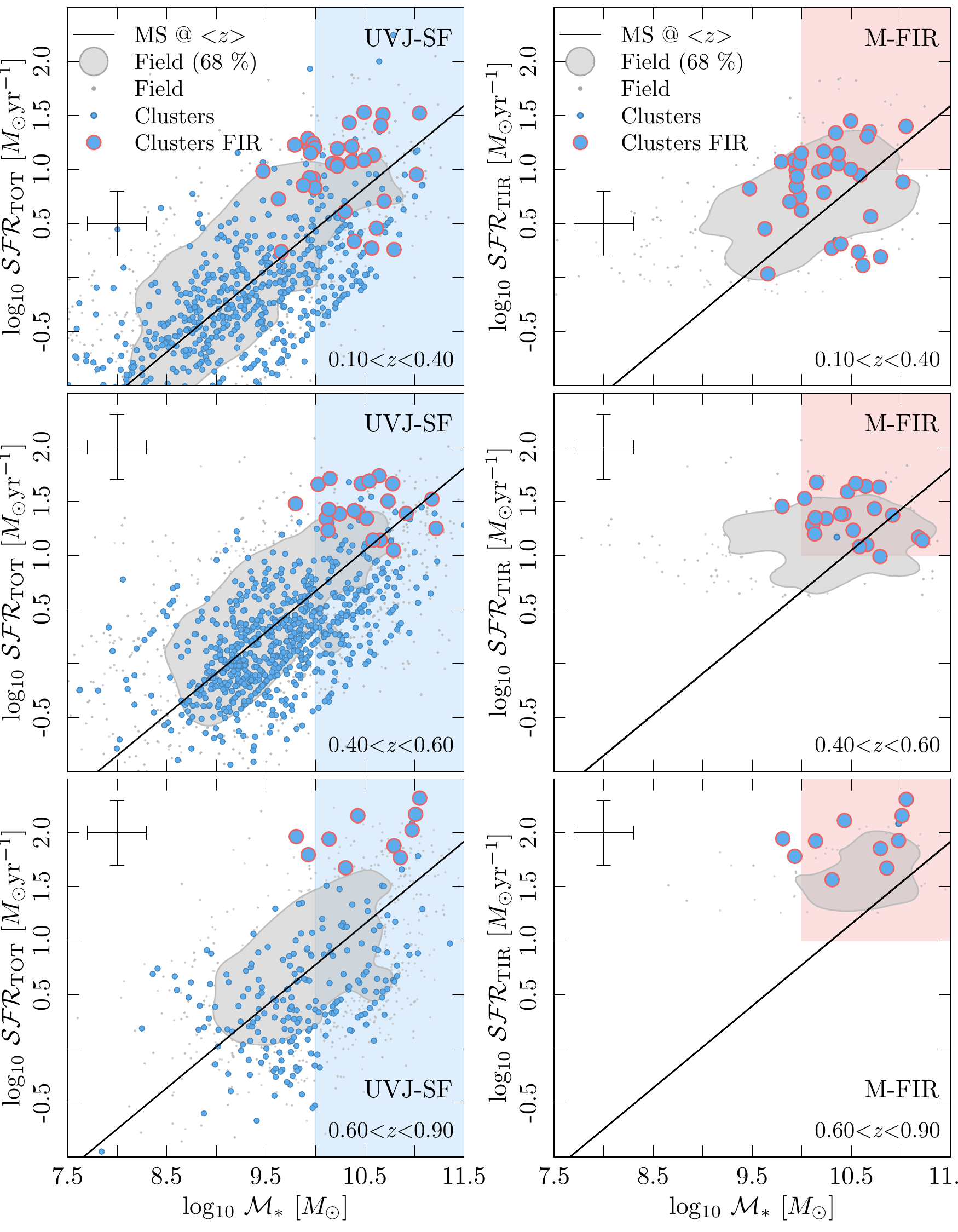}
\caption{$\mathcal{SFR}_{\mathrm{TOT}}$ \textit{vs} $\mathcal{M}_{*}$ relation
for the star-forming cluster members in our study split up in three
increasing $z$ bins (top, middle, and bottom panels). On the (left-)
right-hand panels, we include the ($UVJ$-SF) M-FIR galaxies across
the whole mass range. The $\mathcal{SFR}_{\mathrm{TOT}}$ refers
to the $\mathcal{SFR}_{\mathrm{TIR}}$+$\mathcal{SFR}_{\mathrm{UV}}$
for those galaxies M-FIR detected, and $\mathcal{SFR}_{UV,corr}$
otherwise. Blue points always represent the distribution of clusters
members in both cases. Those galaxies detected in the FIR (i.e.,
\textit{Herschel}) are shown with larger blue points highlighted
with red borders. grey contours represent the distribution 
(68 confidence levels) of field galaxies. We also display the MS by
\protect\citet[][black lines]{2015ApJ...801L..29R} scaled to the median
redshift of the corresponding subsample of cluster members considering
a trend of $s\mathcal{SFR}$ with redshift $\propto(1+z)^{2.8\pm0.1}$
\protect\citep{2012ApJ...747L..31S}. The shaded areas represent
the selection criteria used to build the final samples of $UVJ$-SF and
M-FIR galaxies (i.e. they represent the cut in $\mathcal{M}_{*}$,
and $\mathcal{SFR}_{\mathrm{TIR}}$). \label{fig:MS}}
\end{figure*}

\begin{figure*}
\includegraphics[width=1\linewidth]{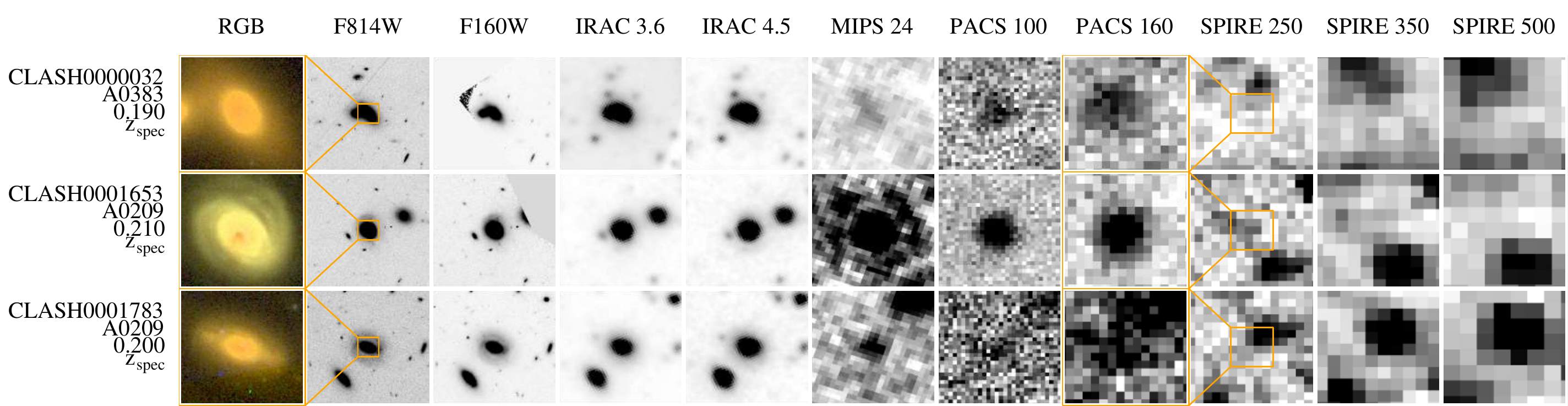}
\caption{Thumbnails of 3 cluster members from the M-FIR sample ordered by 
increasing redshift. From left to right we display a RGB image
(5"$\times$5") created using \textit{HST}/ACS/F814W, F606W, and
F435W, following the methodology by \citet{2004PASP..116..133L},
30"$\times$30" postage stamps in the \textit{HST}/ACS/F814W and
\textit{HST}/WFC3/F160W bands followed by \textit{Spitzer}/IRAC and MIPS,
and \textit{Herschel}/PACS bands, and 90"$\times$90" postage stamps in the
\textit{Herschel}/SPIRE bands, all ordered by increasing wavelength.  When
there is a difference in the sizes of two adjacent frames, we mark with
an orange square the size of the smallest on the largest.  On the left
side we show the ID of the object, the name of the cluster, the redshift
and either if it is photometric or spectroscopic. The 
thumbnails of the rest of the sample can be found as online material.}\label{cromos_1}
\end{figure*}

We divide the samples of field and cluster galaxies into star-forming
and passive using the rest-frame $U$$-$$V$ \textit{vs} $V$$-$$J$
colour-colour space (hereafter, $UVJ$-diagram).  Different works
(e.g., \citealt{2007ApJ...655...51W}, \citealt{2009ApJ...691.1879W})
have evidenced the power of the $UVJ$-diagram
to select \textit{pure} samples of either quiescent and SFGs
(e.g., \citealt{2007ApJ...655...51W}, \citealt{2011ApJ...739...24B}, 
\citealt{2012ApJ...745..179W}, \citealt{2015ApJ...811L..12W}).
\textcolor{black}{ {In particular, we identify passive galaxies (hearafter, $UVJ$-P)
following the recipes by \citet{2009ApJ...691.1879W} for the redshift
bins 0$<$$z$$<$0.5 ($U$$-$$V$$>$0.88$\times$$V$$-$$J$$+$0.69,
$U$$-$$V$$>$1.3, and $V$$-$$J$$<$1.6) and 0.5$<$$z$$<$1.0
($U$$-$$V$$>$0.88$\times$$V$$-$$J$$+$0.59, $U$$-$$V$$>$1.3, and
$V$$-$$J$$<$1.6). Galaxies with rest-frame $U$$-$$V$ and $V$$-$$J$
behaving otherwise are classified as star-forming (hereafter,
$UVJ$-SF). We perform Monte Carlo simulations to assess the
reciprocal contamination between the two types of galaxies considering
the uncertainties in the synthetic photometry. We retrieve $\leq$1\%
differences in the number counts of either category and sample. We
find that in the clusters (field) samples, 25\% (5\%) of SFGs could be
classified as passive given their error bars and 28\% (22\%) of passive
galaxies could be classified as SFGs. We have checked that excluding
the galaxies in the vicinities of the limits between the $UVJ$-P and the
$UVJ$-SF loci do not change the results of our work significantly. This
is probably due to the fact that these \textit{transition} galaxies
present similar properties on either side of the border.}}

In Figure~\ref{fig:uvj}, we show the $UVJ$-diagram for the cluster
and field samples. As we can see, some galaxies detected in the FIR
(i.e., presumably SFGs) are located in the region theoretically populated
by passive galaxies. This contamination has been reported in the past
(see, e.g., \citealt{2016MNRAS.457.3743D}) and evidences the necessity
of a correction of the aforementioned selection criteria. In the final
$UVJ$-SF ($UVJ$-P) samples, we include (exclude) both the
galaxies located in the SFGs locus of the $UVJ$-diagram and those
detected in the M/FIR (see Section~\ref{sect:subsampl}) independently of
their position in the $UVJ$-diagram.  This correction increases
(decreases) 1\% (1\%) and 2\% (5\%) the number of star-forming (passive)
galaxies in the cluster and field samples, respectively.

The $UVJ$-SF ($UVJ$-P) samples built in CLASH-HLS
clusters and the field include 443 (1075) and 4649 (2817)
log$_{10}\mathcal{M}_{*}/M_{\odot}$$>$10 galaxies, respectively.

An alternative methodology to select SFGs uses a threshold of
$s\mathcal{SFR}$ under which a galaxy is considered to be passive
(e.g., \citealt{2009MNRAS.394.1131K}). In Figure~\ref{fig:ms_s}, we
represent the $s\mathcal{SFR}_{\mathrm{TOT}}$-$\mathcal{M}_{*}$ diagrams
for the $UVJ$-SF samples in the redshift bins of 
Figure~\ref{fig:uvj}. We can see that our $UVJ$-SF selection criteria 
corresponds approximately to 
$\mathrm{log}_{10}$\,$s\mathcal{SFR}_{\mathrm{TOT}}$/yr$^{-1}$$\gtrsim$$-10.5$.

On the left-hand half of Figure~\ref{fig:MS}, we display the distribution
of the $UVJ$-SF samples selected in the clusters and the field on
the $\mathcal{SFR}_{\mathrm{TOT}}$-$\mathcal{M}_{*}$ plane.  The blue shaded
area illustrates the effective definition of the $UVJ$-SF samples
considered in the rest of the work. For comparison, we also represent
the MS defined by \citet[][black line]{2015ApJ...801L..29R} scaled to the
median redshift of the bin, assuming and evolution with redshift of the 
$s\mathcal{SFR}$ of the shape $(1+z)^{2.8\pm0.1}$ \citep{2012ApJ...747L..31S}. 
We notice a systematic offset of the distribution of cluster SFGs towards lower
$\mathcal{SFR}$ at fixed $\mathcal{M}_{*}$ (see also Figure~\ref{fig:ms_s}). 
The quantification of this difference can be found in Section~\ref{sfr_ssfr}. 

\subsection{Samples of M- and/or FIR-detected galaxies} \label{sect:subsampl}

In order to build comparable samples of galaxies
(log$_{10}\mathcal{M}_{*}/M_{\odot}$$>$10) detected in the M- and/or
FIR (M-FIR samples), we perform the following steps. First, we select
galaxies with at least a 3$\sigma$ detection in one of the M- and/or
FIR bands available (i.e., MIPS 24$\mu$m, PACS 100 \& 160$\mu$m, and
SPIRE 250, 350 \& 500$\mu$m), and flux larger than the limiting fluxes
at 3$\sigma$ level in the clusters (see Table~\ref{tab:photometry_ir}
for the limiting fluxes at 5$\sigma$ detection level). These galaxies
are represented in the bottom panel of Figure~\ref{fig:limits}. \textcolor{black}{ {Then,
we select only the 50 (1496) clusters (field) galaxies for which
the estimated $\mathcal{SFR}_{TIR}$ is larger than the (conservative)
$\mathcal{SFR}_{\mathrm{TIR}}$ limits obtained for each cluster (black symbols
in the bottom panel of Figure~\ref{fig:limits}).}} Figure~\ref{cromos_1}
shows the thumbnails of the cluster members detected in the M-
and/or FIR. Finally, we consider galaxies
with $\mathcal{SFR}_{\mathrm{TIR}}$$>$$10$$M_{\odot}yr^{-1}$ to obtain 
a comparable set of samples of galaxies throughout the whole redshift range. 
\textcolor{black}{ {This value is larger than the $\mathcal{SFR}_{\mathrm{TIR}}$ limits of our 
sample, except for the four furthest clusters.}} Our
final \textit{M-FIR samples} include 36 cluster members and 974
field galaxies. On the right-hand half of Figure~\ref{fig:MS},
we display the distribution of these samples on the
$\mathcal{SFR}_{\mathrm{TIR}}$-$\mathcal{M}_{*}$ plane. The red shaded
area marks the $\mathcal{M}_{*}$ and $\mathcal{SFR}_{\mathrm{TIR}}$ cuts
performed to define the samples. 

It is worth mentioning that we perform a visual inspection of each
cluster member selected as a M-FIR emitter. We exclude spurious MIPS
24$\mu$m sources without a counterpart in longer wavelengths (e.g.,
sources on Airy ring features), galaxies in the borders of the images
that are selected as counterparts of M/FIR sources with coordinates
outside the area covered by CLASH catalogues, or galaxies suffering from
over-deblending in the CLASH catalogues.

Interestingly, we find 8 BCGs detected in
the M/FIR out of 24 clusters, which corresponds to 33\% of our
sample. This percentage is consistent with the results of the study
conducted by \citet{2012ApJ...747...29R} using HLS data on a sample
of 68 massive galaxy clusters spread out in the redshift range between
0.08$<$$z$$<$1.00. Their sample includes only 12 CLASH+HLS clusters. As
expected, among the BCGs of these 12 systems, we detect traces of obscured
SF in the same two, namely A0383 and MACS1423.  We exclude BCGs from our
samples given their unique SFH and in order to focus our results on the
SF activity of the general cluster galaxy population.

The fraction of active galactic nuclei (AGN) among IR-bright cluster
members has been observed to increase rapidly from 3\% up to 65\% for
galaxies with increasing $\mathcal{L}_{\mathrm{TIR}}$ values varying from
$10^{11}$$L_{\odot}$ to $>$$10^{11.6}$$L_{\odot}$ in clusters within the
redshift range 0.15$<$$z$$<$0.30 \citep{2013ApJ...775..126H}. Given the 
SED-fitting methodology explained and sample selection, 
we exclude from our analysis the 
galaxies whose photometry was fitted to AGN templates. 

The so-called luminous and ultra-luminous infrared galaxies (LIRGs and
ULIRGs, respectively) display $\mathcal{L}_{\mathrm{TIR}}$ in the range of
$10^{11}$$L_{\odot}$$<$$\mathcal{L}_{\mathrm{TIR}}$$<$$10^{12}$$L_{\odot}$
and $\mathcal{L}_{\mathrm{TIR}}$$>$$10^{12}$$L_{\odot}$, respectively,
which correspond to $\mathcal{SFR}_{\mathrm{TIR}}$ from tens to thousands
of $M_{\odot}$yr$^{-1}$. Our M-FIR sample of cluster members includes 25
LIRGs and 1 ULIRGs (within CLJ1226, the highest redshift cluster) and our
M-FIR sample of field galaxies includes 639 LIRGs, and 10 ULIRGs. These
numbers correspond to comparable percentages of LIRGs and ULIRGs within
the M-FIR samples in clusters and field.

\begin{landscape}
\begin{table}
\scriptsize
	\centering
	\caption{Number of galaxies selected with the different criteria 
used to build the final samples of star-forming cluster members 
(\textcolor{black}{ {$\mathcal{R}/R_{200}$$<$0.1}}), and
average SF activity indicators. In particular, we report: (1) ID of the
corresponding field; (2) number of cluster members with a detection at
a level $>$3$\sigma$ in IRAC\,4.5$\mu$m and $\mathcal{M}_{*}>10^{10}
\mathrm{M}_\odot$; we show within parentheses the number of cluster
members without the $\mathcal{M}_{*}$ cut; (3) number of galaxies
selected as star-forming using the UVJ diagram and/or detected
in the MIR and/or FIR ($\mathcal{M}_{*}>10^{10} \mathrm{M}_\odot$),
what we call the \textit{$UVJ$-SF sample}; (4) cluster members with
$\mathcal{M}_{*}>10^{10} \mathrm{M}_\odot$ detected in the MIR and/or FIR
with a $\mathcal{SFR}_{\mathrm{TIR}}>$10$\mathrm{M}_\odot\mathrm{yr}^{-1}$,
what we call \textit{M-FIR sample}; we show the total number
without $\mathcal{SFR}_{\mathrm{TIR}}$ cut within parentheses; (5 \& 6)
fraction of $UVJ$-SF and M-FIR galaxies, respectively, obtained using as
reference the number of cluster members with $\mathcal{M}_{*}>10^{10}
\mathrm{M}_\odot$; (7 \& 8) median and quantiles 16$^{\mathrm{th}}$ and
84$^{\mathrm{th}}$ values of the $\mathcal{M}_{*}$ of the $UVJ$-SF and
M-FIR samples respectively; (9) median and quantiles 16$^{\mathrm{th}}$
and 84$^{\mathrm{th}}$ values of the $\mathcal{SFR}_{\mathrm{TOT}}$ for the
$UVJ$-SF sample obtained as the addition of the $\mathcal{SFR}_{\mathrm{TIR}}$
and the $\mathcal{SFR}_{\mathrm{UV}}$ when the former is available, and
the $\mathcal{SFR}_{\mathrm{UV, corr}}$ in the rest of the cases; (10)
median and quantiles 16$^{\mathrm{th}}$ and 84$^{\mathrm{th}}$ values of the
$\mathcal{SFR}_{\mathrm{TOT}}$ for the M-FIR sample obtained as the addition
of the $\mathcal{SFR}_{\mathrm{TIR}}$ and the $\mathcal{SFR}_{\mathrm{UV}}$;
(11 \& 12) median and quantiles 16$^{\mathrm{th}}$ and 84$^{\mathrm{th}}$
values of the $s\mathcal{SFR}_{\mathrm{TOT}}$ for the $UVJ$-SF and the M-FIR
sample, respectively.}
	\label{tab:samples_c:R1}
	\begin{tabular}{lccccccccccc} 
		\hline
		\rotatebox{0}{Cluster ID} & \multicolumn{1}{c}{Members} & \multicolumn{1}{c}{$UVJ$-SF} & \multicolumn{1}{c}{M-FIR} & \multicolumn{1}{c}{$\mathcal{F}_{UVJ\mathrm{-SF}}$} &  \multicolumn{1}{c}{$\mathcal{F}_{\mathrm{M-FIR}}$} & \multicolumn{1}{c}{$\mathcal{M}_{*,\mathrm{UVJ}-SF}$} & \multicolumn{1}{c}{$\mathcal{M}_{*,\,\mathrm{M-FIR}}$} & \multicolumn{1}{c}{$\mathcal{SFR}_{\mathrm{TOT}, UVJ\mathrm{-SF}}$} & \multicolumn{1}{c}{$\mathcal{SFR}_{\mathrm{TOT, M-FIR}}$} & \multicolumn{1}{c}{$s\mathcal{SFR}_{\mathrm{TOT}, UVJ\mathrm{-SF}}$} & \multicolumn{1}{c}{$s\mathcal{SFR}_{\mathrm{TOT, M-FIR}}$} \\
		(1) & \multicolumn{1}{c}{(2)} & \multicolumn{1}{c}{(3)} & \multicolumn{1}{c}{(4)} & \multicolumn{1}{c}{(5)} & \multicolumn{1}{c}{(6)} & \multicolumn{1}{c}{(7)} & \multicolumn{1}{c}{(8)} & \multicolumn{1}{c}{(9)} & \multicolumn{1}{c}{(10)} & \multicolumn{1}{c}{(11)} & \multicolumn{1}{c}{(12)}\\
		\hline
a0383 & 11 (53) & 2 & 0 (0) & 0.17$\pm$0.11 & -- & 10.70$^{+0.13}_{-0.12}$ & -- & 0.46$^{+0.12}_{-0.11}$ & -- & -10.24$^{+0.01}_{-0.01}$ & -- \\ 
a0209 & 23 (72) & 8 & 0 (1) & 0.35$\pm$0.10 & -- & 10.65$^{+0.39}_{-0.24}$ & -- & 0.68$^{+0.43}_{-0.40}$ & -- & -10.19$^{+0.31}_{-0.09}$ & -- \\ 
a2261 & 30 (192) & 7 & 0 (3) & 0.23$\pm$0.08 & -- & 10.23$^{+0.11}_{-0.02}$ & -- & 0.55$^{+0.06}_{-0.24}$ & -- & -9.73$^{+0.06}_{-0.33}$ & -- \\ 
rbs1748 & 14 (48) & 5 & 0 (0) & 0.36$\pm$0.13 & -- & 10.28$^{+0.13}_{-0.12}$ & -- & 0.06$^{+0.14}_{-0.10}$ & -- & -10.25$^{+0.10}_{-0.04}$ & -- \\ 
a0611 & 18 (34) & 6 & 0 (0) & 0.33$\pm$0.11 & -- & 10.24$^{+0.25}_{-0.13}$ & -- & 0.48$^{+0.42}_{-0.55}$ & -- & -9.85$^{+0.61}_{-0.46}$ & -- \\ 
ms2137 & 6 (11) & 2 & 0 (0) & 0.33$\pm$0.19 & -- & 10.28$^{+0.03}_{-0.03}$ & -- & -0.18$^{+0.02}_{-0.02}$ & -- & -10.46$^{+0.01}_{-0.01}$ & -- \\ 
as1063 & 28 (48) & 15 & 1 (1) & 0.54$\pm$0.09 & 0.04$\pm$0.04 & 10.45$^{+0.16}_{-0.23}$ & 10.37$^{--}_{--}$ & 0.11$^{+1.01}_{-0.31}$ & 1.21$^{--}_{--}$ & -10.29$^{+0.59}_{-0.19}$ & -9.15$^{--}_{--}$ \\ 
macs1931 & 22 (47) & 8 & 0 (0) & 0.36$\pm$0.10 & -- & 10.25$^{+0.37}_{-0.12}$ & -- & 0.48$^{+0.70}_{-0.66}$ & -- & -9.71$^{+0.39}_{-0.72}$ & -- \\ 
macs1115 & 18 (31) & 4 & 0 (0) & 0.22$\pm$0.10 & -- & 10.31$^{+0.14}_{-0.10}$ & -- & 0.01$^{+0.24}_{-0.13}$ & -- & -10.32$^{+0.26}_{-0.15}$ & -- \\ 
rxj1532 & 14 (28) & 4 & 0 (0) & 0.29$\pm$0.12 & -- & 10.46$^{+0.29}_{-0.28}$ & -- & 0.53$^{+0.91}_{-0.29}$ & -- & -9.89$^{+0.82}_{-0.30}$ & -- \\ 
macs1720 & 15 (37) & 4 & 0 (0) & 0.27$\pm$0.11 & -- & 10.47$^{+0.37}_{-0.32}$ & -- & 0.62$^{+0.32}_{-0.24}$ & -- & -9.83$^{+0.06}_{-0.08}$ & -- \\ 
macs0416 & 34 (53) & 12 & 1 (1) & 0.35$\pm$0.08 & 0.03$\pm$0.03 & 10.35$^{+0.21}_{-0.18}$ & 10.43$^{--}_{--}$ & 0.55$^{+0.16}_{-0.20}$ & 1.40$^{--}_{--}$ & -9.85$^{+0.12}_{-0.10}$ & -9.03$^{--}_{--}$ \\ 
macs0429 & 8 (21) & 4 & 0 (0) & 0.50$\pm$0.18 & -- & 10.42$^{+0.50}_{-0.07}$ & -- & 0.86$^{+0.45}_{-0.49}$ & -- & -9.98$^{+0.49}_{-0.09}$ & -- \\ 
macs1206 & 35 (73) & 15 & 1 (1) & 0.43$\pm$0.08 & 0.03$\pm$0.03 & 10.52$^{+0.09}_{-0.35}$ & 11.22$^{--}_{--}$ & 0.67$^{+0.45}_{-0.36}$ & 1.25$^{--}_{--}$ & -9.84$^{+0.38}_{-0.39}$ & -9.97$^{--}_{--}$ \\ 
macs0329 & 13 (34) & 8 & 0 (0) & 0.62$\pm$0.13 & -- & 10.47$^{+0.10}_{-0.22}$ & -- & 0.54$^{+0.33}_{-0.19}$ & -- & -9.88$^{+0.22}_{-0.11}$ & -- \\ 
rxj1347 & 28 (44) & 12 & 0 (0) & 0.43$\pm$0.09 & -- & 10.27$^{+0.62}_{-0.14}$ & -- & 0.32$^{+0.62}_{-0.31}$ & -- & -9.93$^{+0.08}_{-0.35}$ & -- \\ 
macs1311 & 22 (42) & 8 & 1 (1) & 0.36$\pm$0.10 & 0.05$\pm$0.04 & 10.40$^{+0.12}_{-0.28}$ & 10.52$^{--}_{--}$ & 0.68$^{+0.61}_{-0.41}$ & 1.34$^{--}_{--}$ & -9.85$^{+0.72}_{-0.42}$ & -9.18$^{--}_{--}$ \\ 
macs1149 & 42 (82) & 20 & 0 (0) & 0.48$\pm$0.08 & -- & 10.52$^{+0.24}_{-0.26}$ & -- & 0.47$^{+0.66}_{-0.46}$ & -- & -9.92$^{+0.29}_{-0.56}$ & -- \\ 
macs0717 & 57 (72) & 8 & 0 (0) & 0.14$\pm$0.05 & -- & 10.32$^{+0.34}_{-0.28}$ & -- & 0.35$^{+0.14}_{-0.14}$ & -- & -9.97$^{+0.28}_{-0.49}$ & -- \\ 
macs1423 & 26 (30) & 7 & 1 (1) & 0.27$\pm$0.09 & 0.04$\pm$0.04 & 10.47$^{+0.72}_{-0.31}$ & 10.78$^{--}_{--}$ & 0.54$^{+1.13}_{-0.45}$ & 1.66$^{--}_{--}$ & -10.00$^{+0.27}_{-0.10}$ & -9.12$^{--}_{--}$ \\ 
macs2129 & 17 (18) & 4 & 1 (1) & 0.24$\pm$0.10 & 0.06$\pm$0.06 & 10.24$^{+0.27}_{-0.17}$ & 10.14$^{--}_{--}$ & 0.58$^{+0.56}_{-0.33}$ & 1.43$^{--}_{--}$ & -9.75$^{+0.58}_{-0.26}$ & -8.71$^{--}_{--}$ \\ 
macs0647 & 17 (24) & 7 & 0 (0) & 0.41$\pm$0.12 & -- & 10.82$^{+0.27}_{-0.58}$ & -- & 0.63$^{+0.76}_{-0.09}$ & -- & -10.08$^{+0.45}_{-0.13}$ & -- \\ 
macs0744 & 20 (37) & 9 & 0 (0) & 0.45$\pm$0.11 & -- & 10.69$^{+0.11}_{-0.30}$ & -- & 0.68$^{+0.79}_{-0.09}$ & -- & -9.76$^{+0.57}_{-0.37}$ & -- \\ 
clj1226 & 33 (57) & 18 & 0 (0) & 0.55$\pm$0.09 & -- & 10.45$^{+0.34}_{-0.39}$ & -- & 0.49$^{+0.31}_{-0.28}$ & -- & -10.08$^{+0.34}_{-0.16}$ & -- \\ 
\hline
Total & 551 (1188) &197 & 6 (10) \\
		\cline{1-4}
& \multirow{3}{*}{Median}  & \multirow{1}{*}{0.2$<$$z$$<$0.4} & $\mathcal{R}/R_{200}$$<$0.1 & $0.33^{+0.03}_{-0.10}$ & -- & $10.30^{+0.15}_{-0.03}$ & $10.37^{--}_{--}$ & $0.49^{+0.05}_{-0.37}$ & $1.21^{--}_{--}$ & $-10.19^{+0.34}_{-0.06}$ & $-9.15^{--}_{--}$ \\
		\cline{3-12}
& & \multirow{1}{*}{0.4$<$$z$$<$0.6}  & $\mathcal{R}/R_{200}$$<$0.1 & $0.41^{+0.07}_{-0.16}$ & $0.00^{+0.04}_{-0.00}$ & $10.42^{+0.05}_{-0.02}$ & $10.52^{+0.26}_{-0.09}$ & $0.55^{0.08}_{-0.01}$ & $1.40^{+0.03}_{-0.06}$ & $-9.93^{+0.08}_{-0.04}$ & $-9.12^{+0.10}_{-0.06}$ \\
		\cline{3-12}
& & \multirow{1}{*}{0.6$<$$z$$<$0.9} & $\mathcal{R}/R_{200}$$<$0.1 & $0.50^{+0.03}_{-0.03}$ & -- & $10.57^{+0.12}_{-0.12}$ & -- & $0.58^{+0.09}_{-0.09}$ & -- & $-9.92^{+0.16}_{-0.16}$ & -- \\
		\cline{2-12}
\end{tabular}
\end{table}
\end{landscape}

\begin{landscape}
\begin{table}
\scriptsize
	\centering
	\caption{As in Table~\ref{tab:samples_c:R1}, for the galaxies at \textcolor{black}{ {0.1$<$$\mathcal{R}/R_{200}$$<$0.2}}.}
	\label{tab:samples_c:R2}
	\begin{tabular}{lccccccccccc} 
		\hline
		\rotatebox{0}{Cluster ID} & \multicolumn{1}{c}{Members} & \multicolumn{1}{c}{$UVJ$-SF} & \multicolumn{1}{c}{M-FIR} & \multicolumn{1}{c}{$\mathcal{F}_{UVJ\mathrm{-SF}}$} &  \multicolumn{1}{c}{$\mathcal{F}_{\mathrm{M-FIR}}$} & \multicolumn{1}{c}{$\mathcal{M}_{*,\,UVJ\mathrm{-SF}}$} & \multicolumn{1}{c}{$\mathcal{M}_{*,\,\mathrm{M-FIR}}$} & \multicolumn{1}{c}{$\mathcal{SFR}_{\mathrm{TOT}, UVJ\mathrm{-SF}}$} & \multicolumn{1}{c}{$\mathcal{SFR}_{\mathrm{TOT, M-FIR}}$} & \multicolumn{1}{c}{$s\mathcal{SFR}_{\mathrm{TOT}, UVJ\mathrm{-SF}}$} & \multicolumn{1}{c}{$s\mathcal{SFR}_{\mathrm{TOT, M-FIR}}$} \\
		(1) & \multicolumn{1}{c}{(2)} & \multicolumn{1}{c}{(3)} & \multicolumn{1}{c}{(4)} & \multicolumn{1}{c}{(5)} & \multicolumn{1}{c}{(6)} & \multicolumn{1}{c}{(7)} & \multicolumn{1}{c}{(8)} & \multicolumn{1}{c}{(9)} & \multicolumn{1}{c}{(10)} & \multicolumn{1}{c}{(11)} & \multicolumn{1}{c}{(12)}\\
		\hline
a0383 & 10 (55) & 1 & 0 (0) & 0.10$\pm$0.09 & -- & 10.09$^{--}_{--}$ & -- & -0.19$^{--}_{--}$ & -- & -10.28$^{--}_{--}$ & -- \\ 
a0209 & 3 (20) & 1 & 0 (1) & 0.33$\pm$0.27 & -- & 10.59$^{--}_{--}$ & -- & 1.13$^{--}_{--}$ & -- & -9.45$^{--}_{--}$ & -- \\ 
a2261 & 22 (140) & 5 & 0 (1) & 0.23$\pm$0.09 & -- & 10.69$^{+0.14}_{-0.37}$ & -- & 0.51$^{+0.21}_{-0.36}$ & -- & -10.22$^{+0.21}_{-0.03}$ & -- \\ 
rbs1748 & 12 (53) & 1 & 0 (1) & 0.08$\pm$0.08 & -- & 11.02$^{--}_{--}$ & -- & 0.95$^{--}_{--}$ & -- & -10.07$^{--}_{--}$ & -- \\ 
a0611 & 30 (64) & 7 & 1 (2) & 0.23$\pm$0.08 & 0.03$\pm$0.03 & 10.37$^{+0.09}_{-0.15}$ & 10.49$^{--}_{--}$ & 0.59$^{+0.48}_{-0.40}$ & 1.53$^{--}_{--}$ & -9.81$^{+0.64}_{-0.46}$ & -8.96$^{--}_{--}$ \\ 
ms2137 & 13 (25) & 3 & 1 (1) & 0.23$\pm$0.12 & 0.08$\pm$0.07 & 10.22$^{+0.30}_{-0.12}$ & 10.66$^{--}_{--}$ & 0.27$^{+0.78}_{-0.14}$ & 1.41$^{--}_{--}$ & -9.95$^{+0.47}_{-0.02}$ & -9.25$^{--}_{--}$ \\ 
as1063 & 24 (48) & 6 & 0 (0) & 0.25$\pm$0.09 & -- & 10.35$^{+0.13}_{-0.29}$ & -- & -0.14$^{+0.13}_{-0.21}$ & -- & -10.47$^{+0.04}_{-0.05}$ & -- \\ 
macs1931 & 27 (56) & 3 & 0 (0) & 0.11$\pm$0.06 & -- & 10.38$^{+0.12}_{-0.09}$ & -- & 0.24$^{+0.14}_{-0.15}$ & -- & -10.32$^{+0.35}_{-0.02}$ & -- \\ 
macs1115 & 23 (54) & 9 & 1 (1) & 0.39$\pm$0.10 & 0.04$\pm$0.04 & 10.41$^{+0.33}_{-0.22}$ & 10.49$^{--}_{--}$ & 0.76$^{+0.29}_{-0.33}$ & 1.09$^{--}_{--}$ & -9.80$^{+0.58}_{-0.29}$ & -9.41$^{--}_{--}$ \\ 
rxj1532 & 12 (31) & 1 & 0 (1) & 0.08$\pm$0.08 & -- & 10.22$^{--}_{--}$ & -- & 1.03$^{--}_{--}$ & -- & -9.19$^{--}_{--}$ & -- \\ 
macs1720 & 26 (65) & 8 & 1 (1) & 0.31$\pm$0.09 & 0.04$\pm$0.04 & 10.21$^{+0.13}_{-0.12}$ & 11.05$^{--}_{--}$ & 0.36$^{+0.25}_{-0.40}$ & 1.52$^{--}_{--}$ & -9.83$^{+0.24}_{-0.33}$ & -9.53$^{--}_{--}$ \\ 
macs0416 & 24 (63) & 5 & 1 (1) & 0.21$\pm$0.08 & 0.04$\pm$0.04 & 10.64$^{+0.25}_{-0.28}$ & 10.64$^{--}_{--}$ & 0.74$^{+0.62}_{-0.20}$ & 1.74$^{--}_{--}$ & -9.96$^{+0.82}_{-0.19}$ & -8.91$^{--}_{--}$ \\ 
macs0429 & 12 (41) & 5 & 0 (1) & 0.42$\pm$0.14 & -- & 10.51$^{+0.14}_{-0.28}$ & -- & 0.04$^{+0.41}_{-0.13}$ & -- & -10.55$^{+0.57}_{-0.01}$ & -- \\ 
macs1206 & 42 (86) & 9 & 0 (0) & 0.21$\pm$0.06 & -- & 10.16$^{+0.37}_{-0.12}$ & -- & 0.18$^{+0.42}_{-0.38}$ & -- & -10.08$^{+0.43}_{-0.36}$ & -- \\ 
macs0329 & 27 (59) & 6 & 0 (0) & 0.22$\pm$0.08 & -- & 10.28$^{+0.13}_{-0.11}$ & -- & 0.37$^{+0.35}_{-0.26}$ & -- & -10.03$^{+0.45}_{-0.13}$ & -- \\ 
rxj1347 & 29 (67) & 2 & 0 (0) & 0.07$\pm$0.05 & -- & 10.67$^{+0.38}_{-0.38}$ & -- & 1.19$^{+0.16}_{-0.17}$ & -- & -9.48$^{+0.54}_{-0.55}$ & -- \\ 
macs1311 & 27 (61) & 8 & 2 (2) & 0.30$\pm$0.09 & 0.07$\pm$0.05 & 10.33$^{+0.32}_{-0.19}$ & 10.44$^{+0.20}_{-0.20}$ & 1.04$^{+0.44}_{-1.02}$ & 1.61$^{+0.07}_{-0.08}$ & -9.54$^{+0.73}_{-0.78}$ & -8.84$^{+0.27}_{-0.27}$ \\ 
macs1149 & 70 (158) & 16 & 3 (3) & 0.23$\pm$0.05 & 0.04$\pm$0.02 & 10.30$^{+0.38}_{-0.15}$ & 10.25$^{+0.45}_{-0.08}$ & 0.65$^{+0.71}_{-0.36}$ & 1.38$^{+0.01}_{-0.10}$ & -9.67$^{+0.66}_{-0.31}$ & -8.90$^{+0.02}_{-0.43}$ \\ 
macs0717 & 80 (107) & 15 & 6 (6) & 0.19$\pm$0.04 & 0.07$\pm$0.03 & 10.45$^{+0.47}_{-0.29}$ & 10.47$^{+0.29}_{-0.38}$ & 1.14$^{+0.24}_{-0.40}$ & 1.30$^{+0.25}_{-0.16}$ & -9.52$^{+0.48}_{-0.42}$ & -9.27$^{+0.58}_{-0.27}$ \\ 
macs1423 & 17 (29) & 2 & 0 (0) & 0.12$\pm$0.08 & -- & 10.07$^{+0.04}_{-0.05}$ & -- & 0.57$^{+0.13}_{-0.12}$ & -- & -9.49$^{+0.07}_{-0.08}$ & -- \\ 
macs2129 & 33 (38) & 1 & 0 (0) & 0.03$\pm$0.03 & -- & 10.21$^{--}_{--}$ & -- & 0.94$^{--}_{--}$ & -- & -9.27$^{--}_{--}$ & -- \\ 
macs0647 & 27 (54) & 14 & 1 (1) & 0.52$\pm$0.10 & 0.04$\pm$0.04 & 10.30$^{+0.24}_{-0.22}$ & 10.46$^{--}_{--}$ & 0.70$^{+0.68}_{-0.18}$ & 1.66$^{--}_{--}$ & -9.50$^{+0.33}_{-0.32}$ & -8.80$^{--}_{--}$ \\ 
macs0744 & 33 (56) & 6 & 0 (0) & 0.18$\pm$0.07 & -- & 10.66$^{+0.29}_{-0.33}$ & -- & 0.91$^{+0.12}_{-0.02}$ & -- & -9.75$^{+0.53}_{-0.26}$ & -- \\ 
clj1226 & 38 (60) & 15 & 0 (0) & 0.39$\pm$0.08 & -- & 10.32$^{+0.26}_{-0.18}$ & -- & 0.46$^{+0.46}_{-0.18}$ & -- & -9.80$^{+0.28}_{-0.25}$ & -- \\ 
\hline
Total &661 (1490) &149 & 17 (23) \\
		\cline{1-4}
& \multirow{2}{*}{Median}  & \multirow{1}{*}{0.4$<$$z$$<$0.6} & 0.1$<$$\mathcal{R}/R_{200}$$<$0.2  & $0.21^{+0.12}_{-0.13}$ & $0.00^{+0.05}_{-0.00}$ & $10.30^{+0.15}_{-0.02}$ & $10.46^{+0.01}_{-0.02}$ & $0.70^{+0.24}_{-0.12}$ & $1.61^{+0.06}_{-0.22}$ & $-9.54^{+0.04}_{-0.42}$ & $-8.90^{+0.06}_{-0.01}$ \\
		\cline{3-12}
& &  \multirow{1}{*}{0.6$<$$z$$<$0.9}& 0.1$<$$\mathcal{R}/R_{200}$$<$0.2 & $0.29^{+0.07}_{-0.07}$ & -- & $10.49^{+0.17}_{-0.17}$ & -- & $0.68^{+0.23}_{-0.23}$ & -- & $-9.78^{+0.02}_{-0.02}$ & -- \\
		\cline{2-12}
	\end{tabular}
\end{table}
\end{landscape}

\begin{landscape}
\begin{table}
\scriptsize
	\centering
	\caption{As in Table~\ref{tab:samples_c:R1}, for the galaxies at \textcolor{black}{ {0.2$<$$\mathcal{R}/R_{200}$$<$0.3}}.}
	\label{tab:samples_c:R3}
	\begin{tabular}{lccccccccccc} 
		\hline
		\rotatebox{0}{Cluster ID} & \multicolumn{1}{c}{Members} & \multicolumn{1}{c}{$UVJ$-SF} & \multicolumn{1}{c}{M-FIR} & \multicolumn{1}{c}{$\mathcal{F}_{UVJ\mathrm{-SF}}$} &  \multicolumn{1}{c}{$\mathcal{F}_{\mathrm{M-FIR}}$} & \multicolumn{1}{c}{$\mathcal{M}_{*,\,UVJ\mathrm{-SF}}$} & \multicolumn{1}{c}{$\mathcal{M}_{*,\,\mathrm{M-FIR}}$} & \multicolumn{1}{c}{$\mathcal{SFR}_{\mathrm{TOT}, UVJ\mathrm{-SF}}$} & \multicolumn{1}{c}{$\mathcal{SFR}_{\mathrm{TOT, M-FIR}}$} & \multicolumn{1}{c}{$s\mathcal{SFR}_{\mathrm{TOT}, UVJ\mathrm{-SF}}$} & \multicolumn{1}{c}{$s\mathcal{SFR}_{\mathrm{TOT, M-FIR}}$} \\
		(1) & \multicolumn{1}{c}{(2)} & \multicolumn{1}{c}{(3)} & \multicolumn{1}{c}{(4)} & \multicolumn{1}{c}{(5)} & \multicolumn{1}{c}{(6)} & \multicolumn{1}{c}{(7)} & \multicolumn{1}{c}{(8)} & \multicolumn{1}{c}{(9)} & \multicolumn{1}{c}{(10)} & \multicolumn{1}{c}{(11)} & \multicolumn{1}{c}{(12)}\\
		\hline
a0383 & 6 (14) & 1 & 0 (1) & 0.17$\pm$0.15 & -- & 10.79$^{--}_{--}$ & -- & 0.26$^{+0.00}_{-0.01}$ & -- & -10.53$^{--}_{--}$ & -- \\ 
a0209 & 0 (0) & 0 & 0 (0) & -- & -- & -- & -- & -- & -- & -- & -- \\ 
a2261 & 0 (0) & 0 & 0 (0) & -- & -- & -- & -- & -- & -- & -- & -- \\ 
rbs1748 & 0 (0) & 0 & 0 (0) & -- & -- & -- & -- & -- & -- & -- & -- \\ 
a0611 & 5 (11) & 2 & 0 (1) & 0.40$\pm$0.22 & -- & 10.16$^{+0.01}_{-0.00}$ & -- & 0.69$^{+0.25}_{-0.26}$ & -- & -9.48$^{+0.25}_{-0.25}$ & -- \\ 
ms2137 & 10 (24) & 1 & 0 (0) & 0.10$\pm$0.09 & -- & 10.07$^{--}_{--}$ & -- & 1.13$^{--}_{--}$ & -- & -8.94$^{--}_{--}$ & -- \\ 
as1063 & 5 (10) & 2 & 0 (0) & 0.40$\pm$0.22 & -- & 10.64$^{+0.30}_{-0.29}$ & -- & 0.31$^{+0.37}_{-0.37}$ & -- & -10.34$^{+0.08}_{-0.07}$ & -- \\ 
macs1931 & 18 (33) & 4 & 0 (0) & 0.22$\pm$0.10 & -- & 10.18$^{+0.08}_{-0.10}$ & -- & 0.19$^{+0.34}_{-0.20}$ & -- & -9.99$^{+0.36}_{-0.20}$ & -- \\ 
macs1115 & 13 (30) & 2 & 0 (0) & 0.15$\pm$0.10 & -- & 10.15$^{+0.09}_{-0.10}$ & -- & 0.13$^{+0.20}_{-0.20}$ & -- & -10.01$^{+0.29}_{-0.30}$ & -- \\ 
rxj1532 & 9 (25) & 0 & 0 (0) & -- & -- & -- & -- & -- & -- & -- & -- \\ 
macs1720 & 19 (49) & 10 & 4 (4) & 0.53$\pm$0.11 & 0.21$\pm$0.09 & 10.44$^{+0.31}_{-0.16}$ & 10.36$^{+0.17}_{-0.08}$ & 0.70$^{+0.63}_{-0.59}$ & 1.31$^{+0.16}_{-0.18}$ & -9.72$^{+0.63}_{-0.75}$ & -9.10$^{+0.13}_{-0.14}$ \\ 
macs0416 & 9 (25) & 3 & 0 (0) & 0.33$\pm$0.16 & -- & 10.11$^{+0.16}_{-0.02}$ & -- & 0.55$^{+0.19}_{-0.19}$ & -- & -9.80$^{+0.35}_{-0.01}$ & -- \\ 
macs0429 & 12 (51) & 6 & 0 (1) & 0.50$\pm$0.14 & -- & 10.23$^{+0.30}_{-0.12}$ & -- & 0.24$^{+0.73}_{-0.43}$ & -- & -9.85$^{+0.28}_{-0.59}$ & -- \\ 
macs1206 & 17 (35) & 4 & 0 (1) & 0.24$\pm$0.10 & -- & 10.53$^{+0.27}_{-0.32}$ & -- & 0.62$^{+0.33}_{-0.27}$ & -- & -9.86$^{+0.06}_{-0.06}$ & -- \\ 
macs0329 & 27 (85) & 6 & 0 (0) & 0.22$\pm$0.08 & -- & 10.49$^{+0.11}_{-0.38}$ & -- & 0.34$^{+0.39}_{-0.11}$ & -- & -9.81$^{+0.13}_{-0.50}$ & -- \\ 
rxj1347 & 12 (29) & 4 & 0 (0) & 0.33$\pm$0.14 & -- & 10.29$^{+0.39}_{-0.23}$ & -- & 0.37$^{+0.30}_{-0.22}$ & -- & -9.98$^{+0.10}_{-0.05}$ & -- \\ 
macs1311 & 9 (27) & 3 & 0 (0) & 0.33$\pm$0.16 & -- & 10.40$^{+0.48}_{-0.15}$ & -- & 0.59$^{+0.14}_{-0.14}$ & -- & -9.81$^{+0.01}_{-0.35}$ & -- \\ 
macs1149 & 15 (48) & 7 & 1 (1) & 0.47$\pm$0.13 & 0.07$\pm$0.06 & 10.39$^{+0.39}_{-0.30}$ & 10.39$^{--}_{--}$ & 0.51$^{+0.43}_{-0.20}$ & 1.41$^{--}_{--}$ & -9.80$^{+0.52}_{-0.50}$ & -8.98$^{--}_{--}$ \\ 
macs0717 & 13 (17) & 0 & 0 (0) & -- & -- & -- & -- & -- & -- & -- & -- \\ 
macs1423 & 7 (12) & 2 & 1 (1) & 0.29$\pm$0.17 & 0.14$\pm$0.13 & 10.34$^{+0.14}_{-0.14}$ & 10.54$^{--}_{--}$ & 1.23$^{+0.31}_{-0.31}$ & 1.69$^{--}_{--}$ & -9.11$^{+0.17}_{-0.17}$ & -8.86$^{--}_{--}$ \\ 
macs2129 & 17 (22) & 3 & 0 (0) & 0.18$\pm$0.09 & -- & 10.37$^{+0.25}_{-0.22}$ & -- & 0.78$^{+0.21}_{-0.48}$ & -- & -9.95$^{+0.45}_{-0.02}$ & -- \\ 
macs0647 & 12 (36) & 5 & 0 (0) & 0.42$\pm$0.14 & -- & 10.27$^{+0.06}_{-0.23}$ & -- & 0.78$^{+0.28}_{-0.22}$ & -- & -9.31$^{+0.10}_{-0.36}$ & -- \\ 
macs0744 & 32 (57) & 17 & 2 (2) & 0.53$\pm$0.09 & 0.06$\pm$0.04 & 10.43$^{+0.25}_{-0.20}$ & 10.37$^{+0.04}_{-0.05}$ & 0.61$^{+0.97}_{-0.29}$ & 1.92$^{+0.16}_{-0.17}$ & -9.79$^{+0.63}_{-0.40}$ & -8.45$^{+0.12}_{-0.12}$ \\ 
clj1226 & 39 (62) & 15 & 5 (5) & 0.38$\pm$0.08 & 0.13$\pm$0.05 & 10.60$^{+0.38}_{-0.48}$ & 10.98$^{+0.05}_{-0.31}$ & 1.24$^{+0.84}_{-0.43}$ & 2.09$^{+0.14}_{-0.09}$ & -8.95$^{+0.21}_{-0.76}$ & -8.84$^{+0.30}_{-0.07}$ \\ 
\hline
Total &306 (702) &97 & 13 (17) \\
		\cline{1-4}
& Median & \multirow{1}{*}{0.6$<$$z$$<$0.9} & 0.2$<$$\mathcal{R}/R_{200}$$<$0.3 & $0.46^{+0.05}_{-0.05}$ & $0.09^{+0.02}_{-0.02}$ & $10.51^{+0.08}_{-0.09}$ & $10.67^{+0.3}_{-0.3}$ & $0.93^{+0.31}_{-0.31}$ & $2.00^{+0.09}_{-0.09}$ & $-9.37^{+0.42}_{-0.42}$ & $-8.64^{-0.19}_{-0.19}$ \\
		\cline{2-12}
	\end{tabular}
\end{table}
\end{landscape}

\begin{landscape}
\begin{table}
\scriptsize
	\centering
	\caption{As in Table~\ref{tab:samples_c:R1}, for the field samples.}
	\label{tab:samples_f}
	\begin{tabular}{lccccccccccc} 
		\hline
		\rotatebox{0}{Field ID} & \multicolumn{1}{c}{Galaxies} & \multicolumn{1}{c}{$UVJ$-SF} & \multicolumn{1}{c}{M-FIR} & \multicolumn{1}{c}{$\mathcal{F}_{UVJ\mathrm{-SF}}$}  & \multicolumn{1}{c}{$\mathcal{F}_{\mathrm{M-FIR}}$} & \multicolumn{1}{c}{$\mathcal{M}_{*,\,UVJ\mathrm{-SF}}$} & \multicolumn{1}{c}{$\mathcal{M}_{*,\,\mathrm{M-FIR}}$} & \multicolumn{1}{c}{$\mathcal{SFR}_{\mathrm{TOT}, UVJ\mathrm{-SF}}$} & \multicolumn{1}{c}{$\mathcal{SFR}_{\mathrm{TOT, M-FIR}}$} & \multicolumn{1}{c}{$s\mathcal{SFR}_{\mathrm{TOT}, UVJ\mathrm{-SF}}$} & \multicolumn{1}{c}{$s\mathcal{SFR}_{\mathrm{TOT, M-FIR}}$} \\
		(1) & \multicolumn{1}{c}{(2)} & \multicolumn{1}{c}{(3)} & \multicolumn{1}{c}{(4)} & \multicolumn{1}{c}{(5)} & \multicolumn{1}{c}{(6)} & \multicolumn{1}{c}{(7)} & \multicolumn{1}{c}{(8)} & \multicolumn{1}{c}{(9)} & \multicolumn{1}{c}{(10)} & \multicolumn{1}{c}{(11)} & \multicolumn{1}{c}{(12)}\\
		\hline
A0383 & 97 (528) & 67 & 16 (43) & 0.69$\pm$0.05 & 0.16$\pm$0.04 & 10.47$^{+0.44}_{-0.39}$ & 10.60$^{+0.28}_{-0.34}$ & 0.79$^{+0.47}_{-0.61}$ & 1.37$^{+0.36}_{-0.28}$ & -9.74$^{+0.53}_{-0.59}$ & -9.34$^{+0.71}_{-0.29}$ \\ 
A0209 & 114 (648) & 84 & 18 (57) & 0.74$\pm$0.04 & 0.16$\pm$0.03 & 10.47$^{+0.49}_{-0.37}$ & 10.60$^{+0.31}_{-0.41}$ & 0.76$^{+0.42}_{-0.56}$ & 1.24$^{+0.57}_{-0.14}$ & -9.78$^{+0.49}_{-0.59}$ & -9.37$^{+0.70}_{-0.27}$ \\ 
A2261 & 124 (635) & 83 & 18 (52) & 0.67$\pm$0.04 & 0.15$\pm$0.03 & 10.47$^{+0.42}_{-0.32}$ & 10.52$^{+0.39}_{-0.33}$ & 0.75$^{+0.43}_{-0.53}$ & 1.24$^{+0.57}_{-0.14}$ & -9.75$^{+0.44}_{-0.53}$ & -9.31$^{+0.64}_{-0.33}$ \\ 
RBS1748 & 137 (657) & 95 & 25 (65) & 0.69$\pm$0.04 & 0.18$\pm$0.03 & 10.47$^{+0.43}_{-0.30}$ & 10.49$^{+0.42}_{-0.30}$ & 0.80$^{+0.44}_{-0.56}$ & 1.26$^{+0.55}_{-0.15}$ & -9.72$^{+0.47}_{-0.56}$ & -9.28$^{+0.59}_{-0.34}$ \\ 
A0611 & 109 (598) & 71 & 15 (43) & 0.65$\pm$0.05 & 0.14$\pm$0.03 & 10.40$^{+0.46}_{-0.24}$ & 10.47$^{+0.15}_{-0.28}$ & 0.81$^{+0.52}_{-0.50}$ & 1.35$^{+0.30}_{-0.16}$ & -9.60$^{+0.38}_{-0.62}$ & -9.09$^{+0.31}_{-0.24}$ \\ 
MS2137 & 139 (792) & 74 & 15 (40) & 0.53$\pm$0.04 & 0.11$\pm$0.03 & 10.41$^{+0.49}_{-0.25}$ & 10.48$^{+0.39}_{-0.24}$ & 0.79$^{+0.56}_{-0.46}$ & 1.36$^{+0.11}_{-0.19}$ & -9.71$^{+0.47}_{-0.47}$ & -9.23$^{+0.32}_{-0.13}$ \\ 
AS1063 & 160 (789) & 78 & 16 (43) & 0.49$\pm$0.04 & 0.10$\pm$0.02 & 10.38$^{+0.44}_{-0.28}$ & 10.50$^{+0.33}_{-0.25}$ & 0.85$^{+0.36}_{-0.49}$ & 1.33$^{+0.14}_{-0.18}$ & -9.58$^{+0.35}_{-0.58}$ & -9.15$^{+0.16}_{-0.34}$ \\ 
MACS1931 & 160 (462) & 78 & 16 (46) & 0.49$\pm$0.04 & 0.10$\pm$0.02 & 10.38$^{+0.44}_{-0.28}$ & 10.50$^{+0.33}_{-0.25}$ & 0.85$^{+0.36}_{-0.49}$ & 1.33$^{+0.14}_{-0.18}$ & -9.57$^{+0.33}_{-0.59}$ & -9.15$^{+0.16}_{-0.34}$ \\ 
MACS1115 & 151 (674) & 72 & 16 (38) & 0.48$\pm$0.04 & 0.11$\pm$0.03 & 10.38$^{+0.44}_{-0.28}$ & 10.50$^{+0.33}_{-0.25}$ & 0.85$^{+0.36}_{-0.53}$ & 1.33$^{+0.14}_{-0.18}$ & -9.61$^{+0.40}_{-0.57}$ & -9.15$^{+0.16}_{-0.34}$ \\ 
RXJ1532 & 181 (699) & 95 & 22 (64) & 0.52$\pm$0.04 & 0.12$\pm$0.02 & 10.38$^{+0.46}_{-0.27}$ & 10.50$^{+0.33}_{-0.23}$ & 0.85$^{+0.32}_{-0.41}$ & 1.30$^{+0.17}_{-0.17}$ & -9.54$^{+0.30}_{-0.59}$ & -9.21$^{+0.23}_{-0.21}$ \\ 
MACS1720 & 224 (919) & 107 & 31 (59) & 0.48$\pm$0.03 & 0.14$\pm$0.02 & 10.37$^{+0.44}_{-0.27}$ & 10.51$^{+0.31}_{-0.35}$ & 0.95$^{+0.34}_{-0.54}$ & 1.29$^{+0.19}_{-0.19}$ & -9.49$^{+0.38}_{-0.65}$ & -9.28$^{+0.39}_{-0.28}$ \\ 
MACS0416 & 220 (851) & 106 & 31 (58) & 0.48$\pm$0.03 & 0.14$\pm$0.02 & 10.36$^{+0.40}_{-0.26}$ & 10.49$^{+0.28}_{-0.33}$ & 0.95$^{+0.36}_{-0.55}$ & 1.31$^{+0.17}_{-0.21}$ & -9.45$^{+0.35}_{-0.68}$ & -9.22$^{+0.33}_{-0.20}$ \\ 
MACS0429 & 220 (795) & 106 & 31 (60) & 0.48$\pm$0.03 & 0.14$\pm$0.02 & 10.36$^{+0.40}_{-0.26}$ & 10.49$^{+0.28}_{-0.33}$ & 0.95$^{+0.36}_{-0.52}$ & 1.31$^{+0.17}_{-0.21}$ & -9.42$^{+0.32}_{-0.71}$ & -9.22$^{+0.33}_{-0.20}$ \\ 
MACS1206 & 250 (936) & 155 & 36 (55) & 0.62$\pm$0.03 & 0.14$\pm$0.02 & 10.38$^{+0.44}_{-0.26}$ & 10.54$^{+0.33}_{-0.31}$ & 0.89$^{+0.38}_{-0.46}$ & 1.31$^{+0.31}_{-0.18}$ & -9.46$^{+0.29}_{-0.69}$ & -9.18$^{+0.25}_{-0.24}$ \\ 
MACS0329 & 270 (972) & 173 & 43 (68) & 0.64$\pm$0.03 & 0.16$\pm$0.02 & 10.36$^{+0.45}_{-0.25}$ & 10.51$^{+0.34}_{-0.29}$ & 0.91$^{+0.35}_{-0.48}$ & 1.31$^{+0.20}_{-0.20}$ & -9.44$^{+0.29}_{-0.65}$ & -9.19$^{+0.25}_{-0.23}$ \\ 
RXJ1347 & 270 (1066) & 176 & 47 (84) & 0.65$\pm$0.03 & 0.17$\pm$0.02 & 10.36$^{+0.46}_{-0.25}$ & 10.56$^{+0.31}_{-0.33}$ & 0.90$^{+0.37}_{-0.46}$ & 1.31$^{+0.25}_{-0.18}$ & -9.44$^{+0.27}_{-0.67}$ & -9.22$^{+0.26}_{-0.20}$ \\ 
MACS1311 & 408 (1218) & 256 & 67 (83) & 0.63$\pm$0.02 & 0.16$\pm$0.02 & 10.39$^{+0.50}_{-0.27}$ & 10.53$^{+0.41}_{-0.29}$ & 0.95$^{+0.36}_{-0.50}$ & 1.30$^{+0.18}_{-0.21}$ & -9.45$^{+0.32}_{-0.63}$ & -9.21$^{+0.22}_{-0.36}$ \\ 
MACS1149 & 470 (1414) & 296 & 72 (72) & 0.63$\pm$0.02 & 0.15$\pm$0.02 & 10.48$^{+0.48}_{-0.35}$ & 10.56$^{+0.39}_{-0.30}$ & 1.00$^{+0.38}_{-0.55}$ & 1.34$^{+0.16}_{-0.13}$ & -9.60$^{+0.51}_{-0.56}$ & -9.21$^{+0.27}_{-0.48}$ \\ 
MACS0717 & 469 (970) & 297 & 80 (80) & 0.63$\pm$0.02 & 0.17$\pm$0.02 & 10.49$^{+0.50}_{-0.35}$ & 10.60$^{+0.48}_{-0.38}$ & 1.02$^{+0.36}_{-0.52}$ & 1.35$^{+0.19}_{-0.13}$ & -9.52$^{+0.44}_{-0.63}$ & -9.22$^{+0.29}_{-0.47}$ \\ 
MACS1423 & 470 (1036) & 297 & 69 (69) & 0.63$\pm$0.02 & 0.15$\pm$0.02 & 10.48$^{+0.48}_{-0.35}$ & 10.61$^{+0.41}_{-0.34}$ & 1.00$^{+0.38}_{-0.55}$ & 1.35$^{+0.16}_{-0.12}$ & -9.59$^{+0.50}_{-0.57}$ & -9.21$^{+0.27}_{-0.48}$ \\ 
MACS2129 & 539 (1462) & 333 & 41 (41) & 0.62$\pm$0.02 & 0.08$\pm$0.01 & 10.48$^{+0.48}_{-0.33}$ & 10.66$^{+0.40}_{-0.27}$ & 1.02$^{+0.38}_{-0.57}$ & 1.49$^{+0.23}_{-0.07}$ & -9.55$^{+0.49}_{-0.64}$ & -9.06$^{+0.17}_{-0.46}$ \\ 
MACS0647 & 484 (1218) & 311 & 54 (54) & 0.64$\pm$0.02 & 0.11$\pm$0.01 & 10.48$^{+0.48}_{-0.33}$ & 10.55$^{+0.38}_{-0.26}$ & 1.04$^{+0.38}_{-0.60}$ & 1.45$^{+0.26}_{-0.09}$ & -9.55$^{+0.56}_{-0.64}$ & -9.03$^{+0.20}_{-0.29}$ \\ 
MACS0744 & 860 (1706) & 533 & 113 (113) & 0.62$\pm$0.02 & 0.13$\pm$0.01 & 10.43$^{+0.40}_{-0.30}$ & 10.71$^{+0.31}_{-0.31}$ & 1.14$^{+0.43}_{-0.52}$ & 1.61$^{+0.27}_{-0.13}$ & -9.29$^{+0.37}_{-0.66}$ & -9.03$^{+0.29}_{-0.32}$ \\ 
CLJ1226 & 940 (1576) & 706 & 82 (82) & 0.75$\pm$0.01 & 0.09$\pm$0.01 & 10.51$^{+0.46}_{-0.38}$ & 10.77$^{+0.44}_{-0.37}$ & 1.32$^{+0.42}_{-0.60}$ & 1.94$^{+0.14}_{-0.10}$ & -9.20$^{+0.42}_{-0.67}$ & -8.86$^{+0.43}_{-0.32}$ \\ 
\hline
Total &7466 (22621) &4649 & 974 (1469) \\
		\cline{1-4}
& & \multirow{3}{*}{Median} & 0.2$<$$z$$<$0.4 & $0.53^{+0.16}_{-0.04}$ & $0.14^{+0.02}_{-0.03}$ & $10.40^{+0.07}_{-0.02}$ & $10.50^{+0.00}_{-0.00}$ & $0.81^{+0.04}_{-0.01}$ & $1.33^{+0.00}_{-0.03}$ & $-9.62^{+0.05}_{-0.10}$ & $-9.23^{+0.07}_{-0.05}$\\
& & & 0.4$<$$z$$<$0.6 & $0.63^{+0.01}_{-0.06}$ & $0.15^{+0.02}_{-0.02}$ & $10.38^{+0.09}_{-0.02}$ & $10.53^{10.54}_{10.50}$ & $0.95^{+0.00}_{-0.05}$ & $1.31^{+0.04}_{-0.00}$ & $-9.46^{+0.01}_{-0.08}$ & $-9.21^{+0.00}_{-0.01}$\\
& & & 0.6$<$$z$$<$0.9  & $0.69^{+0.04}_{-0.04}$ & $0.11^{+0.02}_{-0.02}$ & $10.47^{+0.04}_{-0.04}$ & $10.73^{+0.03}_{-0.03}$ & $1.23^{+0.09}_{-0.09}$ & $1.77^{+0.16}_{-0.16}$ & $-9.24^{+0.05}_{-0.05}$ & $-9.07^{+0.02}_{-0.02}$ \\
		\cline{3-12}
	\end{tabular}
\end{table}
\end{landscape}


\section{Stellar mass distributions}\label{SMF}
\begin{figure*}
\includegraphics[width=1\linewidth]{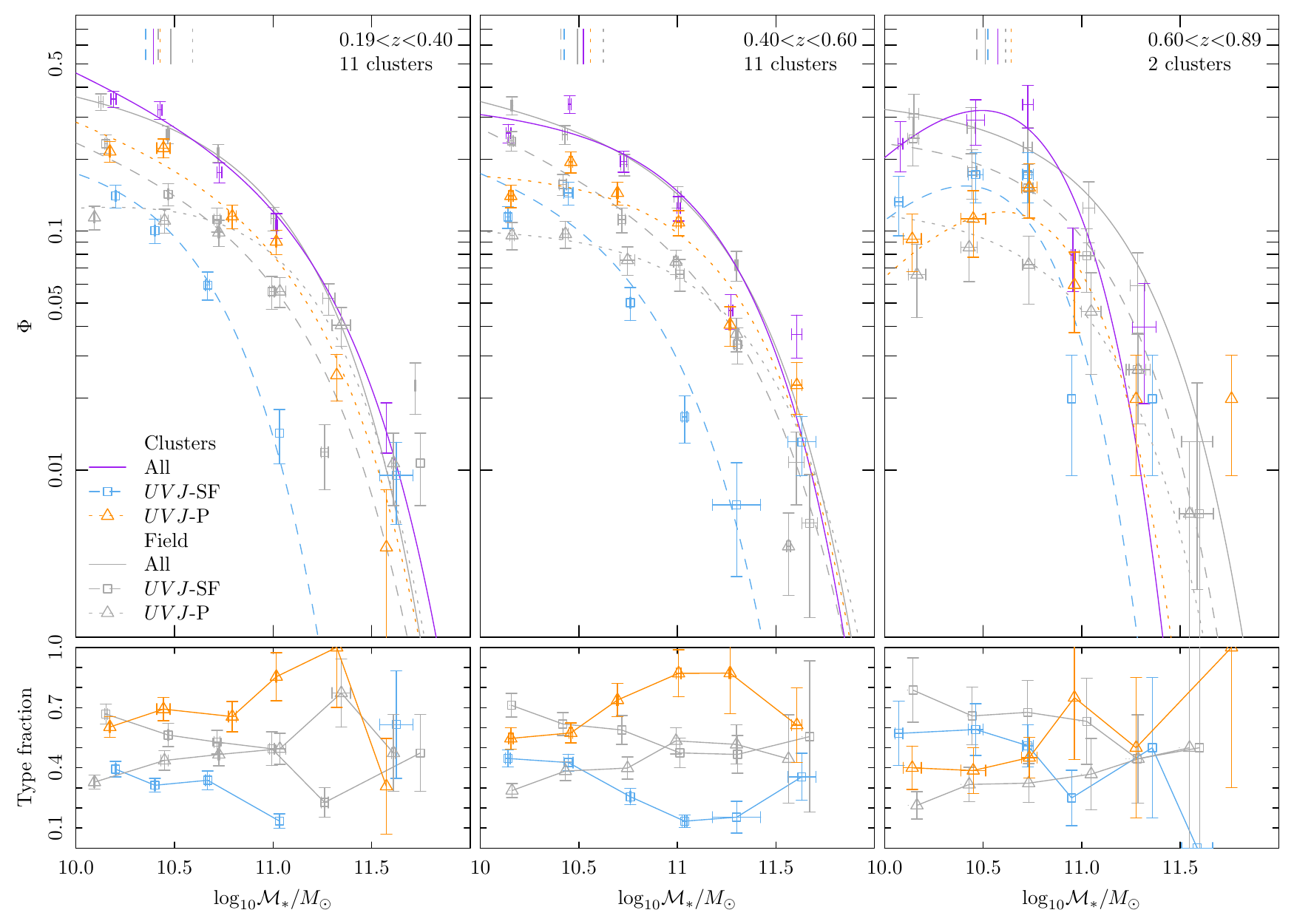}
\caption{\textcolor{black}{ { \textit{Top panels}: Stellar mass distribution 
within $\mathcal{R}/R_{200}$$<$0.1
for cluster and field galaxies (log$_{10}\mathcal{M}_{*}/M_{\odot}$$>$10) 
divided in bins of redshift. On the upper part of each panel, we mark 
the median stellar mass of every sample. \textit{Bottom panels}: Relative fraction of $UVJ$-P and $UVJ$-SF galaxies as a function of stellar mass.}} }\label{fig:MF}
\end{figure*}
\begin{table*}
	\centering
	\caption{\textcolor{black}{ {We report: Median log$_{10}\mathcal{M}_{*}$ (and 1$\sigma$ intervals), best-fitting Schechter parameters (and 1$\sigma$ intervals) and reduced $\chi^2$ for the different samples.}}}
	\label{tab:schechter}
	\begin{tabular}{lllccccc} 
		\hline
		\rotatebox{0}{} & \rotatebox{0}{} & Sample & $<$log$_{10}\mathcal{M}_{*}/M_{\odot}$$>$ & log$_{10}\mathcal{M}^{*}/M_{\odot}$ & $\alpha$ & $\Phi^{*}$ & $\chi^2$ \\
		\hline
		\multirow{6}{*}{0.19$<$$z$$<$0.40} & \multirow{3}{*}{Cluster} & All galaxies & 10.39$^{+0.48}_{-0.22}$ & 11.22$^{+0.56}_{-0.54}$ & -1.3$^{+0.3}_{-0.3}$ & 0.08$^{+0.08}_{-0.08}$ & 1.81\\
		&  & $UVJ$-SF & 10.35$^{+0.30}_{-0.19}$ & 10.55$^{+0.06}_{-0.06}$ & -1.0$^{+0.6}_{-0.2}$ & 0.10$^{+0.01}_{-0.01}$ & 5.52\\
		&  & $UVJ$-P & 10.43$^{+0.49}_{-0.25}$ & 11.14$^{+0.13}_{-0.13}$ & -1.3$^{+0.1}_{-0.1}$ & 0.06$^{+0.03}_{-0.03}$  & 7.18\\
		\cline{3-8}
		& \multirow{3}{*}{Field} & All galaxies & 10.48$^{+0.48}_{-0.38}$ & 11.05$^{+0.12}_{-0.12}$ & -1.1$^{+0.1}_{-0.1}$ & 0.13$^{+0.04}_{-0.04}$ & 5.69\\
		&  & $UVJ$-SF & 10.42$^{+0.47}_{-0.31}$ & 11.11$^{+0.05}_{-0.05}$ & -1.3$^{+0.5}_{-0.5}$ & 0.05$^{+0.00}_{-0.00}$ & 5.61\\
		&  & $UVJ$-P & 10.59$^{+0.47}_{-0.49}$ & 11.09$^{+0.18}_{-0.18}$ & -0.9$^{+0.2}_{-0.2}$ & 0.07$^{+0.02}_{-0.02}$ & 5.41\\
		\hline
		\multirow{6}{*}{0.40$<$$z$$<$0.60} & \multirow{3}{*}{Cluster} & All galaxies & 10.52$^{+0.47}_{-0.33}$ & 11.15$^{+0.30}_{-0.30}$ & -1.0$^{+0.3}_{-0.3}$ & 0.13$^{+0.09}_{-0.09}$ & 3.90\\
		&  & $UVJ$-SF & 10.42$^{+0.35}_{-0.28}$ & 10.83$^{+0.90}_{-0.90}$ & -1.2$^{+0.4}_{-0.4}$ & 0.06$^{+0.05}_{-0.05}$ & 11.47\\
		&  & $UVJ$-P & 10.56$^{+0.47}_{-0.30}$ & 11.22$^{+0.06}_{-0.06}$ & -1.0$^{+0.5}_{-0.5}$ & 0.08$^{+0.00}_{-0.00}$ & 6.23\\
		\cline{3-8}
		& \multirow{3}{*}{Field} & All galaxies & 10.49$^{+0.52}_{-0.35}$ & 11.22$^{+0.02}_{-0.02}$ & -1.2$^{+0.1}_{-0.1}$ & 0.11$^{+0.01}_{-0.01}$ & 6.35\\
		&  & $UVJ$-SF & 10.41$^{+0.53}_{-0.28}$ & 11.30$^{+0.09}_{-0.09}$ & -1.4$^{+0.1}_{-0.1}$ & 0.04$^{+0.01}_{-0.01}$ & 1.49\\
		&  & $UVJ$-P & 10.63$^{+0.49}_{-0.41}$ & 11.32$^{+0.13}_{-0.13}$ & -1.0$^{+0.5}_{-0.5}$ & 0.05$^{+0.01}_{-0.01}$ & 6.07\\
		\hline
		\multirow{6}{*}{0.60$<$$z$$<$0.89} & \multirow{3}{*}{Cluster} & All galaxies & 10.57$^{+0.31}_{-0.38}$ & 10.49$^{+0.26}_{-0.26}$ & \,\,0.0$^{+0.2}_{-0.2}$ & 0.39$^{+0.07}_{-0.07}$ & 1.71\\
		&  & $UVJ$-SF & 10.52$^{+0.27}_{-0.42}$ & 10.41$^{+2.60}_{-2.60}$ & \,\,0.0$^{+0.8}_{-0.8}$ & 0.18$^{+0.46}_{-0.46}$ & 12.29\\
		&  & $UVJ$-P & 10.64$^{+0.32}_{-0.39}$ & 10.60$^{+0.07}_{-0.07}$ & \,\,0.0$^{+0.5}_{-0.5}$ & 0.14$^{+0.02}_{-0.02}$& 9.60\\
		\cline{3-8}
		& \multirow{3}{*}{Field} & All galaxies & 10.51$^{+0.46}_{-0.35}$ & 11.11$^{+0.05}_{-0.05}$ & -1.0$^{+0.1}_{-0.1}$ & 0.14$^{+0.02}_{-0.02}$  & 1.22\\
		& & $UVJ$-SF & 10.47$^{+0.44}_{-0.34}$ & 11.00$^{+0.50}_{-0.50}$ & -1.0$^{+0.5}_{-0.5}$ & 0.10$^{+0.01}_{-0.01}$ & 7.86\\
		& & $UVJ$-P & 10.64$^{+0.50}_{-0.37}$ & 11.00$^{+0.29}_{-0.29}$ & -1.0$^{+0.2}_{-0.2}$ & 0.05$^{+0.03}_{-0.03}$ & 0.58\\
		\hline
	\end{tabular}
\end{table*}

\textcolor{black}{ {As a step prior to the evaluation of the SF within cluster
cores and how it compares to the SF in the field, we explore the stellar
mass function (SMF) of the samples presented in the previous section.
The SMF is a fundamental observable for the study of the evolution of galaxy populations.
Furthermore, overlooking hypothetical differences in the SMF of field and cluster
samples can lead to a misinterpretation of the physics behind the level of SF
quantified in the following sections.}}

\textcolor{black}{ {In the top panels of
Figure~\ref{fig:MF}, we display the SMF for clusters and field
galaxies (log$_{10}\mathcal{M}_{*}/M_{\odot}$$>$10)
divided into bins of redshift. We include only galaxies at
$\mathcal{R}$$<$0.1$R_{200}$, i.e., the $\mathcal{R}$ range homogeneously
covered along the whole redshift range. We exclude the BCGs in our analysis. 
We correct for different cluster richnesses by
randomly re-sampling the galaxy population of each cluster using the
average sample size of each redshift bin. Then, to render the field
and cluster samples statistically comparable, we re-sample each field
drawing randomly the number of galaxies in the corresponding cluster
sample. The uncertainties are estimated from the combination of 500
bootstraps. Then, we model the data by
fitting a Schechter function (\citealt{1976ApJ...203..297S}) to the
SMF. The form of the function is}}
\begin{equation}
\Phi\left(\mathcal{M}_{*}\right) d\mathcal{M}_{*} = \Phi^{*} \left( \frac{\mathcal{M}_{*}}{\mathcal{M}^{*}} \right)^{\alpha} \mathrm{e}^{-\frac{\mathcal{M}_{*}}{\mathcal{M}^{*}}} \frac{d\mathcal{M}_{*}}{\mathcal{M}^{*}},
\end{equation}
\textcolor{black}{ {with $\mathcal{M}^{*}$ being the characteristic mass, $\alpha$ the
low-mass slope, and $\Phi^{*}$ the normalization. The normalization is
evaluated by requiring that the integral of the Schechter function over
the stellar mass range considered equals the fraction of galaxies in
the sample fitted with respect the total sample.
In Table~\ref{tab:schechter} we report the
best-fit parameters. The function provides overall 
reasonable fits, although we report a quite large scatter of the data points 
for some of the samples. This is probably due to the limited number counts
we work with. 
In the bottom panels of Figure~\ref{fig:MF}, we
display the fraction of $UVJ$-P and $UVJ$-SF galaxies in each stellar
mass bin. The plots are not perfectly symmetric because we do not fix
the median value of each mass bin. We do not represent the stellar mass
distribution of the M-FIR sample because its size is not statistically
significant for this analysis. The median value of stellar mass
corresponding to each sample is marked in the upper panels 
of the same figure (see also Table~\ref{tab:schechter}).}} 

\textcolor{black}{ {We compare the best-fitting Schechter parameters with
those published recently by \citet{2018arXiv180700820V} for cluster and
field galaxies at 0.5$<$$z$$<$0.7. We focus our comparison on their
inner $\mathcal{R}$ bin ($\mathcal{R}/R_{200}$$\lesssim$0.4). Their
log$_{10}\mathcal{M}^{*}$ are 11.01$^{+0.02}_{-0.02}$,
11.01$^{+0.01}_{-0.02}$, and 10.70$^{+0.04}_{-0.04}$ for the whole
population, the quiescent, and the star-forming samples of the clusters,
respectively, and 11.18$^{+0.02}_{-0.02}$, 11.06$^{+0.02}_{-0.02}$, and
10.89$^{+0.05}_{-0.05}$ for the same subsamples in the field. We assume
a 0.2~dex conversion from Chabrier (\citealt{2003PASP..115..763C}) to 
Salpeter IMF (\citealt{2009ApJ...699..486C}). Our results for
the clusters and field between 0.4$<$$z$$<$0.6 are compatible with theirs 
except in the case of the cluster $UVJ$-P sample, for which we derive
log$_{10}\mathcal{M}^{*}$$=$11.22$^{+0.06}_{-0.06}$, and the field $UVJ$-SF and
$UVJ$-P populations, for which we derive larger values: 
log$_{10}\mathcal{M}^{*}$$=$11.30$^{+0.09}_{-0.09}$, 
and 11.32$^{+0.13}_{-0.13}$, respectively. Regarding 
$\alpha$, they retrieve $-0.91$$^{+0.02}_{-0.02}$, $-0.83$$^{+0.03}_{-0.02}$,
and $-1.02$$^{+0.06}_{-0.06}$ for the whole population, the quiescent, and
the star-forming samples of the clusters, and $-1.20$$^{+0.02}_{-0.02}$,
$-0.55$$^{+0.03}_{-0.03}$, and $-1.33$$^{+0.03}_{-0.03}$ for the field. In
this case, our results are compatible with theirs within the error bars.}}

\textcolor{black}{ {In the first two redshift bins, there are no large differences between
the SMF of the whole population of galaxies
in the field and the clusters, with values of the slope and the knee
of the Schechter function within the 1$\sigma$ errors (see
Table~\ref{tab:schechter}). This result has been found in previous works
at intermediate and high redshift (e.g., \citealt{2012MNRAS.420.1481V}, \citealt{2013A&A...550A..58V}, \citealt{2013A&A...557A..15V},
\citealt{2016A&A...592A.161N}). On the contrary, the highest redshift
bin displays large differences between the cluster and the field best-fit
Schechter functions. We claim these differences are mainly due to a 
poor sampling of the cluster SMF. In fact,
data points in the stellar mass range including 80\% of the stellar mass
of both cluster and field samples are compatible within the error bars.}}

\textcolor{black}{ {We report hints of a different behaviour of 
the SMFs of field and clusters and their evolution with $z$ when we split 
the galaxy populations in $UVJ$-SF and $UVJ$-P. 
At the lowest redshift, the $UVJ$-P SMF appears to present a steeper $\alpha$
than the field, which is not obvious in the
second redshift bin. This makes the $UVJ$-P SMF
present a shape apparently more similar to the field $UVJ$-SF stellar
mass distributions (excluding normalization differences). 
\citet{2001ApJ...557..117B} also find that while in the field environment 
the SMF of SFGs has much steeper faint-end slope than that for passive galaxies, 
in the clusters, the passive galaxies have also a steep faint-end. 
\citet{2014A&A...571A..80A} find that for the $z$$=$0.44 (our
second redshift bin) cluster MACS1206 (also included in our sample),
the SMF of SFGs is significantly steeper than
the SMF of passive galaxies at the faint
end. This is in agreement with our best-fitting SMFs in the intermediate redshift bin. Furthermore, they find a smaller slopes SMF for passive cluster galaxies in the 
inner core of clusters ($\mathcal{R}/R_{200}$$\lesssim$0.25), than in the 
outskirts.}}

\textcolor{black}{ {However, these differences are not significant in most cases. The
best-fitting values of $\alpha$ and log$_{10}\mathcal{M}^{*}$ for the 
$UVJ$-SF and $UVJ$-P samples in the clusters and in the field are overall
compatible within the error bars. The only significant difference
appears in the value of the log$_{10}\mathcal{M}^{*}$ for the
$UVJ$-SF samples in the lowest redshift bin: 10.55$^{+0.06}_{-0.06}$
and 11.11$^{+0.05}_{-0.05}$ for the clusters and the field,
respectively. Other works have also reported the lack of significant 
differences between the SMF of star-forming and passive galaxies in different environments 
(i.e., \citealt{2013A&A...550A..58V}).
The $UVJ$-SF and $UVJ$-P SMF evolution with redshift
is also mild in terms of the best-fitting Schechter
parameters $\alpha$ and log$_{10}\mathcal{M}^{*}$, and considering
our resolution.
}} 

\textcolor{black}{ {In the first two redshift bins, we find that the galaxy population in
massive clusters is clearly dominated by quiescent galaxies all the
way down to $\mathcal{M}_{*}$$=$$10^{10}$$M_{\odot}$, which is in 
agreement with (e.g.) \citet{2018arXiv180700820V}. The largest
mass bins are dominated by stochasticity given the small number of
galaxies included. \citet{2010ApJ...721..193P} predicts that the SMFs of passive 
and SFGs should cross (crossing mass) at 
log$_{10}\mathcal{M}_{*}/M_{\odot}$$\approx$10.4 and 9.6 
for central ("field") and satellites, respectively, at low redshift.
In our work, the crossing mass for the 
cluster SMFs shows up at log$_{10}\mathcal{M}_{*}/M_{\odot}$$\approx$10 
in the second redshift bin. In the third
redshift bin, the contribution of 
$UVJ$-SF and $UVJ$-P samples to the whole population of clusters 
is $\approx$50\%, with type fractions 
comparable within the error bars. This is comparable with the
$UVJ$-P and $UVJ$-SF type fractions derived by
\citet{2016A&A...592A.161N} for $z$$\sim$1.5. Regarding the field,
lower mass bins (log$_{10}\mathcal{M}_{*}/M_{\odot}$$<$10.6, 10.9, and
10.9 for the first, second, and third redshift bins, respectively) 
are dominated by star-forming galaxies, whereas the contribution of 
$UVJ$-P and $UVJ$-SF galaxies tend to converge and even to be inverse 
towards higher mass bins. Other previous studies (e.g., 
\citealt{2012ApJ...744...88Q}, \citealt{2016A&A...592A.161N},
\citealt{2018ApJ...854...30P}) 
have claimed a rapid increase in the
number density of low- and intermediate-mass 
(log$\mathcal{M}_{*}$/$M_{\odot}$<10--10.6) quiescent
galaxies in denser environments since $z$$\approx$1.5. 
\citet{2018MNRAS.479.2147M}
and \citet{2015MNRAS.447....2M} also find evidence for a higher number
density of quiescent low-mass galaxies in denser environments in our
same redshift range. However, our $\mathcal{M}_{*}$ completeness levels
hampers the analysis of a possible evolution of the distribution of
stellar mass at such low values.}}

\textcolor{black}{ {It is worth noting that numerous works 
(e.g., \citealt{2016A&A...585A.160A}) find that passive
cluster galaxies are better fitted by a double Schechter function,
revealing the existence of two sub-populations of red cluster members
thought to have followed distinct evolutionary paths. 
On the one hand, a population of high mass galaxies thought to be
quenched by processes scaling with stellar mass, and on the other hand,
a population of low-mass galaxies quenched by environmental processes
(\citealt{2010ApJ...721..193P}). These composite SMF of red passive
galaxies have also been observed in the field in works such as, e.g.,
\citet{2009ApJ...707.1595D} and \citet{2012MNRAS.421..621B}. However,
the evidence for these double Schechter functions 
(i.e., an upturn at low stellar masses) is only visible at
log$_{10}\mathcal{M}_{*}/M_{\odot}$$\lesssim$10 (\citealt{2009ApJ...707.1595D}), 
below the mass limit of our work.}}

\section{Quantification of star formation processes within cluster cores}\label{sf_process}

\begin{figure*}
    \centering
        \includegraphics[width=0.85\linewidth]{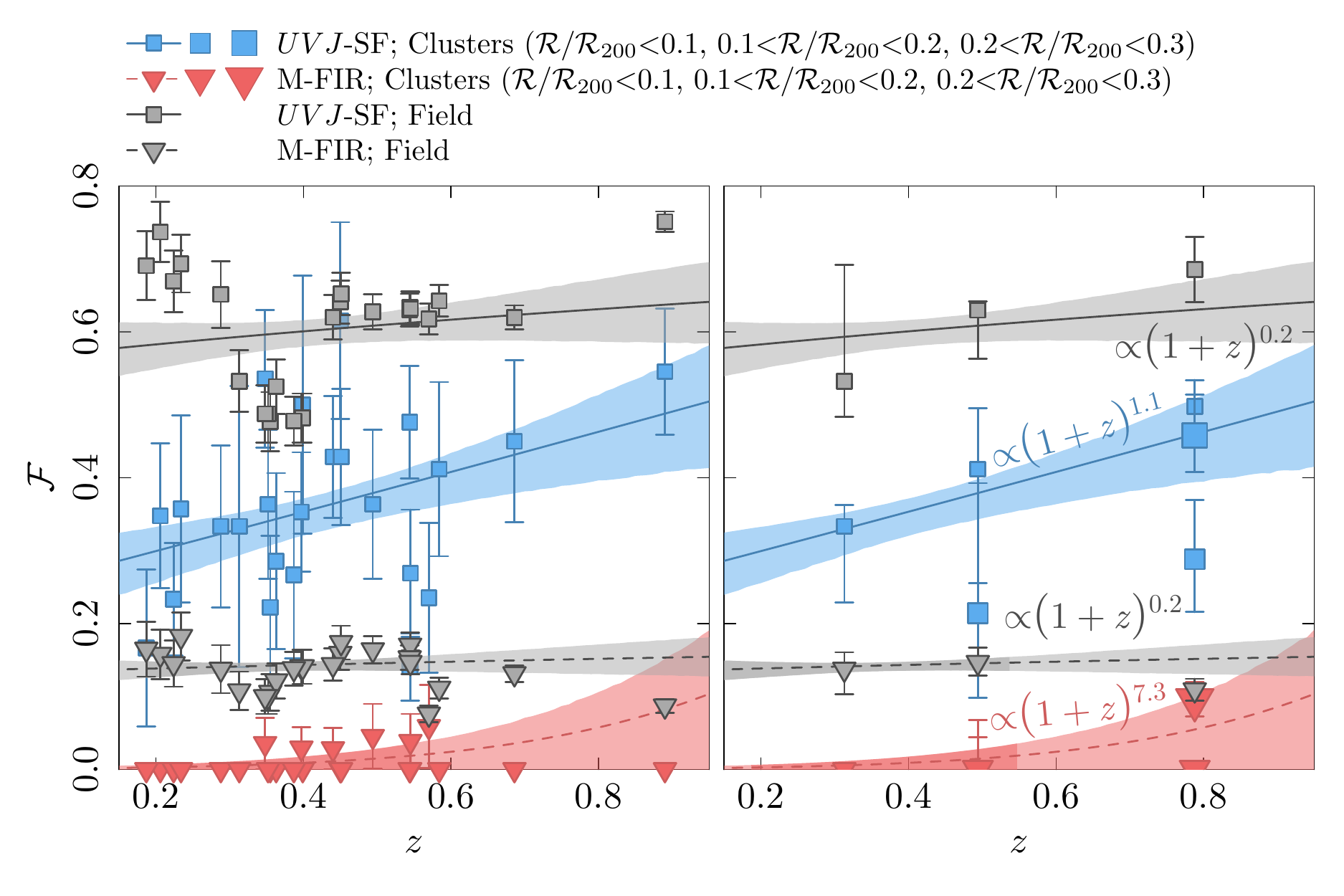}
        \caption{\textcolor{black}{ {$UVJ$-SF and M-FIR fractions.
In the left panel we consider only cluster members at $\mathcal{R}/R_{200}$$<$0.1. In the right-hand panels we show the average of the individual values at $\mathcal{R}/R_{200}$$<$0.1, 0.1$<$$\mathcal{R}/R_{200}$$<$0.2, and 
0.2$<$$\mathcal{R}/R_{200}$$<$0.3, in three redshift bins
(0.2$<$$z$$<$0.4, 0.4$<$$z$$<$0.6, 0.6$<$$z$$<$0.9). In both panels,
we show the best fit of a trend with redshift of the shape $\alpha(1+z)^\beta$ and corresponding 1$\sigma$ confidence intervals (continuous and dashed lines, and corresponding shaded areas, respectively). Darker red shaded areas represent the redshift range used for the fit of the M-FIR samples of clusters and field. For each cluster for which the M-FIR sample is empty, we represent the corresponding data point on the bottom-horizontal axis.}}} \label{fig:trends_f}
\end{figure*}

In this section, we present a quantification of the SF
activity hosted by cluster  members and field galaxies with
log$_{10}\mathcal{M}_{*}/M_{\odot}$$>$10, as traced by the UV and the M- and FIR.

\begin{table*}
	\centering
	\caption{\textcolor{black}{ {Best-fitting parameters derived from the fit of the evolution 
with redshift of the $\mathcal{F}$, and median
$\mathcal{SFR}_{\mathrm{TOT}}$ and $s\mathcal{SFR}_{\mathrm{TOT}}$ for all the
$UVJ$-SF and M-FIR samples in the clusters and in the field. For the clusters
we include the results only for $\mathcal{R}/R_{200}$$<$0.1.
The function fitted is a power-law of the shape $\alpha(1+z)^{\beta}$. The units of $\beta$ are
$M_{\odot}$yr$^{-1}$ and yr$^{-1}$ in the case of the fit of
$\mathcal{SFR}$ and $s\mathcal{SFR}$, respectively. The reduced $\chi^{2}$
for each case are shown in the last column. The fits of the $UVJ$-SF samples are performed using the data points spread out the whole redshift range. In the case of the M-FIR we fit only reported redshift ranges.}}}
	\label{tab:fits}
	\begin{tabular}{lllcccc} 
		\hline
		\rotatebox{0}{Quantity} & Environment & Subsample & $z$-range & $\alpha$ & $\beta$ & $\chi^{2}$\\
		\hline
		\multirow{4}{*}{$\mathcal{F}$} & \multirow{2}{*}{Cluster ($\mathcal{R}/R_{200}$$<$0.1)} & $UVJ$-SF & 0.19-0.89 & 0.25$\pm$0.05 & 1.1$\pm$0.6 & 2.11\\
		& & M-FIR & 0.19-0.57 & 0.00$\pm$0.00 & 7.3$\pm$5.8 & 0.19\\
		\cline{2-7}
		&\multirow{2}{*}{Field} & $UVJ$-SF & 0.19-0.89 & 0.56$\pm$0.06 & 0.2$\pm$0.3 & 7.27\\
		& & M-FIR & 0.19-0.57 & 0.13$\pm$0.02 & 0.2$\pm$0.5 & 0.93\\
		\hline
		\multirow{4}{*}{$\mathcal{SFR}_{\mathrm{TOT}}$} & \multirow{2}{*}{Cluster ($\mathcal{R}/R_{200}$$<$0.1)} & $UVJ$-SF & 0.19-0.89 & 1.82$\pm$0.71 & 1.3$\pm$1.0 & 21.75\\
		& & M-FIR & 0.34-0.57 & 2.67$\pm$3.24 & 5.9$\pm$2.8 & 0.11\\
		\cline{2-7}
		&\multirow{2}{*}{Field} & $UVJ$-SF & 0.19-0.57 & 3.36$\pm$0.20 & 2.6$\pm$0.2 & 1.53\\
		& & M-FIR & 0.19-0.57 & 18.10$\pm$1.37 & 0.4$\pm$0.2 & 0.29 \\
		\hline
		\multirow{4}{*}{$s\mathcal{SFR}_{\mathrm{TOT}}$} & \multirow{2}{*}{Cluster ($\mathcal{R}/R_{200}$$<$0.1)} & $UVJ$-SF & 0.19-0.89 & (0.67$\pm$0.22)$\times10^{-10}$ & 1.2$\pm$0.9 & 50.37\\
		& & M-FIR & 0.34-0.57 & (0.51$\pm$8.70)$\times10^{-9}$ & 0.0$\pm$9.5 & 1.31 \\
		\cline{2-7}
		&\multirow{2}{*}{Field} & $UVJ$-SF & 0.19-0.89 & (1.24$\pm$0.17)$\times10^{-10}$ & 2.4$\pm$0.4 & 2.69\\
		& & M-FIR & 0.19-0.57 & (4.56$\pm$0.60)$\times10^{-10}$ & 0.8$\pm$0.4 & 0.60\\
		\hline
	\end{tabular}
\end{table*}

\subsection{Star-forming galaxy fraction}\label{fraction}

Figure~\ref{fig:trends_f} (left hand panel) shows the
fraction ($\mathcal{F}$) of $UVJ$-SF and M-FIR galaxies
($\mathcal{F}_{UVJ\mathrm{-SF}}$ and $\mathcal{F}_{\mathrm{M-FIR}}$,
respectively; Section~\ref{sect:subsampl}) in the clusters
($\mathcal{R}/R_{200}$$<$0.1) and in the field. Error bars
are obtained using the margin of error of a percentage\footnote{The
confidence interval of a point sample estimate of the population
proportion at 1$\sigma$ can be derived considering a standard normal
distribution with the expression $\sqrt{p\,(1-p)/n}$, where $n$ is
the size of the sample and p is the proportion. Both of them must
satisfy the condition that $n\,p \geq$ 5 and $n\left(1 - p\right)
\geq$ 5.} assuming a standard normal distribution. On the right
panel, we show the median $\mathcal{F}$ and quantiles 16$^{th}$ and
84$^{th}$ (in the shape of error bars) in the same redshift bins
of Figure~\ref{fig:MS}. \textcolor{black}{ {We also include with larger symbols the
fractions obtained at 0.1$<$$\mathcal{R}/R_{200}$$<$0.2
and 0.2$<$$\mathcal{R}/R_{200}$$<$0.3, at the corresponding
redshift bins. In all cases, the median and quantiles are obtained
using the bootstrap methodology.}}

To quantify the trends of $\mathcal{F}$ with redshift, we fit to the
data points (fraction for each individual cluster within 
$\mathcal{R}/R_{200}$$<$0.1) a function with the 
shape $\alpha(1+z)^{\beta}$, where
$\alpha$ corresponds to the value of $\mathcal{F}$ at $z$$=$0, and
$\beta$ describes its evolution with redshift (with larger values of
$\beta$ meaning a steeper trend).  This methodology is also applied by
(e.g.) \citet{2013ApJ...775..126H} and \citet{2014MNRAS.437..437A}. The
corresponding curves and 1$\sigma$ confidence intervals (generated
using Monte Carlo simulations) are over-plotted in 
Figure~\ref{fig:trends_f} with a coloured line and a shaded area
around it, respectively. Table~\ref{tab:fits} shows the $\alpha$ and
$\beta$ values of the best-fit. \textcolor{black}{ {In the case of the M-FIR samples, we fit 
only the clusters with a $\mathcal{SFR}_{TIR}$ limit below 
10$M_{\odot}$yr$^{-1}$ ($z$$<$0.570) to derive the redshift trend.}}

The first information we can derive from
Figure~\ref{fig:trends_f} is that, as expected, the
$\mathcal{F}$ within clusters is much smaller than in the field for both
$UVJ$-SF and M-FIR samples. On average, $\mathcal{F}_{UVJ\mathrm{-SF}}$
in clusters seems to be approximately $1/2$ the value in the
field. \textcolor{black}{ {The $\mathcal{F}_{\mathrm{M-FIR}}$ in clusters drop down to values not
significantly different to zero. Assuming the same fraction of
M-FIR galaxies among the SFGs in clusters and field, the expected
average $\mathcal{F}_{\mathrm{M-FIR}}$ for the former would be $\sim$5\%,
which seems reasonably consistent with our results. Therefore, we
cannot say there is a smaller fraction of highly star-forming galaxies
($\mathcal{SFR}_{TIR}$$>$10$\mathrm{M}_{\odot}$yr$^{-1}$) and/or dusty
systems in the inner cores of clusters at intermediate redshifts.}}

Figure~\ref{fig:trends_f} also displays different evolutions of
$\mathcal{F}$ for clusters and field with $z$. The latter displays mild
increasing trends for $\mathcal{F}_{UVJ\mathrm{-SF}}$
and $\mathcal{F}_{\mathrm{M-FIR}}$, which vary with $\beta$$=$0.2$\pm$0.3
$\beta$$=$0.2$\pm$0.5, respectively. $\mathcal{F}$ remains $\sim$60\%
for the $UVJ$-SF samples between $z$$=$0.19-0.89. Flat/mild trends
for the fraction of the star-forming population of galaxies in
the field at intermediate redshifts ($z$$<$1) are also found by
\citealt{2011ApJ...739...24B} and \citealt{2017ApJ...837...16D}. In
particular, the latter gives 70\% of fraction of SFGs which is comparable
with our results, although there is a larger offset between these numbers
and the 40\% given by the former. These differences are likely due
to the sample selection criteria. The fraction of M-FIR galaxies remain
also constant ($\sim$0.15) in the same redshift range. The decreasing trend 
of the data points at $z$$>$0.570 (not fitted) is due to the fact that the 
minimum $\mathcal{SFR}_{\mathrm{TIR}}$ detectable for this clusters 
is larger than the value used to select M-FIR galaxies.

If we now focus on the clusters, we can see that, despite
the cluster-to-cluster variations (which reach $\sim$0.3), we
identify for both $UVJ$-SF and M-FIR samples a trend resembling the
\citet{1984ApJ...285..426B} effect, in which the fraction of SFGs in
clusters is observed to increase with redshift. In this case, the trends
are fitted with $\beta$$=$$1.1\pm0.6$ and $\beta$$=$$7.3\pm5.8$ for the
$UVJ$-SF and M-FIR samples, respectively. The fraction of $UVJ$-SF
galaxies within clusters increases from 28\% at $z$$\sim$0.2
to 47\% at $z$$\sim$0.9, while the fraction of M-FIR galaxies
grows from 0\% to 9\% in the same period. 
These values are in agreement with previous studies. For instance, \citet{2009ApJ...704..126H}
find that the fraction of massive galaxies with
$\mathcal{L}_{\mathrm{TIR}}$$>$5$\times$10$^{10}$$\mathcal{L}_{\mathrm{\odot}}$
and $\mathcal{R}$$<$$R_{200}$ varies from
$\sim$3\% at $z$$=$0.02 to $\sim$10\% at $z$$=$0.3 with
$\beta$$=$5.7$^{+2.1}_{-1.8}$. The fraction varies between
$\sim$1\% at $z$$=$0.15 and $\sim$4\% at $z$$=$0.3 considering
only $\mathcal{R}$$\lesssim$0.3$R_{200}$. 
Finally, the contribution of M-FIR galaxies to the whole SFGs
population ($UVJ$-SF sample) remains $\sim$23\% in the field, and
varies from 0\% to 19\% in the clusters between $z$$\sim$0.2 and
$z$$\sim$0.9. \citet{2016ApJ...827L..25M} reports very little evolution
of the ratio of dusty and non-dusty star-forming galaxies as a function
of stellar mass throughout this same redshift range.

\textcolor{black}{ {The average values of $\mathcal{F}_{UVJ\mathrm{-SF}}$ and
$\mathcal{F}_{\mathrm{M-FIR}}$ do not present a clear trend with
$\mathcal{R}$. In fact, all of them are compatible with the curve fitted
to the fractions at $\mathcal{R}/R_{200}$$<$0.1. However,
the distribution of SFGs in these high density environments has been
observed to increase with the projected cluster-centric radius by
(e.g.) \citealt{2016ApJ...825...72A}, and \citealt{2015ApJ...806..101H}.
This could be the result of a combination of factors such as cluster to 
cluster variations and an intrinsic negligible trend with redshift at 
$\mathcal{R}/R_{200}$$<$0.3.}}

\subsection{\textcolor{black}{ {Environmental quenching efficiency}}}\label{QE}

\textcolor{black}{ {The environmental quenching efficiency ($\mathcal{QE}_{\mathrm{env}}$; \citealt{2008MNRAS.387...79V}, 
\citealt{2010ApJ...721..193P}, \citealt{2016MNRAS.456.4364B}) is defined as 
\begin{equation}
\mathcal{QE}_{\mathrm{env}} = (\mathcal{F}_{\mathrm{P, cluster}} - \mathcal{F}_{\mathrm{P, field}}) / \mathcal{F}_{\mathrm{SF, field}},
\end{equation}
\noindent where $\mathcal{F}_{\mathrm{P, cluster}}$ and 
$\mathcal{F}_{\mathrm{P, field}}$ are the 
fraction of passive galaxies in the cluster and field, respectively, and 
$\mathcal{F}_{\mathrm{SF, field}}$ is the fraction of SFGs in the field.} }

\begin{figure}
\includegraphics[width=1\linewidth]{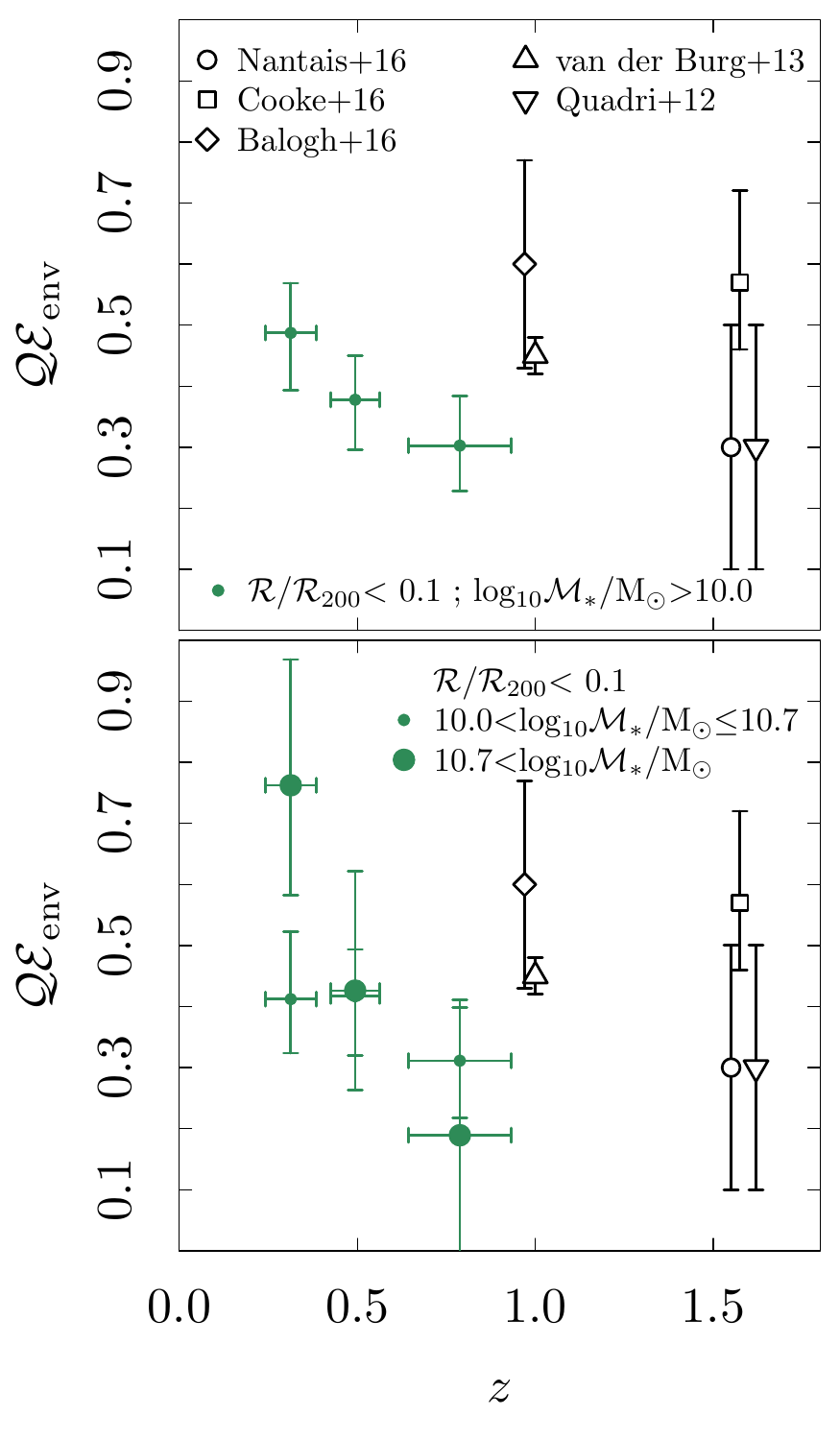}
\caption{\textcolor{black}{ { \textit{Top panel}: Environmental quenching efficiency for galaxies with log$\mathcal{M}_{*}$/$\mathrm{M}_{\odot}$$>$10 in three $z$ bins. \textit{Bottom panel}: Environmental quenching efficiency calculated for two mass bins (10.0$<$log$\mathcal{M}_{*}$/$\mathrm{M}_{\odot}$$\leq$10.7 and log$\mathcal{M}_{*}$/$\mathrm{M}_{\odot}$$>$10.7). In all cases, error bars are calculated propagating the errors of the fractions, which were obtained through bootstrap (500 realizations) in the initial cluster members and field galaxies samples. For comparison purposes, we include the $\mathcal{QE}_{\mathrm{env}}$ values given by \protect\citet{2012ApJ...744...88Q}, \protect\citet{2013A&A...557A..15V}, \protect\citet{2016MNRAS.456.4364B}, \protect\citet{2016ApJ...816...83C}, \protect\citet{2016A&A...592A.161N}.}}}\label{fig:QE}
\end{figure}

\textcolor{black}{ {In Figure~\ref{fig:QE}, we show the $\mathcal{QE}_{\mathrm{env}}$ in
the cluster cores ($\mathcal{R}/R_{200}$$<$0.1) grouped in
three redshift bins (0.2$<$$z$$<$0.4, 0.4$<$$z$$<$0.6, 0.6$<$$z$$<$0.9).
We derive $\mathcal{QE}_{\mathrm{env}}$ values of 0.49$^{+0.09}_{-0.08}$,
0.38$^{+0.08}_{-0.07}$, and 0.30$^{+0.07}_{-0.08}$ at $z$$\sim$0.31,
0.49, and 0.79, respectively.  These values are smaller than those
presented by \citet{2017MNRAS.465L.104N} at 0.87$<$$z$$<$1.63 for
galaxies with log$\mathcal{M}_{*}/\mathrm{M}_{\odot}$$\geq10.3$. Our
value of $\mathcal{F}_{UVJ\mathrm{-SF}}$ for clusters (field) in the
highest redshift bin is 0.50$^{+0.03}_{-0.03}$ (0.69$^{+0.04}_{-0.04}$)
which leads to smaller values of the passive fraction than their
0.88$^{+0.04}_{-0.03}$.  Our results at $z$$\sim$0.8 are also
smaller than other works
such as \citet{2016MNRAS.456.4364B} at redshift $z$$\sim$1 for
the same values of stellar mass. It is worth noting that these works calculate the
$\mathcal{QE}_{\mathrm{env}}$ within cluster-centric distances of 1\,Mpc
or $R_{200}$, while we focus on the inner cluster core, where
the fraction of passive galaxies is expected to be larger.}}

\textcolor{black}{ {The dependence of the $\mathcal{QE}_{\mathrm{env}}$
with stellar mass is under debate. While some works (e.g.,
\citealt{2010ApJ...721..193P}, \citealt{2018arXiv180700820V}) claim
environmental quenching to be independent of mass quenching, others
(e.g., \citealt{2014ApJ...782...33L}, \citealt{2017ApJ...847..134K})
have detected an increasing trend of the $\mathcal{QE}_{\mathrm{env}}$
with stellar mass. The bottom panel of Figure~\ref{fig:QE}
shows the values of $\mathcal{QE}_{\mathrm{env}}$ obtained for
galaxies at $\mathcal{R}$$<$0.1$R_{200}$ in two stellar
mass bins (10.0$<$log$\mathcal{M}_{*}/\mathrm{M}_{\odot}$$<$10.7
and 10.7$<$log$\mathcal{M}_{*}/\mathrm{M}_{\odot}$). As we can
see, only in the first redshift bin the $\mathcal{QE}_{\mathrm{env}}$ appears
significantly larger for the more massive galaxies. This 
$\mathcal{QE}_{\mathrm{env}}$ appears larger also if we split the sample at 
lower masses, but the significance of the result decreases. 
\citet{2016ApJ...825..113D} claims that environmental
quenching efficiency is almost independent of stellar mass at $z$$<$1,
except for galaxies with log$\mathcal{M}_{*}/\mathrm{M}_{\odot}$$>$10.9,
that high density environments could quench more efficiently.}}

\subsection{Average $\mathcal{SFR}$ and $s\mathcal{SFR}$}\label{sfr_ssfr}

\begin{figure*}
    \centering
        \includegraphics[width=1\linewidth]{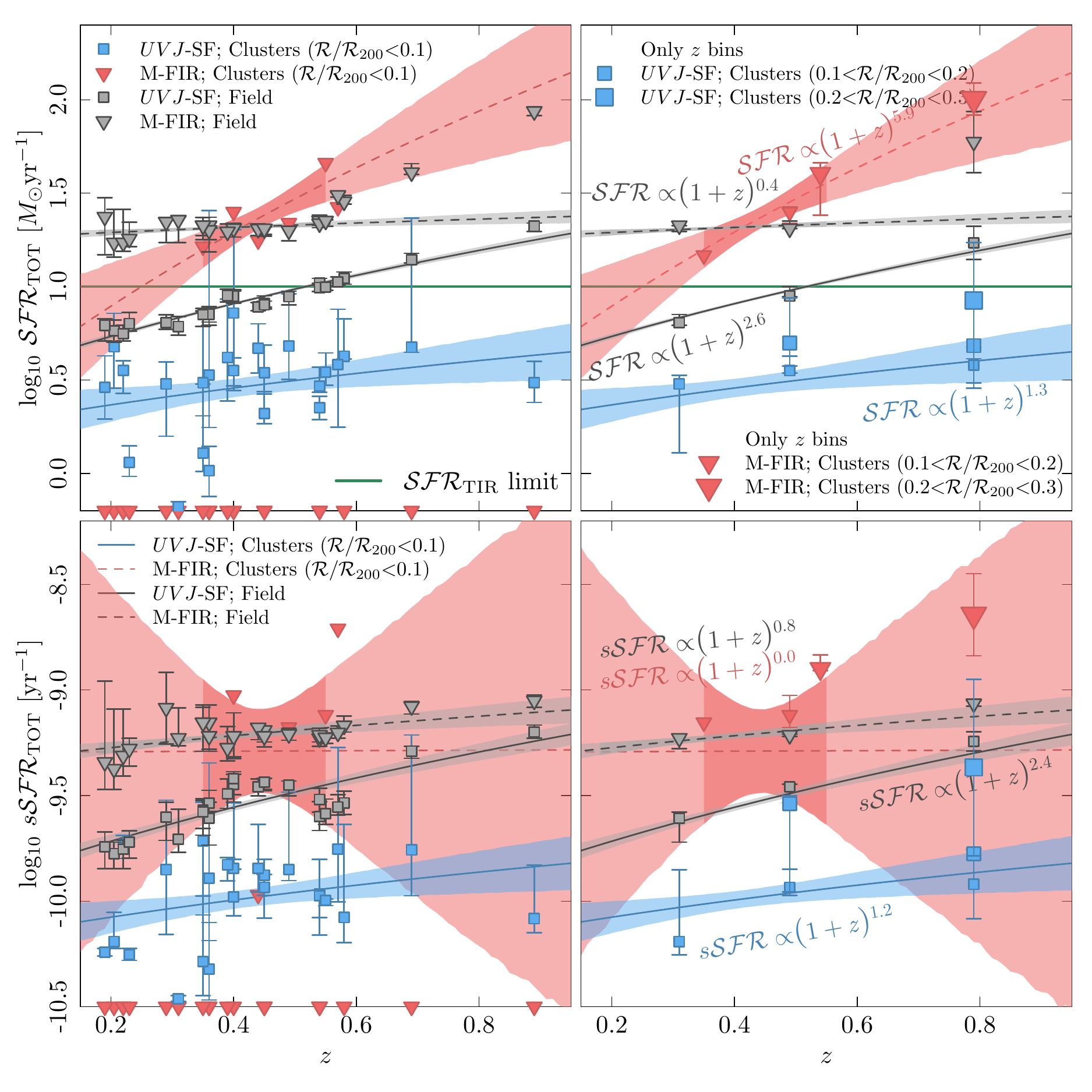}
        \caption{\textcolor{black}{ { 
\textit{Top panel}: median $\mathcal{SFR}$ for the $UVJ$-SF and M-FIR samples. \textit{Bottom panel}: median
$s\mathcal{SFR}_{\mathrm{TOT}}$ for the $UVJ$-SF and M-FIR samples. Representation as in Figure~\protect\ref{fig:trends_f}.}}} \label{fig:trends_sfr}
\end{figure*}

A complementary quantification of the SF activity in clusters tackles the
question whether beyond the decrease in $\mathcal{F}$ shown in 
Figure~\ref{fig:trends_f}, the impact of the cluster environment
modifies the distribution of the rates at which the remaining SFGs form
stars. In Figure~\ref{fig:trends_sfr} (top and bottom left-hand panels),
we display, as a function of redshift, the median $\mathcal{SFR}$
and $s\mathcal{SFR}$ of each cluster ($\mathcal{R}/R_{200}$$<$0.1) 
and field sample of $UVJ$-SF and M-FIR galaxies. 
The error bars are determined using the bootstrap
technique to derive the 1$\sigma$ confidence intervals, and thus,
they represent the spread in the $\mathcal{SFR}$ and $s\mathcal{SFR}$
of each subsample, not the intrinsic error of the estimation of these
parameters ($\sim$0.3\,dex). In the corresponding right-hand panels we
display the median values and confidence intervals in three redshift bins. 
\textcolor{black}{ { We also include the median values obtained at 
0.1$<$$\mathcal{R}/R_{200}$$<$0.2 and 
0.2$<$$\mathcal{R}/R_{200}$$<$0.3, when possible.}} 

To quantify the trends of the average $\mathcal{SFR}$ and $s\mathcal{SFR}$
with redshift, we again fit the median values (of the individual clusters)
using a function of the shape $\alpha(1+z)^\beta$. \textcolor{black}{ { Regarding the M-FIR
samples, we only fit those data points corresponding
to clusters at $z$$<$0.57 where at least a galaxy is detected in
the M- or FIR. Effectively, the fit is performed only between
0.34$<$$z$$<$0.57 (darker shaded area in Figure~\ref{fig:properties})}}.
The best-fit parameters are shown in Table~\ref{tab:fits}. We also
include a corresponding 1$\sigma$ confidence intervals of the fit
(generated using Monte Carlo simulations) as a shaded area around each
best-fit curve. The confidence intervals are not representative of the
dispersion of the $\mathcal{SFR}$ and $s\mathcal{SFR}$ distributions,
typically $\sim$0.3\,dex.

Regarding the $UVJ$-SF samples, Figure~\ref{fig:trends_sfr} clearly shows an
offset between the field and the clusters, with the latter displaying
$\mathcal{SFR}$ and $s\mathcal{SFR}$ on average $\sim$0.3~dex lower.
This offset cannot be explained by the differences between the mass
distribution of field and clusters samples (see Section~\ref{SMF}).
This can be seen in Figure~\ref{fig:ms_s} and Figure~\ref{fig:MS}, 
where the offsets in $\mathcal{SFR}$ and $s\mathcal{SFR}$ are visible at 
fixed $\mathcal{M}_{*}$.

Figure~\ref{fig:trends_sfr} also displays a clear increasing trend with $z$ of the
$\mathcal{SFR}$ for both field and cluster $UVJ$-SF samples ($\beta$$=$2.6$\pm$0.2
and $\beta$$=$1.3$\pm$1.0, respectively). The average 
$\mathcal{SFR}$ and $s\mathcal{SFR}$ do not show a strong differential 
evolution relative to the field but a systematic offset.
Analogous trends are found for the $s\mathcal{SFR}$, with $\beta$$=$2.4$\pm$0.4
and $\beta$$=$1.2$\pm$0.9 for the field and the clusters,
respectively. This also suggests that there is not a significant
evolution of the $\mathcal{M}_{*}$ distributions driving the variation
in $s\mathcal{SFR}$, at least at log$_{10}\mathcal{M}_{*}/M_{\odot}$$>$10. 
A hypothetical impact of the stellar mass distributions of the cluster 
and field samples would translate into a different behaviour of the variation 
of the average values of $\mathcal{SFR}$ and $s\mathcal{SFR}$ with environment, which is 
something we do not observe. 

The high cut in $\mathcal{SFR}_{\mathrm{TIR}}$ we use to build the M-FIR
galaxy samples translates into a mild increasing trend with $z$ of the
median value of the average $\mathcal{SFR}$ ($s\mathcal{SFR}$) for the
M-FIR galaxies in the field, which varies with $\beta$$=$$0.4$$\pm$$0.2$
($\beta$$=$$0.8$$\pm$$0.4$). Within the cluster cores, we derive
field-like values of $\mathcal{SFR}$ and $s\mathcal{SFR}$. Also, due to
the mentioned $\mathcal{SFR}_{\mathrm{TIR}}$ constraint we are not able to
explore whether the M-FIR samples behave in the same way as the $UVJ$-SF
samples. The M-FIR galaxies with suppressed SF are simply missed by the
selection function.

A number of works have also identified an offset between the
average $\mathcal{SFR}$ ($s\mathcal{SFR}$) in the clusters
and in the field (e.g., \citealt{2009ApJ...705L..67P},
\citealt{2010ApJ...710L...1V}, \citealt{2015ApJ...806..101H},
\citealt{2013ApJ...775..126H}, \citealt{2016ApJ...816L..25P}). Among
them, \citet{2014MNRAS.437..437A} find that blue cluster galaxies
($\mathcal{M}_{*}$$\geq$1.3$\times$10$^{10}$$M_{\odot}$) present
systematically lower average $s\mathcal{SFR}_{\mathrm{TIR}}$ up to
$z$$\sim$1.4. Their results, derived through a stacking analysis
on \textit{Herschel}/SPIRE 250$\mu$m imaging of 270 massive galaxy
clusters between $z$$\sim$0.3 and 1.5, quantify the average level
of SF of the whole star-forming cluster galaxy population, rather
than the typical rate of SF of FIR-detected galaxies. In fact, the
average $s\mathcal{SFR}$ they retrieve for clusters at $z$$\sim$0.5
and $z$$\sim$0.8 ($\sim$-9.70 and $\sim$-9.50, respectively) are
comparable with ours, as well as their 0.2-0.3~dex differences with
the field. This systematic suppression of the level of star-forming
activity within rich environments is created by the existence of a
numerous population of transition galaxies located in the lower part
of the well-studied MS of SFGs (e.g., \citealt{2016ApJ...816L..25P},
\citealt{2018MNRAS.473.5617C}).  Also, \citet{2013ApJ...775..126H} find
a 0.2\,dex suppression of the $s\mathcal{SFR}$ in SFGs 
with log$_{10}\mathcal{M}_{*}/M_{\odot}$$>$10 and 
$\mathcal{SFR}$$>$3\,$M_{\odot}$yr$^{-1}$
within $\mathcal{R}_{200}$ at 0.15$<$$z$$<$0.3. 

\textcolor{black}{ { If we now focus on the trend with $\mathcal{R}$ in the two last
redshift bins, we can see how the average $\mathcal{SFR}$ and
$s\mathcal{SFR}$ increase significantly for $UVJ$-SF galaxies at
0.2$<$$\mathcal{R}/R_{200}$$<$0.3, reaching field-like
values. This is probably due to the fact that we are reaching the region
slightly beyond 0.3$R_{200}$, where most of the prototypes of
galaxies violently interacting with the ICM are found (e.g., jellyfish
galaxies, \citealt{2016AJ....151...78P}; see \citealt{2006PASP..118..517B}
and references therein).  The average values of $\mathcal{SFR}$
for the cluster M-FIR remain overall compatible with the field
values. Instead, the $s\mathcal{SFR}$ depart from the field trend at
larger $\mathcal{R}$. However, limited number counts of this sample 
do not allow to extract robust conclusions about this sample.}}

\subsection{Star formation dependence on individual cluster properties: 
cool-core and BCG's star formation}

\begin{figure*}
\centering
\includegraphics[width=\linewidth]{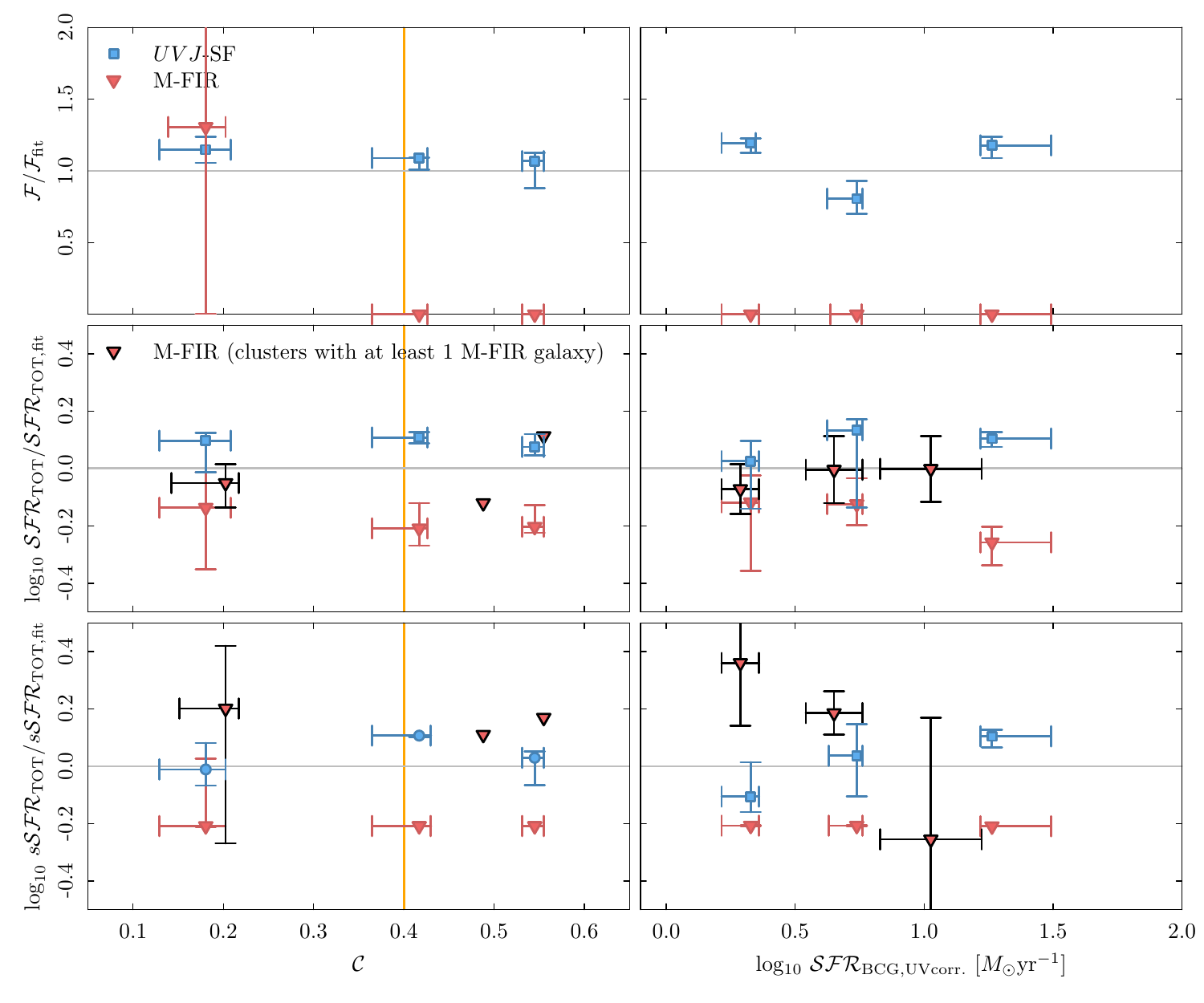}
\caption{Median values of the $\mathcal{F}$, $\mathcal{SFR}_{\mathrm{TOT}}$, 
and $s\mathcal{SFR}_{\mathrm{TOT}}$ (top, middle, and bottom
panel, respectively) normalized to the values predicted by the
cluster trends in Figure~\ref{fig:trends_f} and~\ref{fig:trends_sfr}  
at the corresponding
redshifts \textit{vs} the $\mathcal{C}$ coefficient given by
\protect\citet{2015ApJ...805..177D} (indicator of the presence of a CC;
left hand panels), and the $\mathcal{SFR}_{\mathrm{UV}}$ of the BCG (right
panels; \protect\citet{2015ApJ...805..177D}) corrected for extinction. 
The vertical yellow line on the left-hand panels represent
the value of $\mathcal{C}$ over which the CC are normally located
\protect\citet{2015ApJ...805..177D}. We present the averaged values in
three equally populated bins of each x-axis parameter. Values derived for
the $UVJ$-SF and M-FIR are shown with blue and red symbols, respectively. We
use circles to represent the results including all the clusters. In
the case of the clusters where no galaxy was selected in the M-FIR we
use an average $\mathcal{SFR}_{\mathrm{TOT}}$$=$10$M_{\odot}$yr$^{-1}$
(the $\mathcal{SFR}_{\mathrm{TIR}}$ limit of our study), and an average
log$_{10}s\mathcal{SFR}_{\mathrm{TOT}}$$=$$-9.5$. We use triangles to represent
the averages found using only clusters with obscured SF activity in
their core (at least 1 M-FIR detected galaxy).}\label{fig:properties}
\end{figure*}

In the previous subsections, we have analysed the SF properties
of $\mathcal{M}_{*}$-limited samples of star-forming cluster members
detected and undetected in the M- and/or FIR. Even though we are able
to identify a trend of the SF indices with redshift, the scatter in the
average properties is large. These cluster-to-cluster
variations have been observed frequently in the past, and some works have
attempted to quantify them (e.g., \citealt{2016ApJ...825...72A}). This
scatter is likely due to a combination of stochastic processes, such as
galaxy mergers (probably, the limited area covered by our study worsens
this effect), and differences in the properties of the clusters, such
as the dynamical state (e.g., \citealt{2015MNRAS.450..646S}). In this
section, we aim at exploring this latter. 

Despite the fact of being selected to be largely relaxed, there is
disagreement in the literature on the dynamical state of CLASH
sample members (see \citealt{2016MNRAS.460..569R} and references
therein). Given that we are focusing our study on the inner cores of
clusters, we use as a proxy of the dynamical state of these systems
the presence of a CC and the SF activity undergone by their BCGs.
\citet{2012ApJ...747...29R} found these observables to be strongly
correlated, which suggests that the SF activity of the BCGs is influenced
by the cluster-scale cooling process. In fact, star-forming BCGs seem
to be exclusively found in the centers of CC-clusters. However, the
separation between cool- and not-cool-core clusters is challenging. In
this work, we use as an indicator of the presence of this feature the
parameter $\mathcal{C}$, as defined by \citet{2016ApJ...819...36D},
which is a measure of the concentration of the X-ray emission.
More precisely, it gives the ratio between the light within a
circular aperture with a 100\,kpc radius and the total light enclosed
within a circular aperture with a 500\,kpc radius. For CC-clusters,
$\mathcal{C}$ values are likely $>$0.4 (\citealt{2016ApJ...819...36D}).
Among the 24 CLASH+HLS clusters, 12 qualify this criterion. As
we previously mentioned we find 8 M/FIR-emitter BCGs. Two of them
already identified by \citet[][the remaining 6 are not included in
their sample]{2012ApJ...747...29R}. Among the 8, 7 are characterized
by $\mathcal{C}$$>$0.4 ($\mathcal{C}_{\mathrm{AS1063}}$$=$0.19$\pm$0.03).
In turn, the formation of a CC appears also to be linked to the dynamical
states of the clusters, with relaxed clusters exhibiting more likely
CC than un-relaxed systems. Although some works have identified distant
clusters hosting a CC, their strength at $z$$>$0.7 appears significantly
lower due to the expected higher cluster merger rate and their more
immature evolutionary state (\citealt{2008A&A...483...35S}).

Figure~\ref{fig:properties} displays, for both the $UVJ$-SF and the
M-FIR samples ($\mathcal{R}/R_{200}$$<$0.1), the relation
between the three quantities we use to analyze the SF activity in
clusters (i.e., $\mathcal{F}$, $\mathcal{SFR}$, and $s\mathcal{SFR}$)
and both the parameter $\mathcal{C}$ and the $\mathcal{SFR}_{\mathrm{UV}}$
of the BCGs extinction corrected ($\mathcal{SFR}_{\mathrm{UV corr, BCG}}$)
provided by \citet{2015ApJ...805..177D}. In order to remove the global
trends with redshift of the average $\mathcal{F}$, $\mathcal{SFR}$
and $s\mathcal{SFR}$ that could have an impact on the results, we
remove them by normalizing these quantities to the values predicted by
the trends fitted for the clusters in the previous subsection at the
corresponding redshifts.
In each panel of Figure~\ref{fig:properties}, we show the median
in three bins of the corresponding x-axis parameter populated
by the 33\% of the clusters sample. Error bars represent the
confidence intervals derived through a bootstrap methodology. In
the case of the M-FIR samples, we show with highlighted triangles
(black border) the median values of the clusters which contain at
least 1 object. We use red triangles for the medians calculated
considering upper-limits $\mathcal{SFR}$$=$10$M_{\odot}$yr$^{-1}$
(our $\mathcal{SFR}_{\mathrm{TIR}}$ limit for the M-FIR samples) and
$s\mathcal{SFR}_{\mathrm{TOT}}$$=3\times10^{-10}$yr$^{-1}$ for those
clusters where no M-FIR galaxy is found.

\textcolor{black}{ {If we focus on the upper panels of Figure~\ref{fig:properties}, we
see that the bins of larger $\mathcal{C}$ are marginally dominated
by less SFGs. However, the large error bars corresponding to the
average of the M-FIR samples in the first $\mathcal{C}$ bin makes
the trend not significant for this subsamples. In the middle and
bottom panels, we do not find a clear correlation between the average
$\mathcal{SFR}$ or the $s\mathcal{SFR}$ and either $\mathcal{C}$ or
log$_{10}$$\mathcal{SFR}_{\mathrm{UV corr, BCG}}$.}}
 
\section{Discussion}\label{discussion}

It has long been claimed that galaxies quench more efficiently
in clusters than in the field (e.g., \citealt{1984ApJ...285..426B},
\citealt{2007MNRAS.376.1425G}, \citealt{2009ApJ...704..126H},
\citealt{2013ApJ...775..126H}). The general interpretation of this
suppression of SF is that environmental processes favour the removal
of gas reservoirs from galaxies. In fact, this average deficit of
gas in cluster members has been observationally confirmed in 
star-forming cluster spirals by, e.g., \citet{2013A&A...557A.103J}. In
agreement with this framework, our results clearly display a lack of SF
activity in massive cluster cores with respect the field at intermediate
redshifts in terms of both the fraction of SFGs and the rates at which
they form stars. 

The observed significant systematic $\sim$0.3~dex offsets between 
clusters and field average $\mathcal{SFR}$ and $s\mathcal{SFR}$ do not 
appear to be the result of differences in the SMF of the galaxy samples studied. 
Supporting this, \citet{2015MNRAS.450.2749G} find that 
galaxies of a given mass have different star formation histories depending on their environment, 
and therefore, it is not the distributions of galaxy masses in clusters 
the origin of the observed dependence of the SF with the environment.
Given that the population of star-forming galaxies within
massive clusters at the intermediate redshifts probed is thought to
be dominated by infalling field galaxies \citep{1995MNRAS.274..153K},
if the quenching of these galaxies were dominated by the same processes
that turn  galaxies off in the field (leading to the global SF decline
in the universe since $z$$\sim$1-2; \citealt{2014ARA&A..52..415M}) the 
fraction of SFGs should decrease proportionally in both environments 
(\citealt{2009ApJ...704..126H}). Given the different evolution with redshift 
we derive for $\mathcal{F}_{UVJ\mathrm{-SF}}$ in clusters and field, we can 
say that we are witnessing the imprint of the impact of environment on the 
evolution of cluster galaxies ($\mathcal{M}_{*}$$>$10$^{10}$$\mathrm{M}_{\odot}$).

Our results appear to support the observed evolution of the 
environmental quenching efficiency (\citealt{2008MNRAS.387...79V}, 
\citealt{2010ApJ...721..193P}, \citealt{2016MNRAS.456.4364B}), defined as the 
fraction of passive cluster galaxies which would be still star-forming if
they were in the field (\citealt{2017MNRAS.465L.104N}), with a major 
rise since $z$$\sim$2 (e.g., \citealt{1984ApJ...285..426B}, 
\citealt{2007MNRAS.376.1425G}, \citealt{2009ApJ...704..126H}, 
\citealt{2013ApJ...775..126H}, \citealt{2016ApJ...825...72A}).

It is straightforward to wonder what are the processes intrinsic
to high density environments that drive the aforementioned
galaxy transformation. Some of the most commonly invoked are:
\textit{strangulation} (\citealt{1980ApJ...237..692L}), which
consists on the removal of the loosely bound hot halo gas reservoirs
by the ICM on long time-scales ($>$1\,Gyr); the removal of the
interstellar medium through interactions with the ICM on moderate/short
time-scales ($\lesssim$1\,Gyr) RPS (\citealt{1972ApJ...176....1G},
\citealt{2016A&A...591A..51S}); either galaxy-galaxy or galaxy-cluster
gravitational interactions, grouped together under the name
\textit{harassment} (\citealt{1996Natur.379..613M}). The SFGs
infalling into high density environments at z$\lesssim$1 are very likely
influenced by a combination of these dynamical gas removal processes (see
\citealt{2006PASP..118..517B}, \citealt{2016ApJ...833..178V}). Merger
events are probably less frequent in cluster cores at these redshifts,
where the high relative velocities hamper reaching the fraction of
encounters observed in the field. However, there is growing evidence
(e.g., \citealt{2013ApJ...779..138B}, \citealt{2013ApJ...773..154L},
\citealt{2015MNRAS.447L..65S}, \citealt{2016ApJ...825...72A},
\citealt{2016MNRAS.456.4364B}) that at higher redshifts,
mergers play the major role in quenching infalling SFGs due to
high galaxy space densities and low relative velocities (e.g.,
\citealt{2011ApJ...732...33B}).

The small scatter ($\sim$0.3\,dex) found for the MS of
SFGs in field samples (e.g., \citealt{2007ApJ...660L..43N},
\citealt{2015ApJ...801L..29R}) is usually interpreted as the consequence
of a quenching mechanism that is capable of moving rapidly (0.1\,Gyr
time-scales) the galaxies out (downward) of the MS. For this reason,
the downward offset of the MS found in our work and in other
previous studies in clusters (e.g., \citealt{2013ApJ...775..126H},
\citealt{2016ApJ...816L..25P}) has frequently been interpreted
as the imprint of different environmentally-driven quenching
mechanisms that could turn off infalling SFGs slowly (e.g.,
\citealt{2013ApJ...775..126H}), thus, populating the region below the MS
with galaxies on their way to be quenched. The work by \citet{2015ApJ...806..101H},
based on the analysis of the actual orbits
of infalling galaxies in the 75 most massive clusters in the
Millennium Simulation \citep{2005Natur.435..629S} 
support the slow quenching scenario with
time-scales $\sim$ 0.7-2~Gyr. 
The most frequently proposed mechanism
for slow quenching in high density environments is strangulation. In
this evolving scenario, the decline in star formation is very likely due to
\textit{overconsumption} \citep{2014MNRAS.442L.105M}, 
the exhaustion of a gas reservoir through
star formation and expulsion via modest outflows in the absence of
cosmological accretion. \citet{2016A&A...590A.108M} also propose it
as the explanation for the higher metallicities found in the accreted
cluster galaxies of MACS0416. It has also been invoked to explain the
increasing distribution of SFGs with the projected cluster-centric radius
(e.g., \citealt{2016ApJ...825...72A}, \citealt{2015ApJ...806..101H}).

However, numerous studies have found observational evidence of
rapid quenching mechanisms, such as RPS, that can
remove the gas of an infalling galaxy in time-scales of the order
of the cluster crossing time ($\lesssim$1\,Gyr; 
e.g., \citealt{2013MNRAS.432..336W}), playing a significant role
building the populations of passive galaxies in clusters at different
redshifts. Also, some models of galaxy strangulation (e.g.,
\citealt{2016A&A...596A..11B} and references therein) and numerical
simulations (e.g., \citealt{2014MNRAS.442L.105M}) predict extremely
long time-scales in order to reproduce the observed lack of SF activity
in cluster members, while for instance \citet{2016A&A...596A..11B}
claim that only RPS is able to significantly quench
SF activity in galaxies perturbed by high density environments. The 
contribution of RPS in the core of clusters is thought 
to be important given the high relative velocities and higher densities 
of the ICM (e.g., \citealt{1972ApJ...176....1G}). However, this phenomenon 
operates efficiently for extreme cases of infall in which the orbital 
velocity is particularly high and the galaxy inclination is perpendicular 
to the direction of motion \citep{1999MNRAS.308..947A}. 
Furthermore, RPS can present a fluctuating behaviour which means that 
galaxies suffering from stripping can present
a wide range of properties, as observed by \citet{2016ApJ...833..178V}
and \citet{2017ApJ...837..126V}.

As an alternative to the slow/fast dichotomy frequently discussed,
\citet{2013MNRAS.432..336W} propose a delayed-then-rapid quenching
scenario, in which the satellites $\mathcal{SFR}$s evolve unaffected
for 2-4~Gyr after infall, and are eventually quenched rapidly, with an
e-folding time of $<$0.8~Gyr. This scenario has been frequently embraced 
to conciliate the observations of smaller fractions of SFGs in clusters and 
values of $\mathcal{SFR}$ comparable to those in the field at the same redshift.

In addition, \citet{2013MNRAS.432..336W} propose the quenching
time-scales do not depend on the halo mass. Interestingly,
they claim that up to half of quenched satellites in massive
clusters is the result of quenching in infalling groups, namely,
\textit{pre-processing}. Other authors have highlighted the importance
of this phenomenon to explain the properties of galaxy populations of
intermediate redshift clusters (e.g., \citealt{2015ApJ...806..101H},
\citealt{2015ApJ...812..153O}). The cluster-centric distances we probe
in this work ($\mathcal{R}/R_{200}$$<$0.3) do not allow the assessment
of pre-processing.

In this context, our results
favour slow quenching mechanisms (e.g., strangulation) to be dominating the
evolution of the observed $UVJ$-SF cluster core galaxies with
log$_{10}\mathcal{M}_{*}/M_{\odot}$$>$10 
throughout the last 8\,Gyr. This is because these samples 
appear to be heavily populated by transition galaxies observed while they 
quench (\citealt{2016ApJ...816L..25P}). However, we cannot rule out the
contribution of fast processes such as RPS to the enhanced fraction of
quenched galaxies observed. We also note that our methodology cannot
directly select galaxies quenching on short time-scales, such as PSB
(e.g., \citealt{2004ApJ...601..197P}, \citealt{2007ApJ...661..750T},
\citealt{2014ApJ...796...65M}, \citealt{2017ApJ...838..148P}), as this
would require spectral information, which we lack for more than half of
our clusters sample.

\section{Summary \& Conclusions}\label{summary}

We have presented a detailed analysis of the SF activity within 24 massive
clusters cores at 0.2$\lesssim$$z$$\lesssim$0.9 targeted by the HLS and CLASH surveys. 
The deep multi-wavelength photometric dataset
on these fields cover the whole rest-frame UV-to-FIR
regimes. In particular, we have made use of the CLASH catalogues, which
contain photometry measured on \textit{HST} ACS/WFC (F435W, F475W,
F606W, F625W, F775W, F814W, and F850LP), WFC3/UVIS (F225W, F275W,
F336W, and F390W), and WFC3/IR (F105W, F110W, F125W, F140W, and F160W)
imaging.  Then, we have combined these catalogues with others built on
\textit{Spitzer} IRAC (3.6, 4.5, 5.8, and 8.0 $\mu$m) and MIPS (24$\mu$m)
bands, and \textit{Herschel} PACS (100, and 160 $\mu$m) and SPIRE (250,
350, and 500 $\mu$m), deblending the former in the position of the CLASH
catalogues and selecting the most probable UV/optical counterpart for the
sources in the rest MIR and FIR bands. Finally, we have also gathered
the spectroscopic information available on these fields, mainly released
by CLASH-VLT and GLASS surveys. Consequently, we have derived high
quality photometric redshifts ($\sigma_{\mathrm{NMAD}}$$=$0.04, and 8\%
of outliers) fitting the UV-to-NIR photometry with the \textsc{EAZY}code. We have selected
cluster members by applying either a spectroscopic redshift criterion or
a probabilistic methodology that takes into account the whole information
included in the PDF of the photometric redshift estimation. We 
have used the $z_{\mathrm{phot}}$ derived and the \textsc{Rainbow}
Cosmological Database software package to fit, on the one hand, the
optical/NIR photometry (CLASH \& IRAC), and on the other hand, the FIR
photometry (MIPS \& \textit{Herschel}). In this way, we have estimated
the physical properties of the cluster members such as their $\mathcal{M}_{*}$ 
and the rates at which they form stars (as traced
by the UV and FIR emission independently). 
With the aim of building up analogous field samples with which compare the
results on clusters, we have applied the same analysis and selection
criteria on three CANDELS fields. Finally, we have used samples of SFGs
($\mathcal{M}_{*}$ $>$$10^{10}$$M_{\odot}$) selected using the UVJ-diagram
($UVJ$-SF samples) to evaluate and compare the SF processes in high density
environments and the field. Furthermore, we have used samples of galaxies
($\mathcal{M}_{*}$ $>$$10^{10}$$M_{\odot}$) detected in the MIR and/or
FIR with $\mathcal{SFR}_{\mathrm{TIR}}$$>$10$M_{\odot}$yr$^{-1}$ (M-FIR
samples) to explore the obscured SF activity. 
Taking advantage of the rich dataset available, we have based our 
results on the quantification
of the total SF, defined as either the sum of the SF traced by the
rest-frame UV emission and the FIR, or the un-obscured SF (traced only
by the rest-frame UV) corrected for the dust extinction with our own
optimized recipe.

The main results and conclusions of our work can be summarized 
in the following points: 

\begin{itemize}
\item The SF activity in the inner regions of intermediate-$z$ 
clusters appears to be suppressed in terms of both the
fraction of SFGs and the rate at which they turn gas into stars.

\item \textcolor{black}{ {We derive average fractions of $UVJ$-SF galaxies a factor $\sim$2
smaller in cluster ($\mathcal{R}/R_{200}$$<$0.1) than in the field 
across. The average fraction of M-FIR cluster 
members ($\mathcal{R}/R_{200}$$<$0.1) is negligible but compatible with 
a factor $\sim$2 smaller in clusters.}} 

\item We identify increasing trends of $\mathcal{F}_{UVJ\mathrm{-SF}}$ and 
$\mathcal{F}_{\mathrm{M-FIR}}$ with $z$, which evolve faster within clusters 
($\beta$$=$1.1$\pm$0.6 and $\beta$$=$7.3$\pm$5.8, respectively,
at $\mathcal{R}/R_{200}$$<$0.1) than in the field 
($\beta$$=$0.2$\pm$0.3 and $\beta$$=$0.2$\pm$0.5, respectively). 

\item $UVJ$-SF cluster members ($\mathcal{R}/R_{200}$$<$0.1) 
present $\mathcal{SFR}$ and $s\mathcal{SFR}$ typically $\sim$0.3\,dex smaller than $UVJ$-SF field galaxies. 
Average $\mathcal{SFR}$ and $s\mathcal{SFR}$ values evolve similarly (within the error bars) in 
clusters, with $\beta$$=$1.3$\pm$1.0 and $\beta$$=$1.2$\pm$0.9, respectively.
The evolution in the field is described by $\beta$$=$2.6$\pm$0.2 
and $\beta$$=$2.4$\pm$0.4, respectively. Due to the high $\mathcal{SFR}_{\mathrm{TIR}}$s completeness value given \textit{Spitzer}/MIPS\,24$\mu$m and \textit{Herschel} imaging used in this study, we can not explore 
whether is there a different trend between
field and clusters dusty SFGs in the average $\mathcal{SFR}$ and $s\mathcal{SFR}$.

\item We find increasing SF activity with cluster-centric distance out to
$\mathcal{R}/R_{200}$$=$0.3 in terms of the average $\mathcal{SFR}$
and $s\mathcal{SFR}$ of the $UVJ$-SF sample. No clear trend is found, however, for the fraction of SFGs.

\item We do not find an obvious relationship between SF activity in clusters and the presence of a CC or a BCG forming stars actively.
  
\end{itemize}

Our results evidence the impact of the cluster environment on
the evolution of its inhabitants and favour a dominant role of physical
processes quenching galaxies slowly. The mechanism typically invoked in
these cases is strangulation. This process appears to be responsible for
the shift of the average $\mathcal{SFR}$/$s\mathcal{SFR}$ exhibited by 
SFGs in high density environments since $z$$\sim$0.9, which is 
interpreted as the evidence of the existence of a large population of 
transition galaxies below the MS, on their way to be quenched. 
However, we can not rule 
out the impact of other processes occurring at shorter time-scales, 
such as RPS, which could be partially responsible for a fraction
of the SFGs missing in this clusters.

We release the multi-wavelength photometry, photometric redshifts, and
physical properties of the star-forming cluster members associated to
this paper through the \textsc{Rainbow} Cosmological Database.

\section*{Acknowledgements}

The authors thank Fran\c{c}oise Combes, 
Carlos L\'opez-Sanjuan, Dieter Lutz, Bianca
Poggianti and Alvio Renzini for their suggestions to improve
this work.  We acknowledge funding from the INAF PRIN-SKA 2017
program 1.05.01.88.04. L.R.-M. acknowledges funding support from
the Universit\`a degli studi di Padova - Dipartimento di Fisica e
Astronomia ``G. Galilei''. GR and CM acknowledge support from an INAF 
PRIN-SKA 2017 grant. P.G.P.-G. acknowledges funding support
from the Spanish Government MINECO under grants AYA2015-70815-ERC
and AYA2015- 63650-P. A.C.E. acknowledges support from STFC grant
ST/P00541/1. A.M. acknowledges funding from the INAF PRIN-SKA 2017
program 1.05.01.88.04. Finally, we thank the anonymous referee for the 
valuable comments and suggestions, which led to a substantial improvement 
of this paper. Analyses were performed in R 3.4.0 \citep{R}.




\bibliographystyle{mnras}
\bibliography{bibliography} 




\appendix
\begin{figure*}
\includegraphics[width=1\linewidth]{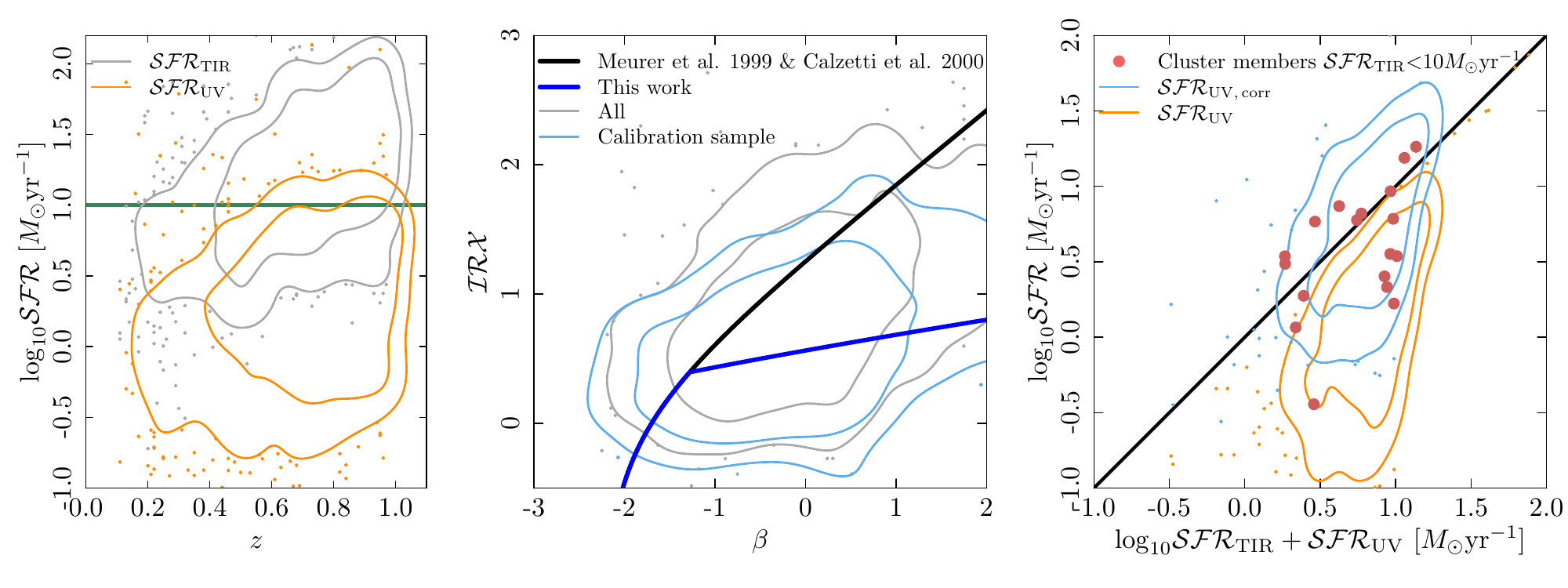}
\caption{\textit{Left panel}: $\mathcal{SFR}$$_{\mathrm{TIR}}$
(grey contours) and $\mathcal{SFR}$$_{\mathrm{UV}}$ (orange
contours) versus redshift for all the 1548 M/FIR-detected
galaxies in CANDELS fields with UVJ colours corresponding to SFGs,
log$_{10}\mathcal{M}_{*}/M_{\odot}$$>$10, and 0.1$<z<$1.0. The green
line represents the selection criteria for the selection of the
calibration sample. \textit{Central panel}: $\mathcal{IRX}$-$\beta$
relation for the galaxies in CANDELS fields with UVJ colours
corresponding to SFGs, log$_{10}\mathcal{M}_{*}/M_{\odot}$$>$10,
and 0.1$<z<$1.0 (grey contours), and the 525 galaxies with
$\mathcal{SFR}$$_{\mathrm{TIR}}$$<$10\,$M_{\odot}$yr$^{-1}$
of which the calibration sample is made of (blue contours). We
represent our calibration with a blue line. The black line is the
$\mathcal{IRX}$-$\beta$ fit from \protect\citet{1999ApJ...521...64M}
modified with a \citet{2000ApJ...533..682C} extinction law to the UV
wavelength we consider in our study (2800\AA). \textit{Right panel}:
Comparison between $\mathcal{SFR}_{\mathrm{TIR}}+\mathcal{SFR}_{\mathrm{UV}}$
and the $\mathcal{SFR}$$_{\mathrm{UV, corr.}}$ corrected for dust
extinction using our own calibration (Equation~\ref{eq:calibration},
blue contours). For comparison we show the distribution of values of
$\mathcal{SFR}$$_{UV}$ previous to the dust extinction correction (orange
contours). To evaluate the behaviour of our UV correction in the clusters,
we represent the comparison between the $\mathcal{SFR}$$_{\mathrm{TOT}}$
and the $\mathcal{SFR}$$_{\mathrm{UV, corr.}}$ of the cluster members with
$\mathcal{SFR}_{\mathrm{TIR}}$$<$10$M_{\odot}$yr$^{-1}$.\label{fig:corr}}
\end{figure*}

\section{Data available on the CANDELS fields}

In the following subsections we briefly enumerate the photometric and 
spectroscopic data on the CANDELS fields which is used in our analysis. 

\subsection{GOODS-S}\label{goodss}

We use the multi-wavelength catalogue on the CANDELS/GOODS-S field
published by \citet{2013ApJS..207...24G}, which combines the CANDELS
\textit{HST}/WFC3 F105W, F125W, and F160W bands with data from UV ($U$
band from both CTIO/MOSAIC and VLT/VIMOS), optical (\textit{HST}/ACS
F435W, F606W, F775W, F814W, and F850LP), and infrared (\textit{HST}/WFC3
F098M, VLT/ISAAC $Ks$, VLT/HAWK-I $Ks$, and \textit{Spitzer}/IRAC 3.6,
4.5, 5.8, 8.0\,$\mu$m) observations. The catalogue is based on source
detection in the WFC3 F160W band. Applying the methodology described
in Section~\ref{multi-wavelength-photo} we complement the catalogue with
MIR photometry in \textit{Spitzer}/MIPS 24\,$\mu$m and 70\,$\mu$m from
\citep{2008ApJ...675..234P} and FIR photometry from the GOODS-Herschel
\citep{2011A&A...533A.119E} and PEP \citep{2013A&A...553A.132M}
surveys, including PACS 100 and 160\,$\mu$m, and SPIRE 250, 350, and
500\,$\mu$m. The spectroscopic data are gathered from 
the VIMOS VLT deep survey (\citealt{2004A&A...428.1043L}), 
\citet{2004ApJS..155..271S}, the K20 survey (\citealt{2005A&A...437..883M}),
and other surveys such as those carried out by (e.g.) 
\citet{2008A&A...482...21C}, \citet{2008A&A...478...83V}. 
See \citet{2013ApJS..207...24G} for the details.

\subsection{GOODS-N}\label{goodsn}
The multi-wavelength catalogue used on CANDELS/GOODS-N is built and
described by Barro et al. (in prep.) and includes UV to far IR
and radio data.  In particular, UV data from GALEX (PI C. Martin),
ground-based optical data from $U$ to $z$ bands taken by the \textit{Kitt
Peak} telescope and from the \textit{Subaru}/Suprime-Cam as part of the
Hawaii Hubble Deep Field North project (\citealt{2004AJ....127..180C}),
25 medium-bands from the GTC SHARDS (\citealt{2013ApJ...762...46P})
survey, $J$, $H$, and $Ks$ imaging from the Subaru MOIRCS deep survey
(\citealt{2009ApJ...702.1393K}) and CFHT/WIRCam $Ks$ photometry (Lin in
prep.); IRAC 3.6, 4.5, 5.8, 8.0\,$\mu$m maps, maps from Spitzer-GOODS
(\citealt{2003sptz.prop..196D}), SEDS (\citealt{2013ApJS..209...22A}) and SCANDELS
(\citealt{2015ApJS..218...33A}); MIPS data from FIDEL (PI: M. Dickinson);
Herschel from the GOODS-Herschel \citep{2011A&A...533A.119E} and PEP
\citep{2013A&A...553A.132M} surveys, including PACS 100 and 160\,$\mu$m,
and SPIRE 250, 350, and 500\,$\mu$m. The spectroscopic redshifts used are 
a compilation based primarily on ACS-GOODS redshift survey 
(\citealt{2004AJ....127.3137C}; \citealt{2008ApJ...689..687B}), 
the Team Keck Redshift Survey (\citealt{2004AJ....127.3121W}), and 
the DEEP3 galaxy redshift survey (\citealt{2011ApJS..193...14C}).

\subsection{COSMOS}\label{cosmos}

We use the multi-wavelength catalogue on the CANDELS/COSMOS field
published by \citet{2017ApJS..228....7N}, which combines the CANDELS
\textit{HST}/WFC3 F105W, F125W, and F160W bands with data from
\textit{HST}/ACS F606W and F814W, CFHT/MegaPrime in the $u^{*}$, $g^{*}$,
$r^{*}$, $i^{*}$, and $z^{*}$ bands, from the \textit{Subaru}/Suprime-Cam
in the $B$, $g^{+}$, $V$, $r^{+}$, $i^{+}$ and $z^{+}$, along with
twelve intermediate and two narrow bands ($\sim$4000-8500\,\AA), from the
VLT/VISTA in the $Y$, $J$, $H$ and $Ks$ bands, \textit{Mayall}/NEWFIRM
$J_1$, $J_2$, $J_3$, $H_1$, $H_2$, $K$, and \textit{Spitzer}/IRAC 3.6,
4.5, 5.8, 8.0\,$\mu$m bands. Again, we combine this catalogue with MIR
photometry in \textit{ Spitzer}/MIPS 24\,$\mu$m and 70\,$\mu$m from
\citet{2007ApJS..172...86S} and FIR photometry including PACS 100 and
160\,$\mu$m from PEP program \citep{2011A&A...532A..90L}, and SPIRE 250,
350, and 500\,$\mu$m from HerMES \citep{2012MNRAS.424.1614O}.
Among the spectroscopic surveys gathered we highlight the 
VIMOS Ultra Deep Survey (\citealt{2015A&A...576A..79L}), zCOSMOS (PI: S. Lilly). 

\section{UV correction}\label{uv_cor}

The ratio of the $\mathcal{L}_{\mathrm{TIR}}$ to
$\mathcal{L}_{\mathrm{UV}}$, usually referred as $\mathcal{IRX}$, is
tightly related to the dust attenuation in a galaxy. This is because dust
absorbs and scatters mainly UV photons obscuring and 
reddening the galaxy SED at wavelengths $\lesssim$1$\mu$m. Then, it re-emits
the absorbed energy in the IR, at wavelengths $\sim$1-1000$\mu$m. Since
the work of \citet{1999ApJ...521...64M} on local starburst galaxies
(i.e., extreme SFGs), the relation between the $\mathcal{IRX}$ and the
slope of the UV ($\beta$) has been frequently used to estimate the UV
dust attenuation of galaxies. In practice, this relation is calibrated
for local blue galaxies for which FIR observations is available (e.g.,
\citealt{1997AJ....113..162C}, \citealt{1999ApJ...521...64M}) and then, it
is used to correct the UV luminosity from extinction up to high redshifts
(\citealt{1999ApJ...521...64M}). However, important deviations from
these relations have been observed (e.g.) for galaxies forming stars
at a lower rates or at different redshifts. Lately, different studies
have explored in detailed the physical origin of variations in the
$\mathcal{IRX}$-$\beta$ relation (e.g., \citealt{2017MNRAS.472.2315P}). In
this context, we aim at deriving an optimized dust attenuation correction
(i.e. $\mathcal{IRX}$-$\beta$ relation) that we can apply to those
star-forming cluster members fainter than our observational limits in
MIPS and/or Herschel, and therefore presumably less star-forming than
the starbursts on which the calibrations in the literature are defined.

Following a similar approach to \citet{2016MNRAS.457.3743D}, we basically
derive a $\mathcal{IRX}$-$\beta$ relation for a sample of SFGs which
are faint M/FIR emitters. In particular, we take advantage of the deep
coverage on CANDELS fields (GOODS and COSMOS) to select a subsample of
SFGs fainter than the CLASH+HLS fields observational limits in MIPS
and/or \textit{Herschel} bands. We only consider galaxies classified
as SFGs using an UVJ-diagram, located in the redshift range between
0.1 and 1.0, and with $\mathcal{M}_{*}/M_{\odot}$$>$10.
In Figure~\ref{fig:corr} (left panel) we display the
distribution with redshift of $\mathcal{SFR}$$_{\mathrm{TIR}}$
and $\mathcal{SFR}_{\mathrm{UV}}$ of these galaxies (obtained
following Equation~\ref{eq:SFR_TIR} and~\ref{eq:SFR_UV},
respectively). The calibration sample includes the 1548 galaxies with
$\mathcal{SFR}_{\mathrm{TIR}}$$<$10$M_{\odot}\mathrm{yr}^{-1}$ (green
horizontal line).

Once the sample is defined, we compute the UV slope for each galaxy using
a linear interpolation between 1500\,\AA\,and 2800\,\AA\,in the best-fit
templates given by \textsc{Rainbow} (Section~\ref{rainbow}). The typical
uncertainty in the $\beta$ values is $\sim$20\%.  Then, we compute their
$\mathcal{IRX}$ as the ratio of their $\mathcal{SFR}$$_{\mathrm{TIR}}$
and $\mathcal{SFR}$$_{\mathrm{UV}}$.  In Figure~\ref{fig:corr} (central
panel) we display the $\mathcal{IRX}$-$\beta$ space for the whole field
sample of M/FIR emitters ($\mathcal{M}_{*}/M_{\odot}$$>$10 
and 0.2$<$$z$$<$1.0; grey contours), and the calibration sample of
faint M/FIR emitters (blue contours). Then, we fit the points in
the $\mathcal{IRX}$-$\beta$ plane for our calibration sample with
a linear function. We derive the following best fit expression:
\begin{equation}\label{eq:calibration} \mathrm{A}_{\mathrm{UV}}
= (1.76 \pm 0.04) + (0.20 \pm 0.02) \beta 
\end{equation} 
Again, following the approach by \citet{2016MNRAS.457.3743D}, 
we apply the \citealt{1999ApJ...521...64M}
$\mathcal{IRX}$-$\beta$ relation ($A_{1600}$$=$4.43
+ 1.99$\beta$) for $\beta$ values lower than the point in which
our fit intercepts the relation by \citet{1999ApJ...521...64M}, 
$\beta$$=$-1.7, and Equation~\ref{eq:calibration} for higher $\beta$ values.

To assess the efficiency of our calibration, we quantify the
scatter of the difference between the $\mathcal{SFR}_{\mathrm{TOT}}$
derived as the addition of $\mathcal{SFR}_{\mathrm{TIR}}$ and
$\mathcal{SFR}_{\mathrm{UV}}$, and the $\mathcal{SFR}_{\mathrm{TOT}}$ computed
as the $\mathcal{SFR}_{\mathrm{UV}}$ corrected for dust extinction for our
calibration sample (right panel in Figure~\ref{fig:corr}).  The values
vary between -0.38 and 0.26~dex with a median of -0.02~dex. Using the
calibration by \citet{1999ApJ...521...64M} instead would have lead to a
median absolute deviation of 0.53~dex. Given that we use the calibration
built on field galaxies to correct also the $\mathcal{SFR}_{\mathrm{UV}}$
of the cluster members not detected in the M/FIR, we compare
how the calibration behaves for those faint M/FIR cluster members
($\mathcal{SFR}_{\mathrm{TIR}}$$<$10$M_{\odot}\mathrm{yr}^{-1}$).  In the
right panel of Figure~\ref{fig:corr}, we see that the dust extinction
correction behaves similarly in the field and the clusters. For the
latter, the median absolute deviation is -0.05~dex, and the differences
vary between -0.54 and 0.23~dex.

\section{catalogues}\label{cats}

\textcolor{black}{ {This appendix details the entries of the catalogues released.}}

\begin{table*}
	\centering
	\caption{Multiwavelength photometry}
	\label{tab:cat_phot}
	\begin{tabular}{ll} 
		\hline
		Entry name & Description \\
		\hline
		object               & ID of the source in the parent catalogue. This ID is not the CLASH catalogue ID. \\
		flux                 & [$\mu$Jy]\\
		err\_flux             & [$\mu$Jy]\\
		\hline
	\end{tabular}
\end{table*}

\begin{table*}
	\centering
	\caption{Flags for the MIPS counterpart identification.}
	\label{tab:cat_counterp}
	\begin{tabular}{ll} 
		\hline
		Entry name & Description \\
		\hline
		object               & ID of the source in the parent catalogue.\\
		MIPS\_n\_counterparts  & Total number of (selection band) counterparts candidates for the MIPS24 source.\\
		MIPS\_ID\_order      & ID of the MIPS24 counterpart flagged with the likelihood.\\
	                             & The most probable counterpart is flagged with a `\_1'. \\
		MIPS\_order           & The order of likelihood of being the right counterpart of the MIPS source.\\
		MIPS\_discriminator   & Quantity used to determine the counterpart likelihood order. \\
		MIPS\_fMIPS24 & MIPS24 flux [$\mu$Jy] used for the MIPS24 counterpart identification. \\
		MIPS\_err\_fMIPS24     & MIPS24 flux error [$\mu$Jy] used for the MIPS24 counterpart  identification.\\
		MIPS\_fIRAC80 & IRAC80 flux [$\mu$Jy] used for the MIPS24 counterpart identification.\\
		MIPS\_err\_fIRAC80     & IRAC80 flux error [$\mu$Jy] used for the MIPS24 counterpart identification.\\
		MIPS\_fIRAC36         & IRAC36 flux [$\mu$Jy] used for the MIPS24 counterpart identification.\\
		MIPS\_err\_fIRAC36     & IRAC36 flux error [$\mu$Jy] used for the MIPS24 counterpart identification.\\
		MIPS\_distance        & Distance between the MIPS24 source and the counterpart candidate.\\
		MIPS24\_snr\_cuts      & Flag regarding the SNR cuts applied in MIPS24: \\
					& 0 no-flux, 1 flux $>$ SNR limit, -1 flux $<$ SNR limit.\\
		n\_MIPS24\_psf0.25/0.5/1/2   & Number of sources in the parent catalogue. \\ 
		n\_MIPS24\_wcs0.25/0.5/1/2   & Number of sources in the parent catalogue. \\
		n\_MIPS\_MIPS24\_psf0.25/0.5/1/2   & Number of MIPS sources within the MIPS24 PSF. \\
		n\_MIPS\_MIPS24\_wcs0.25/0.5/1/2   & Number of MIPS sources within the MIPS24 WCS accuracy. \\
		n\_IRAC\_MIPS24\_psf0.25/0.5/1/2   & Number of IRAC sources within the MIPS24 PSF. \\
		n\_IRAC\_MIPS24\_wcs0.25/0.5/1/2   & Number of IRAC sources within the MIPS24 WCS accuracy. \\
		\hline
	\end{tabular}
\end{table*}

\begin{table*}
	\centering
	\caption{Flags for the PACS counterpart identification.}
	\label{tab:cat_counterp}
	\begin{tabular}{ll} 
		\hline
		Entry name & Description \\
		\hline
		object               &   ID of the source in the parent catalogue.\\
		PACS\_ID\_order        &   ID of the PACS counterpart flagged with the likelihood. \\
					& The most probable counterpart is flaged with a `\_1'.\\
		PACS\_discriminator   &   Quantity used to determine the counterpart likelihood order.\\
		PACS\_fPACS160        &   PACS160 flux [$\mu$Jy] used for the PACS counterpart identification.\\
		PACS\_err\_fPACS160    &   PACS160 flux error [$\mu$Jy] used for the PACS counterpart identification.\\
		PACS\_fPACS100        &   PACS100 flux [$\mu$Jy] used for the PACS counterpart identification.\\
		PACS\_err\_fPACS100    &   PACS100 flux error [$\mu$Jy] used for the PACS counterpart identification.\\
		PACS\_fMIPS24         &   MIPS24 flux [$\mu$Jy] used for the PACS counterpart identification.\\
		PACS\_err\_fMIPS24     &   MIPS24 flux error [$\mu$Jy] used for the PACS counterpart identification.\\
		PACS\_fIRAC80         &   IRAC80 flux [$\mu$Jy] used for the PACS counterpart identification.\\
		PACS\_err\_fIRAC80     &   IRAC80 flux error [$\mu$Jy] used for the PACS counterpart identification.\\
		PACS\_fIRAC36         &   IRAC36 flux [$\mu$Jy] used for the PACS counterpart identification.\\
		PACS\_err\_fIRAC36     &   IRAC36 flux error [$\mu$Jy] used for the PACS counterpart identification.\\
		PACS\_distance        &   Distance between the PACS and the counterpart candidate.\\
		PACS\_order           &   The order of likelihood of being the right counterpart of the PACS source.\\
		PACS\_n\_counterparts  &   Total number of counterparts candidates for the PACS source.\\
		PACS100\_snr\_cuts     &   Flag regarding the SNR cuts applied in PACS100: \\
					& 0 no-flux, 1 flux $>$ SNR limit, -1 flux $<$ SNR limit.\\
		PACS160\_snr\_cuts     &   Flag regarding the SNR cuts applied in PACS160: \\
					& 0 no-flux, 1 flux $>$ SNR limit, -1 flux $<$ SNR limit.\\
		n\_PACS100\_psf0.25/0.5/1/2             &   Number of sources in the parent catalogue within the PACS100 PSF.\\
		n\_PACS160\_psf0.25/0.5/1/2             &   Number of sources in the parent catalogue within the PACS160 PSF.\\
		n\_PACS100\_wcs0.25/0.5/1/2             &   Number of sources in the parent catalogue within the PACS100 WCS accuracy.\\
		n\_PACS160\_wcs0.25/0.5/1/2             &   Number of sources in the parent catalogue within the PACS160 WCS accuracy.\\
		n\_PACS\_PACS100\_psf0.25/0.5/1/2        &   Number of PACS sources within the PACS100 PSF.\\
		n\_PACS\_PACS160\_psf0.25/0.5/1/2        &   Number of PACS sources within the PACS160 PSF.\\
		n\_PACS\_PACS100\_wcs0.25/0.5/1/2        &   Number of PACS sources within the PACS100 WCS accuracy.\\
		n\_PACS\_PACS160\_wcs0.25/0.5/1/2        &   Number of PACS sources within the PACS160 WCS accuracy.\\
		n\_MIPS\_PACS100\_psf0.25/0.5/1/2        &   Number of MIPS sources within the PACS100 PSF.\\
		n\_MIPS\_PACS160\_psf0.25/0.5/1/2        &   Number of MIPS sources within the PACS160 PSF.\\
		n\_MIPS\_PACS100\_wcs0.25/0.5/1/2        &   Number of MIPS sources within the PACS100 WCS accuracy.\\
		n\_MIPS\_PACS160\_wcs0.25/0.5/1/2        &   Number of MIPS sources within the PACS160 WCS accuracy.\\
		n\_IRAC\_PACS100\_psf0.25/0.5/1/2        &   Number of IRAC sources within the PACS100 PSF.\\
		n\_IRAC\_PACS160\_psf0.25/0.5/1/2        &   Number of IRAC sources within the PACS160 PSF.\\
		n\_IRAC\_PACS100\_wcs0.25/0.5/1/2        &   Number of IRAC sources within the PACS100 WCS accuracy.\\
		n\_IRAC\_PACS160\_wcs0.25/0.5/1/2        &   Number of IRAC sources within the PACS160 WCS accuracy.\\
		\hline
	\end{tabular}
\end{table*}

\begin{table*}
	\centering
	\caption{Flags for the SPIRE counterpart identification.}
	\label{tab:cat_counterp}
	\begin{tabular}{ll} 
		\hline
		Entry name & Description \\
		\hline
		object               &   ID of the source in the parent catalogue.\\
		SPIRE\_ID\_order       &   ID of the SPIRE counterpart flagged with the likelihood. \\ 
					& The most probable counterpart is flaged with a `\_1'.\\
		SPIRE\_discriminator  &   Quantity used to determine the counterpart likelihood order.\\
		SPIRE\_fSPIRE500      &   SPIRE500 flux [$\mu$Jy] used for the SPIRE counterpart identification.\\
		SPIRE\_err\_fSPIRE500  &   SPIRE500 flux error [$\mu$Jy] used for the SPIRE counterpart identification.\\
		SPIRE\_fSPIRE350      &   SPIRE350 flux [$\mu$Jy] used for the SPIRE counterpart identification.\\
		SPIRE\_err\_fSPIRE350  &   SPIRE350 flux error [$\mu$Jy] used for the SPIRE counterpart identification.\\
		SPIRE\_fSPIRE250      &   SPIRE250 flux [$\mu$Jy] used for the SPIRE counterpart identification.\\
		SPIRE\_err\_fSPIRE250  &   SPIRE250 flux error [$\mu$Jy] used for the SPIRE counterpart identification.\\
		SPIRE\_fPACS160       &   PACS160 flux [$\mu$Jy] used for the SPIRE counterpart identification.\\
		SPIRE\_err\_fPACS160   &   PACS160 flux error [$\mu$Jy] used for the SPIRE counterpart identification.\\
		SPIRE\_fPACS100       &   PACS100 flux [$\mu$Jy] used for the SPIRE counterpart identification.\\
		SPIRE\_err\_fPACS100   &   PACS100 flux error [$\mu$Jy] used for the SPIRE counterpart identification.\\
		SPIRE\_fMIPS24        &   MIPS24 flux [$\mu$Jy] used for the SPIRE counterpart identification.\\
		SPIRE\_err\_fMIPS24    &   MIPS24 flux error [$\mu$Jy] used for the SPIRE counterpart identification.\\
		SPIRE\_fIRAC80        &   IRAC80 flux [$\mu$Jy] used for the SPIRE counterpart identification.\\
		SPIRE\_err\_fIRAC80    &   IRAC80 flux error [$\mu$Jy] used for the SPIRE counterpart identification.\\
		SPIRE\_fIRAC36        &   IRAC36 flux [$\mu$Jy] used for the SPIRE counterpart identification.\\
		SPIRE\_err\_fIRAC36    &   IRAC36 flux error [$\mu$Jy] used for the SPIRE counterpart identification.\\
		SPIRE\_distance       &   Distance between the SPIRE and the counterpart candidate.\\
		SPIRE\_order          &   The order of likelihood of being the right counterpart of the SPIRE source.\\
		SPIRE\_n\_counterparts &   Total number of counterparts candidates for the SPIRE source.\\
		SPIRE250\_snr\_cuts    &   Flag regarding the SNR cuts applied in SPIRE250: \\
					&  0 no-flux, 1 flux $>$ SNR limit, -1 flux $<$ SNR limit.\\
		SPIRE350\_snr\_cuts    &   Flag regarding the SNR cuts applied in SPIRE350: \\ 
					& 0 no-flux, 1 flux $>$ SNR limit, -1 flux $<$ SNR limit.\\
		SPIRE500\_snr\_cuts    &   Flag regarding the SNR cuts applied in SPIRE500: \\
					& 0 no-flux, 1 flux $>$ SNR limit, -1 flux $<$ SNR limit.\\
		n\_SPIRE250\_psf0.25/0.5/1/2            &   Number of sources in the parent catalogue within the SPIRE250 PSF.\\
		n\_SPIRE350\_psf0.25/0.5/1/2            &   Number of sources in the parent catalogue within the SPIRE350 PSF.\\
		n\_SPIRE500\_psf0.25/0.5/1/2            &   Number of sources in the parent catalogue within the SPIRE500 PSF.\\
		n\_SPIRE250\_wcs0.25/0.5/1/2             &   Number of sources in the parent catalogue within the SPIRE250 WCS accuracy.\\
		n\_SPIRE350\_wcs0.25/0.5/1/2             &   Number of sources in the parent catalogue within the SPIRE350 WCS accuracy.\\
		n\_SPIRE500\_wcs0.25/0.5/1/2             &   Number of sources in the parent catalogue within the SPIRE500 WCS accuracy.\\
		n\_SPIRE\_SPIRE250\_psf0.25/0.5/1/2      &   Number of SPIRE sources within the SPIRE250 PSF.\\
		n\_SPIRE\_SPIRE350\_psf0.25/0.5/1/2      &   Number of SPIRE sources within the SPIRE350 PSF.\\
		n\_SPIRE\_SPIRE500\_psf0.25/0.5/1/2      &   Number of SPIRE sources within the SPIRE500 PSF.\\
		n\_SPIRE\_SPIRE250\_wcs0.25/0.5/1/2      &   Number of SPIRE sources within the SPIRE250 WCS accuracy.\\
		n\_SPIRE\_SPIRE350\_wcs0.25/0.5/1/2      &   Number of SPIRE sources within the SPIRE350 WCS accuracy.\\
		n\_SPIRE\_SPIRE500\_wcs0.25/0.5/1/2      &   Number of SPIRE sources within the SPIRE500 WCS accuracy.\\
		n\_PACS\_SPIRE250\_psf0.25/0.5/1/2       &   Number of PACS sources within the SPIRE250 PSF.\\
		n\_PACS\_SPIRE350\_psf0.25/0.5/1/2       &   Number of PACS sources within the SPIRE350 PSF.\\
		n\_PACS\_SPIRE500\_psf0.25/0.5/1/2       &   Number of PACS sources within the SPIRE500 PSF.\\
		n\_PACS\_SPIRE250\_wcs0.25/0.5/1/2       &   Number of PACS sources within the SPIRE250 WCS accuracy.\\
		n\_PACS\_SPIRE350\_wcs0.25/0.5/1/2       &   Number of PACS sources within the SPIRE350 WCS accuracy.\\
		n\_PACS\_SPIRE500\_wcs0.25/0.5/1/2       &   Number of PACS sources within the SPIRE500 WCS accuracy.\\
		n\_MIPS\_SPIRE250\_psf0.25/0.5/1/2       &   Number of MIPS sources within the SPIRE250 PSF.\\
		n\_MIPS\_SPIRE350\_psf0.25/0.5/1/2       &   Number of MIPS sources within the SPIRE350 PSF.\\
		n\_MIPS\_SPIRE500\_psf0.25/0.5/1/2       &   Number of MIPS sources within the SPIRE500 PSF.\\
		n\_MIPS\_SPIRE250\_wcs0.25/0.5/1/2       &   Number of MIPS sources within the SPIRE250 WCS accuracy.\\
		n\_MIPS\_SPIRE350\_wcs0.25/0.5/1/2       &   Number of MIPS sources within the SPIRE350 WCS accuracy.\\
		n\_MIPS\_SPIRE500\_wcs0.25/0.5/1/2       &   Number of MIPS sources within the SPIRE500 WCS accuracy.\\
		n\_IRAC\_SPIRE250\_psf0.25/0.5/1/2       &   Number of IRAC sources within the SPIRE250 PSF.\\
		n\_IRAC\_SPIRE350\_psf0.25/0.5/1/2       &   Number of IRAC sources within the SPIRE350 PSF.\\
		n\_IRAC\_SPIRE500\_psf0.25/0.5/1/2       &   Number of IRAC sources within the SPIRE500 PSF.\\
		n\_IRAC\_SPIRE250\_wcs0.25/0.5/1/2       &   Number of IRAC sources within the SPIRE250 WCS accuracy.\\
		n\_IRAC\_SPIRE350\_wcs0.25/0.5/1/2       &   Number of IRAC sources within the SPIRE350 WCS accuracy.\\
		n\_IRAC\_SPIRE500\_wcs0.25/0.5/1/2       &   Number of IRAC sources within the SPIRE500 WCS accuracy.\\
		\hline
	\end{tabular}
\end{table*}

\begin{table}
	\centering
	\caption{Redshift and properties}
	\label{tab:cat_z}
	\begin{tabular}{ll} 
		\hline
		Entry name & Description \\
		\hline
		object         & ID of the galaxy in the parent catalogue. \\
		z\_phot        & \textsc{EAZY}$z_{\mathrm{phot}}$. \\
		z\_spec        & Spectroscopic redshift. \\
		flag           & Quality of the $z_{\mathrm{spec}}$. Values $>$2 mean reliable. \\
		stellar\_mass  & Stellar mass in $\mathrm{M}_{\odot}$. \\
		L\_TIR         & Total IR luminosity (8-1000$\mu$m) in L$_{\odot}$, from the best-fit template (\citealt{2007ApJ...657..810D}). \\
		SFR\_UV        & Star formation rate [$\mathrm{M}_{\odot}$yr$^{-1}$] from the rest-frame monochromatic luminosity at 2800\,\AA. \\
		SFR\_UV\_corr  & Star formation rate [$\mathrm{M}_{\odot}$yr$^{-1}$] from the rest-frame monochromatic luminosity at 2800\,\AA. \\
				& corrected by extinction using $\mathcal{A}_{\mathrm{UV}}$$=$(1.76$\pm$0.04)+(0.20$\pm$0.02)$\beta$. \\
		SFR\_TIR       & Star formation rate [$\mathrm{M}_{\odot}$yr$^{-1}$] from the L\_TIR. \\
		$U$            & Rest-frame $U$ absolute magnitude from best-fit template. \\
		$V$            & Rest-frame $V$ absolute magnitude from best-fit template. \\
		$J$            & Rest-frame $J$ absolute magnitude from best-fit template. \\
		\hline
	\end{tabular}
\end{table}


\bsp	
\label{lastpage}
\end{document}